\documentclass[%
 reprint,
superscriptaddress,
nofootinbib,
 amsmath,amssymb,
 aps,
 prx,
floatfix,
]{revtex4-2}

\usepackage{graphicx}
\usepackage{dcolumn}
\usepackage{bm}
\usepackage{bbm}
\usepackage{multirow}

\usepackage[english]{babel}
\usepackage{amsmath}
\usepackage{amsfonts}
\usepackage{amssymb}
\usepackage{mathtools}
\usepackage[utf8]{inputenc}
\usepackage{nicefrac}
\usepackage[normalem]{ulem}
\usepackage[svgnames]{xcolor}
\usepackage{booktabs}
\usepackage[percent]{overpic}

\definecolor{linkColor}{rgb}{0,0.3,0.7}
\usepackage[colorlinks=true,
            allcolors=linkColor,
            pdfborder={0 0 0},
            pdfencoding = auto
            ]{hyperref}
\usepackage[capitalise]{cleveref}
\usepackage{orcidlink}
\usepackage{siunitx}
\addto\captionsenglish{}

\renewcommand{\cite}{\citep}  
\usepackage{tikz}
\usepackage{pgfplots}
\pgfplotsset{compat=1.18}
\usetikzlibrary{patterns}
\usepgfplotslibrary{fillbetween}
\usetikzlibrary{patterns,arrows.meta}
\usepgfplotslibrary{groupplots}
\definecolor{mcrdBlue}{RGB}{102,153,204}

\renewcommand{\d}{\mathrm{d}}

\begin{document}

\title{Duality Between Chemical Potential Dynamics and Reaction–Diffusion Systems}

\author{Daniel Zhou}
\affiliation{Arnold Sommerfeld Center for Theoretical Physics and Center for NanoScience, Department of Physics, Ludwig-Maximilians-Universit\"at M\"unchen, Theresienstra\ss e 37, D-80333 M\"unchen, Germany}
\author{Erwin Frey}
\email{frey@lmu.de}
\affiliation{Arnold Sommerfeld Center for Theoretical Physics and Center for NanoScience, Department of Physics, Ludwig-Maximilians-Universit\"at M\"unchen, Theresienstra\ss e 37, D-80333 M\"unchen, Germany}
\affiliation{Max Planck School Matter to Life, Hofgartenstraße 8, D-80539 Munich, Germany}

\date{\today}
	
\begin{abstract} 
Pattern formation in soft, active, and biological matter is described
by two ostensibly distinct continuum frameworks: phase-field theories
driven by chemical-potential gradients, and mass-conserving
reaction--diffusion (McRD) dynamics governed by local interconversion
kinetics.
The two produce similar phenomenology, from coexistence and coarsening to
length-scale selection and traveling waves, yet existing correspondences
have remained at the level of analogy or confined to
linear and weakly nonlinear regimes. Here we establish a constructive,
equation-level duality between them valid in the nonlinear,
far-from-equilibrium regime. Mass-conserving reaction--diffusion is the
broader class: every chemical-potential field theory with conserved order
parameters embeds as the slow dynamics on an attracting manifold of an
McRD system; conversely,
every McRD with attractive nullcline admits an exact
chemical-potential representation in the fast-interconversion limit, with
the constitutive relation set by the reaction nullcline.
The construction resolves the generic non-invertibility of the
density--chemical-potential relation in phase-separating regimes by
embedding it as an attracting manifold in an extended two-field
description with conserved total density. Crucially, gradient stiffness is
faithfully represented: it maps onto an intrinsic reaction--diffusion
length set by the auxiliary field, yielding a diagonal-diffusion normal
form whose interface profile matches the original Cahn--Hilliard model by
construction.
The duality yields an
explicit dictionary for phase coexistence, with the Maxwell equal-area
construction exactly equivalent to the reactive turnover-balance condition
of McRD. It extends to weakly nonconservative dynamics---unifying
reaction-arrested coarsening and mesa splitting---and to multicomponent
theories with broken Maxwell symmetry.
As a concrete analytical payoff, the dual sharp-interface picture yields a
closed-form velocity law for traveling waves in nonreciprocal
Cahn--Hilliard dynamics, in quantitative agreement with simulations. More
broadly, free-energy
structure and reaction kinetics, gradient stiffness and intrinsic
reaction--diffusion length, Maxwell construction and turnover balance are
exactly equivalent rather than analogous, and each framework gives
analytical access to phenomena inaccessible from the other.
\end{abstract}
\maketitle

\section{Introduction}
\label{sec:introduction}

\paragraph*{Two frameworks, one phenomenology.---}
Mesoscale pattern formation in soft and biological matter is commonly described in two distinct continuum languages. 
One is the language of \emph{chemical-potential dynamics} and field theories: 
conservation laws and constitutive relations are encoded in chemical potentials and currents, as in time-dependent Ginzburg--Landau~\citep{Hohenberg_Halperin:1977} and Cahn--Hilliard~\citep{Bray:1994, Cahn_Hilliard:1958} dynamics, and in their active extensions where the underlying free-energy or reciprocity structure may fail~\citep{Marchetti:2013,Bowick_Marchetti:2022,Cates_Nardini:2025,Gompper:2020}. 
The other is the language of \emph{reaction--diffusion} systems, in which spatiotemporal organization emerges from local interconversion kinetics coupled by diffusion~\citep{Cross_Hohenberg:1993,Korolev_Nelson:2010, Frey:2010,Edelstein-Keshet.etal2013,Goryachev.Leda2017,Frey_Weyer:2026}. 
These frameworks are usually treated as conceptually disjoint. 
Chemical-potential dynamics comes with a thermodynamic vocabulary---chemical equilibrium, Maxwell constructions, and interfacial tension---whereas reaction--diffusion models are typically formulated as kinetic interconversion or population dynamics schemes without an underlying potential structure. 
Yet, across a wide range of systems, the two descriptions produce remarkably similar mesoscale scenarios: 
domain formation and phase separation, coarsening and its modification by turnover, length-scale selection, and persistent traveling or rotating patterns. 
The structural origin of this overlap, and the equation-level dictionary that would translate between the two languages, has remained out of reach.

\smallskip

\paragraph*{Mass-redistribution potential as a chemical-potential analogue.---}
On the reaction--diffusion side, a particularly important class is that of \emph{mass-conserving reaction--diffusion} (McRD) systems, in which two diffusive species locally interconvert while conserving the total density. 
These models provide minimal descriptions of intracellular protein pattern formation and related mass-redistribution instabilities
\citep{Ishihara.etal2007,Mori.etal2008,Otsuji.etal2007,Goryachev.Pokhilko2008,Halatek_Frey:2018,Brauns_Frey:2020,Frey_Weyer:2026}.
Their nonlinear dynamics and stationary states are organized by the \emph{mass-redistribution potential}, an emergent scalar field that plays the role of a chemical potential without being linked to any free-energy functional~\citep{Brauns_Frey:2020}. 
Together with the associated flux-balance geometry of nullclines and reaction manifolds, it takes over the structural role that thermodynamic potentials and Maxwell constructions play on the phase-field side.

\smallskip

\paragraph*{How thermodynamic structure is broken.---}
The equilibrium phase-field paradigm is a gradient flow in which chemical potentials are generated by a single free-energy functional and transport is governed by a symmetric, positive-semidefinite mobility matrix. 
Departures from this thermodynamic structure can occur in two logically independent ways at the deterministic level. 
First, the transport sector can violate Onsager reciprocity, meaning that cross-coupled fluxes are no longer described by a symmetric mobility matrix.
This breaks the gradient-flow structure: the dynamics no longer arises as descent on a single free-energy functional, although the symmetric (dissipative) part of the mobility may still admit a Lyapunov functional~\citep{Onsager:1931a,Onsager:1931b,Casimir:1945,DeGroot_Mazur:1962}.
Second, the constitutive sector can become non-variational, meaning that the chemical potentials cannot be obtained from any single scalar functional.
For purely local constitutive laws this corresponds to a failure of Maxwell-type integrability conditions~\citep{Callen:1985}, while for gradient-dependent field theories the appropriate generalization is the loss of self-adjointness of the linear response operator, or the emergence of a nonzero functional curl in field space \citep{Graham_Haken:1971,Risken:1972}.
Active phase-field models provide canonical realizations of these non-thermodynamic sectors: the failure of equilibrium structure can be traced either to non-symmetric transport coefficients or to explicitly non-variational contributions to the chemical potentials \citep{Cates_Tailleur:2015a,Wittkowski.etal2014,Tjhung_Cates:2018,Saha_Golestanian:2020,You_Marchetti:2020,Tailleur.Cates2008}. 
The recurring appearance of phase-separation phenomenology in McRD systems, and the parallel utility of reaction--diffusion intuition for active phase-field models, point to a structural relationship between the two languages rather than mere analogy.

\smallskip

\paragraph*{Precedents and their limitations.---}
There are several precedents connecting reaction--diffusion models and field-theoretical descriptions, but they typically apply only in restricted regimes and therefore do not provide a general constructive translation between the nonlinear partial differential equations of phase separation and those of mass-redistribution dynamics. 
An early deterministic example is the Schl\"ogl model~\citep{Schlogl:1972}, where bistable local kinetics combined with diffusion yields front dynamics closely related to Allen--Cahn relaxation of a nonconserved order parameter~\citep{Allen_Cahn:1979}. 
More generally, near pattern-forming instabilities, reaction--diffusion equations reduce to universal amplitude equations ---real or complex Ginzburg--Landau, Swift--Hohenberg, and, for systems with two conservation laws exhibiting a large-scale oscillatory instability, the nonreciprocal Cahn--Hilliard equation---establishing systematic but inherently local correspondences to field-theory normal forms~\citep{Frohoff_Thiele:2023}.
Within deterministic mass-conserving reaction--diffusion models, special subclasses of kinetics admit a Lyapunov functional and become isomorphic to dissipative Model~C dynamics~\citep{Morita_Ogawa:2010, Roth_Frey:2026}.
Taken together, these results establish important overlaps, but they rely on non-conserved order parameters, special kinetics, near-threshold reductions, pre-existing free-energy structures, or specific system classes.
They share a common ceiling: the correspondences hold at the linear or weakly nonlinear level, where shared dispersion relations place systems in the same Cross-Hohenberg class \citep{Cross_Hohenberg:1993} but the equivalence extends no further. They do not yield a global mapping that preserves the nonlinear, far-from-equilibrium steady states of phase separation and its nonequilibrium generalizations.
A key motivation for going beyond such special cases is that recent work on intracellular patterns has identified effective, foam-like phenomenology in McRD systems---including an emergent interfacial tension---despite their intrinsically nonequilibrium nature \citep{Weyer_Frey:2026}.
Such findings suggest that quantities that appear thermodynamic in phase-field theories (chemical equilibrium, Maxwell selection, tension-like coefficients) may have precise counterparts encoded in the dynamical geometry of mass redistribution.

\smallskip

\paragraph*{Constructive duality via manifold unfolding.---}
We show that two-component-per-species reaction--diffusion systems
form a broader class within which chemical-potential field theories of
Cahn--Hilliard form, namely those with passive square-gradient
interfacial terms, including their multicomponent and nonreciprocal
extensions, and with or without reactive turnover, embed constructively
as the slow dynamics on attracting manifolds. The construction respects
conservation structure: strictly mass-conserving Cahn--Hilliard dynamics
dualizes to a mass-conserving reaction--diffusion system, while
Cahn--Hilliard with weak reactive turnover dualizes to a
reaction--diffusion system with correspondingly broken conservation.
Conversely, every two-component-per-species reaction--diffusion system
with attractive nullcline, mass-conserving or weakly nonconservative,
admits in the fast-interconversion limit an exact chemical-potential
representation whose effective constitutive relation is read off from
the reaction nullcline. For networks with more components per species,
the construction establishes the existence of bulk chemical potentials,
while the corresponding gradient structure takes a more general form
that we do not address here. The duality holds beyond the linear and
weakly nonlinear regimes accessible to existing reductions, mapping the
nonlinear, far-from-equilibrium steady states of the two descriptions
onto each other.

The basic obstruction to a direct reformulation is that, in phase-separating regimes, the local constitutive relation between density and chemical potential is non-invertible, so that a single chemical potential corresponds to multiple admissible densities. We resolve this non-invertibility by \emph{unfolding} the constitutive relation into an extended two-field description. Specifically, we embed the chemical-potential curve as a one-dimensional submanifold in a two-dimensional state space and impose a fast local interconversion dynamics that makes this submanifold attracting. After a short transient that brings trajectories close to the manifold, the subsequent evolution proceeds near this constraint and the conserved sum of the two fields follows the original chemical-potential dynamics, up to controlled corrections set by the finite relaxation rate toward the manifold.
The interplay between fast interconversion and diffusion sets an intrinsic interfacial length, providing a kinetic origin for the gradient-stiffness coefficient of the phase-field description.
Crucially, the construction is not a change of variables. Away from the attracting manifold the extended system carries genuinely additional degrees of freedom, and different choices of off-manifold continuation yield distinct but equivalent dual representatives in the fast-interconversion limit. We exploit this freedom to select a particularly transparent normal form in which both fields behave as diffusing ``chemical species'' coupled by local conversion: a standard mass-conserving reaction--diffusion system.

\smallskip 

\paragraph*{A dictionary for coexistence and nonequilibrium extensions.---}
The duality yields an explicit dictionary between the two languages. Uniform chemical potential in the phase-field description corresponds to a uniform mass-redistribution potential in the dual reaction--diffusion description. Maxwell-type constructions that determine the chemical potential value at equilibrium reappear as an interfacial turnover-balance (solvability) condition, which fixes the unique value of the mass-redistribution potential instead~\citep{Brauns_Frey:2020}.
The framework extends beyond equilibrium gradient-flow dynamics: the construction does not require a free-energy functional. It applies to multicomponent theories and, in particular, to chemical-potential dynamics in which the equilibrium thermodynamic structure is broken --- either through nonvariational constitutive sectors or through nonreciprocal transport coefficients. In such regimes, the dual reaction--diffusion description is often the more economical starting point: the dynamics is naturally encoded in local interconversion kinetics, without recourse to a free-energy functional.
The dictionary is not merely interpretive but operational: it yields results that are out of reach from either description alone, including an analytical expression for the traveling-wave velocity of nonreciprocal Cahn--Hilliard systems beyond the harmonic regime.
Interfacial notions such as tension-like coefficients and curvature corrections then appear as consequences of the dictionary, rather than as its starting point.

\smallskip

\paragraph*{Outline.---}
Section~\ref{sec:duality-transformation_CH} constructs the dual reaction--diffusion system for Cahn--Hilliard dynamics by unfolding the constitutive relation $\mu(\phi)$ onto an attracting manifold, exploiting the off-manifold representational freedom to obtain a diagonal-diffusion normal form in which the square-gradient stiffness is traded for an intrinsic reaction--diffusion length; numerical simulations of spinodal decomposition and a phase-space view on the $(\phi,\eta)$ plane confirm quantitative agreement with Cahn--Hilliard and show that the surface tension is encoded geometrically as the offset between the nullcline and the flux-balance subspace. Section~\ref{sec:maxwell} establishes the coexistence dictionary (Table~\ref{tab:dictionary}): the Maxwell equal-area construction is exactly equivalent to a reactive turnover-balance condition, and the osmotic and Young--Laplace pressures of the Cahn--Hilliard description appear in the dual as geometric objects on the flux-balance subspace, without reference to any free-energy functional. Section~\ref{sec:broken_mass-conservation} extends the duality to weakly nonconservative dynamics and shows, via a sharp-interface analysis, that Cahn--Hilliard with reactive turnover and its non-mass-conserving reaction--diffusion dual share the same plateau-scale screened-Poisson problem, so that arrested coarsening and large-mesa splitting are mirrored one-to-one across the two descriptions. Section~\ref{sec:multi-component_systems} generalizes the construction to multicomponent theories with broken Maxwell symmetry or Onsager reciprocity, and uses the dual picture to derive, in quantitative agreement with simulations, a closed-form implicit velocity law for traveling waves in the nonreciprocal Cahn--Hilliard model. Section~\ref{ref:discussion_outlook} discusses scope and outlook. Technical reductions, the proof of the converse direction, and numerical details are collected in the Appendices.

\section{Constructive duality between Cahn--Hilliard and mass-conserving reaction--diffusion dynamics}
\label{sec:duality-transformation_CH}

In this section we construct a mass-conserving reaction--diffusion dual of Cahn--Hilliard dynamics.
The core idea is to ``unfold'' the chemical potential $\mu(\phi)$, whose non-invertibility in the phase-separating regime precludes a single-field reaction--diffusion formulation, by embedding it as a one-dimensional manifold in an extended two-field phase space and enforcing this manifold dynamically through fast interconversion kinetics.
The construction conserves the total density exactly, so that the slow dynamics remains purely transport-driven, while the auxiliary degree of freedom provides a fast internal relaxation mode whose competition with diffusion generates an intrinsic interfacial length scale.
We proceed in three steps: we first embed the equation of state and write a dual reaction--diffusion system that reduces to Cahn--Hilliard dynamics on the constitutive manifold (Sec.~\ref{sec:dual_system}); we then exploit the non-uniqueness of the dual away from this manifold to bring it into a normal form with diagonal diffusion (Sec.~\ref{sec:normal_form}); and finally we show how the explicit interface-stiffness term can be removed while preserving the interfacial profile---not merely its width (Sec.~\ref{sec:kappa_removal}).

\subsection{Construction of a dual reaction--diffusion system}
\label{sec:dual_system}

Consider a one-component system described by a conserved scalar field $\phi(\boldsymbol{x},t)$ evolving under Cahn--Hilliard dynamics.
The dynamics is a mass-conserving gradient flow driven by a Ginzburg--Landau free-energy functional,
\begin{equation}
    \mathcal{F}[\phi]
    = \int \mathrm{d}^d x\,
      \Big[f(\phi)+\frac{\kappa}{2}\lvert\nabla\phi\rvert^2\Big],
\end{equation}
where $f(\phi)$ is the local free-energy density and the square-gradient term penalizes spatial variations of~$\phi$ \citep{Cahn_Hilliard:1958}.
The stiffness parameter $\kappa>0$ controls the interfacial width and surface tension.  
The associated total chemical potential is
\begin{equation}
\label{eq:chemical_potential_CH_model}
    \mu_{\mathrm{tot}}[\phi]
    \equiv \frac{\delta\mathcal{F}}{\delta\phi}
    = \mu(\phi)-\kappa\,\nabla^2\phi,
\end{equation}
where $\mu(\phi)=f'(\phi)$ is the local chemical potential.  For
constant mobility~$M$ the conserved dynamics reads
\begin{equation}
\label{eq:CH_model}
    \partial_t\phi
    = M \,\nabla^2\mu(\phi)-M \kappa\,\nabla^4\phi.
\end{equation}

\paragraph*{Parametric embedding.---}
\begin{figure}[!t]
\centering
\begin{tikzpicture}
\begin{axis}[
    width=0.5\textwidth,
    height=0.45\textwidth,
    axis lines=middle,
    xlabel={$m$},
    ylabel={$\zeta$},
    xmin=-3.5, xmax=3.5,
    ymin=-1.8, ymax=1.8,
    ticks=none,
    clip=true,
]
\pgfmathsetmacro{\aa}{2.0}
\pgfmathsetmacro{\bb}{0.7}
\pgfmathsetmacro{\phimax}{2.0}
\pgfmathsetmacro{\phifold}{sqrt(\aa/(3*\bb))}
\pgfmathsetmacro{\mfold}{(\aa+1)*\phifold - \bb*\phifold^3}
\pgfmathsetmacro{\mufold}{-\aa*\phifold + \bb*\phifold^3}
\addplot[
  blue,
  opacity=0.35,
  line width=3.0pt,
  line cap=round
] coordinates {(0,-\mufold) (0,\mufold)};
\addplot[
    thick,
    domain=-\phimax:\phimax,
    samples=650,
]
({(\aa+1)*x - \bb*x^3},{-\aa*x + \bb*x^3});
\addplot[
    only marks,
    mark=*,
    mark size=1.7pt,
]
coordinates {
(\mfold,\mufold)
(-\mfold,-\mufold)
};
\addplot[dashed, thin] coordinates {(\mfold,\mufold) (0,\mufold)};
\addplot[dashed, thin] coordinates {(-\mfold,-\mufold) (0,-\mufold)};
\node[anchor=north east] at (axis cs:3.75,1.55)
{$\gamma(\phi)=(m(\phi),\mu(\phi)))$};
\node[anchor=north east] at (axis cs:-0.75,-1.2)
{$A(m,\zeta)=0$};
\end{axis}
\end{tikzpicture}
\caption{
\textbf{Parametric embedding} of the local equation of state
$\mu(\phi)$ into the extended $(m,\zeta)$-plane.  The curve
$\gamma(\phi)=(m(\phi),\mu(\phi))$ is shown for a standard $\phi^4$
chemical potential $\mu(\phi)=-r\,\phi+u\,\phi^3$ with auxiliary
coordinate $m(\phi)=\phi-\mu(\phi)$.  This embedding converts the
multivalued relation $\mu(\phi)$ into a single-valued geometric curve.
Filled circles mark the fold (spinodal) points where $\mu'(\phi)=0$;
dashed lines project them onto the $\zeta$-axis, and the highlighted
segment indicates the non-invertible portion of the equation of state.}
\label{fig:mu-m-embedding}
\end{figure}
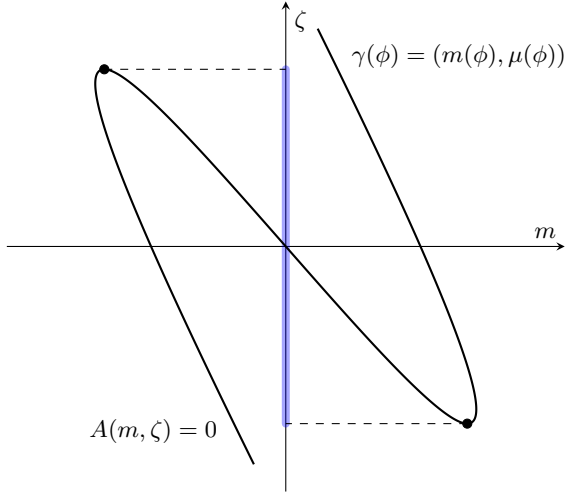

If $\mu(\phi)$ were globally invertible one could use $\mu$ as the dynamical variable and rewrite the dynamics as a (generically nonlinear) reaction--diffusion equation.  
For phase-separating systems, however, $f(\phi)$ is not everywhere convex, so $\mu(\phi)$ is multivalued and such an inversion fails.

To resolve this degeneracy we introduce an auxiliary coordinate $m$ and embed the graph of $\mu(\phi)$ as a regular curve
\begin{equation}
    \gamma(\phi)
    = \bigl(m(\phi),\,\mu(\phi)\bigr)
\end{equation}
in an extended $(m,\zeta)$-plane, where $\phi$ serves as the curve parameter (Fig.~\ref{fig:mu-m-embedding}).  
Regularity means that $\gamma$ is injective and $\gamma'(\phi)\neq(0,0)$, so that each point on the curve corresponds to a unique value of~$\phi$.  
Turning points of $\mu(\phi)$, where $\mu'(\phi)=0$, become fold points on~$\gamma$; their projection onto the $\zeta$-axis marks the spinodal interval (shaded in Fig.~\ref{fig:mu-m-embedding}).

At this stage $\phi$ is merely a parameter along $\gamma$ and carries no
meaning away from the curve.  To promote it to a scalar field on the
full $(m,\zeta)$-plane---a prerequisite for writing a conservation law
for~$\phi$---we define
\begin{equation}
    \phi(m,\zeta) \equiv \zeta + m,
\end{equation}
which by construction reduces to the original curve parameter
on~$\gamma$.  Evaluating on the curve, $\zeta=\mu(\phi)$, then fixes
the auxiliary coordinate as
\begin{equation}
    m(\phi) = \phi - \mu(\phi).\label{eq:naive_embed}
\end{equation}
This extension is \emph{not} unique; alternative choices lead to equivalent
embeddings.

\smallskip

\paragraph*{Constraint function and physical manifold.---}
The curve $\gamma$ defines the one-dimensional \emph{physical manifold}
in the extended $(m,\zeta)$-space: points on $\gamma$ are precisely
those compatible with the constitutive relation $\zeta=\mu(\phi)$.  We
represent this manifold as the zero level set of a constraint function
\begin{equation}
    A(m,\zeta) \equiv \zeta - \mu(\zeta+m),
\end{equation}
which vanishes identically on $\gamma$ and measures the departure from
the embedded equation of state elsewhere.

\smallskip

\paragraph*{Dual reaction--diffusion dynamics.---}
We now endow the embedding with dynamics by prescribing a
reaction--diffusion system for the extended fields $(m,\zeta)$.  The
total density $\phi=\zeta+m$ is conserved and evolves only by transport,
while the constraint violation $A(m,\zeta)$ acts as an interconversion
rate that transfers density between $\zeta$ and $m$ at fixed~$\phi$.  A
natural starting point is
\begin{subequations}
\label{eq:dual-system}
\begin{align}
    \partial_t\zeta
      &= M \,\nabla^2
         \bigl[\zeta-\kappa\,\nabla^2(\zeta+m)\bigr]
         \nonumber \\
         &\quad- D_m\,\nabla^2 m - A(m,\zeta), \\
    \partial_t m
      &= + D_m\,\nabla^2 m + A(m,\zeta),
\end{align}
\end{subequations}
where $D_m$ is the diffusion constant of the auxiliary field.  By
construction the reaction and diffusion terms cancel upon
summation, so that $\phi$ obeys the continuity equation
\begin{equation}
    \partial_t\phi
    = M \,\nabla^2
      \bigl[\zeta-\kappa\,\nabla^2\phi\bigr].
    \label{eq:phi_exact}
\end{equation}
This has the structure of Cahn--Hilliard dynamics but is not equivalent
to it, because $\zeta$ is an independent field rather than the
constitutive chemical potential~$\mu(\phi)$.  Equivalence is restored
only when the constraint $A=0$ is enforced, reducing $\zeta$ to
$\mu(\phi)$.

\smallskip

\paragraph*{Slow-manifold reduction.---}
To control the strength of this enforcement we parametrize
\begin{equation}
    A(m,\zeta)
    = \frac{1}{\tau}\bigl[\zeta-\mu(\zeta+m)\bigr],
    \qquad \tau>0,
\end{equation}
so that $\tau$ sets the relaxation time toward the constitutive
manifold.  In the regime $\tau \to 0$ the interconversion rapidly damps
departures from $A=0$, which becomes an attracting slow manifold: after
a fast transient of duration $\sim\tau$ one has
\begin{equation}
    \zeta = \mu(\phi) + \mathcal{O}(\tau).
    \label{eq:zeta-slaving}
\end{equation}
Substituting into Eq.~\eqref{eq:phi_exact} yields
\begin{equation}
    \partial_t\phi
    = M \,\nabla^2\mu(\phi)
      - M \kappa\,\nabla^4\phi
      + \mathcal{O}(\tau),
\end{equation}
recovering Cahn--Hilliard dynamics with controlled off-manifold
corrections.  For small but finite $\tau$, trajectories display a brief
fast transient toward $A=0$ followed by slow evolution near the
constitutive manifold; this transient can place the system on the same
long-time attractor as the corresponding Cahn--Hilliard dynamics, yet
generally at a different point, producing a spatial or temporal phase
shift even when the attractor set is shared.

Thus the dual construction embeds the original conserved dynamics into a
higher-dimensional reaction--diffusion model: the auxiliary field
provides a fast relaxation that makes $A=0$ an attracting slow manifold,
and the reduced flow on this manifold recovers Cahn--Hilliard dynamics.

\subsection{Reduction to a reaction--diffusion normal form}
\label{sec:normal_form}

The dual construction is not unique away from the constitutive manifold:
because only the \emph{sum} of the two component equations fixes the
$\phi$-dynamics, diffusion and gradient contributions can be
redistributed between them without changing the reduced slow-manifold
evolution.  We exploit this freedom to symmetrise the fourth-order term.
In the dual system, Eq.~\eqref{eq:dual-system}, the interfacial
contribution enters as $-M \kappa\nabla^4(\zeta+m)$ in the
$\zeta$-equation alone.  Adding $+M \kappa\nabla^4 m$ to the
$\zeta$-equation and subtracting the same term from the $m$-equation
preserves the surface-tension contribution $-M \kappa\nabla^4\phi$ in
the $\phi$-equation while distributing the fourth-order derivative
equally between the two fields.  The rearrangement only affects the fast
relaxation transverse to the constraint manifold; to leading order in the
stiff limit one still has $\zeta=\mu(\phi)$, so the slow evolution of
$\phi$ coincides with Cahn--Hilliard dynamics.

This rearrangement still leaves a cross-diffusion term
$-D_m\nabla^2 m$ in the $\zeta$-equation, preventing us from
interpreting $\zeta$ and $m$ as concentrations governed by a diagonal
diffusion matrix.  To remove this coupling, we make use of the non-uniqueness of the embedding. Instead of the naive ansatz, Eq.~\eqref{eq:naive_embed}, we construct an auxiliary reaction $\hat{A}$ that both incorporates the thermodynamic information, as well as let the resulting system be diagonalized.
\footnote{Note that in general, this choice and the demand that $c+m$ be conserved fixes the equations. However, there is at least one more physically interpretable choice for both, namely enforcing that $\zeta$ plays the role of the mass-redistribution potential. We will not take this route here.}
We rename $\zeta\to c$ to stress its role as a concentration field and make the ansatz
\begin{subequations}
\label{eq2:normal-form}
\begin{align}
    \partial_t c
      &= D_c\,\nabla^2 c
         - M \kappa\,\nabla^4 c
         - \hat{A}(c,m), \\
    \partial_t m
      &= D_m\,\nabla^2 m
         - M \kappa\,\nabla^4 m
         + \hat{A}(c,m),
\end{align}
\end{subequations}
where $D_m$ and $D_c$ are positive diffusion constants and the
modified reaction term is
\begin{equation}
    \hat{A}(c,m)
    = \frac{1}{\tau}\,\Bigl[c - \hat{\mu}(c+m)\Bigr].
    \label{eq:hat_A}
\end{equation}
The function $\hat{\mu}$ and the constant $D_c$ are still free; we now
fix them by requiring that $\phi$ obeys Cahn--Hilliard dynamics on the
slow manifold.  
To this end, we sum Eqs.~\eqref{eq2:normal-form} to
obtain the evolution of the conserved total field
\begin{equation}
    \partial_t\phi
    = D_c\,\nabla^2 c + D_m\,\nabla^2 m
      - M \kappa\,\nabla^4\phi.
\end{equation}
On the slow manifold $\hat{A}=0$ we have $c=\hat{\mu}(\phi)$ and
$m=\phi-\hat{\mu}(\phi)$, so that eliminating $(c,m)$ in favour of
$\phi$ gives
\begin{equation}
    \partial_t\phi
    = D_m\,\nabla^2\phi
      + (D_c-D_m)\,\nabla^2\hat{\mu}(\phi)
      - M \kappa\,\nabla^4\phi.
    \label{eq2:phi-reduced}
\end{equation}
For this to coincide with the Cahn--Hilliard dynamics
(Eq.~\ref{eq:CH_model}), two conditions must hold.  First, the contrast
in diffusion constants must equal the mobility,
\begin{equation}
    D_c - D_m = M .
\end{equation}
Second, the modified chemical potential must absorb the residual
$D_m\,\nabla^2\phi$ term, which requires
\begin{equation}
    \hat{\mu}(\phi)
    = \mu(\phi)
      - \frac{D_m}{M }\,(\phi - \phi_{\mathrm{eq}}),
    \label{eq:def-of-mu-hat}
\end{equation}
where $\phi_{\mathrm{eq}}$ is an arbitrary reference value, since only
gradients of $\hat{\mu}$ enter.  With these choices the extra
linear-diffusion term cancels identically, and the reduced
$\phi$-dynamics on the slow manifold reproduces Cahn--Hilliard exactly.

Taken together, Eqs.~\eqref{eq2:normal-form} constitute a genuine
two-species reaction--diffusion system---the \emph{normal form} of the
original model---whose reaction nullcline $c=\hat{\mu}(\phi)$ encodes
the thermodynamic equilibrium condition.  However, the fourth-order
terms $-M \kappa\nabla^4 c$ and $-M \kappa\nabla^4 m$, which are
essential for the slow-manifold reduction to Cahn--Hilliard dynamics,
prevent a direct interpretation of
Eqs.~\eqref{eq2:normal-form} as standard mass-conserving 
reaction--diffusion equations with purely diffusive transport. 
In the next subsection we show how these terms can be sent to zero while retaining an interface profile---not just the width---identical to that of the underlying
Cahn--Hilliard model.

\subsection{Reaction--diffusion dual and the role of $\kappa$}
\label{sec:kappa_removal}

In the Cahn--Hilliard model the square-gradient penalty sets the
interfacial width, scaling as ${\ell_\kappa\sim\sqrt{\kappa}}$;
taking ${\kappa\to0}$ therefore makes interfaces sharp and the continuum
dynamics singular.  In the dual system, however, the auxiliary field $m$
provides an alternative source of regularisation: $m$ diffuses with
coefficient $D_m$ and relaxes by interconversion on the time scale
$\tau$, and the competition of these two processes introduces an
intrinsic reaction--diffusion length
\begin{equation}
    \ell_m \equiv \sqrt{D_m\,\tau}.
\end{equation}
The strategy is therefore to set ${\kappa=0}$ in the dual equations and
treat $m$ as a fast variable slaved to the slowly evolving conserved
field~$\phi$.  Eliminating $m$ in the fast-reaction regime will generate
an \emph{effective} fourth-order contribution in the reduced description
for $\phi$, even though no such term appears explicitly in the dual
equations; we then choose $D_m$ and $\tau$ so that its coefficient
matches the interfacial stiffness of the original Cahn--Hilliard
dynamics.  In this way, the explicit gradient regularisation is traded
for an intrinsic length scale generated by the auxiliary
reaction--diffusion sector.

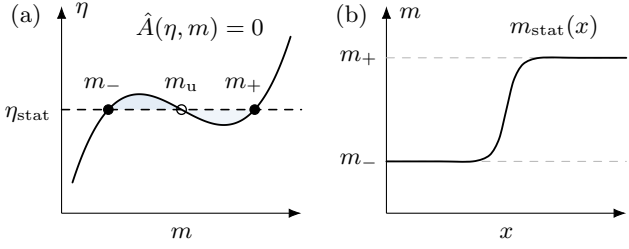
\begin{figure}[!t]
\centering

{\begin{tikzpicture}

\begin{axis}[
  xshift=-0.03\columnwidth,
  name=left,
  width=0.55\columnwidth,
  height=0.5\columnwidth, 
  axis lines=left,
  axis line style={-Latex, line width=0.4pt},
  every axis plot/.append style={line cap=round},
  xlabel={$m$},
  ylabel={},
  xtick=\empty, ytick=\empty,
  enlargelimits=false,
  clip=false,
  xmin=-2.2, xmax=2.2,
  ymin=0.0,  ymax=1.0
]

\addplot[
  line width=0.7pt,
  smooth,
  samples=400,
  domain=-2:2,
  name path=nullcline
] {0.50 + 0.08*(x^3 - 1.8*x)};
\node[anchor=west, font=\small] at (axis cs:-1.0,0.9) {$\hat A (\eta,m)=0$};

\addplot[
  dashed,
  line width=0.6pt,
  domain=-2.2:2.2,
  name path=fbs
] {0.50};
\node[anchor=south west, font=\small] at (axis cs:-3.4,0.4) {$\eta_\mathrm{stat}$};

\addplot[draw=none, fill=mcrdBlue, fill opacity=0.18] fill between[
  of=nullcline and fbs,
  soft clip={domain=-1.342:0}
];
\addplot[draw=none, fill=mcrdBlue, fill opacity=0.10] fill between[
  of=fbs and nullcline,
  soft clip={domain=0:1.342}
];

\addplot[only marks, mark=*, mark size=1.8pt]
coordinates {(-1.342,0.50) (1.342,0.50)};
\node[anchor=south, font=\small] at (axis cs:-1.442,0.54) {$m_-$};
\node[anchor=south, font=\small] at (axis cs: 1.142,0.54) {$m_+$};

\addplot[only marks, mark=o, mark size=1.8pt, line width=0.5pt]
coordinates {(0,0.50)};
\node[anchor=south, font=\small] at (axis cs:0,0.55) {$m_\mathrm{u}$};

\node[anchor=south, font=\small] at (axis description cs:+0.085,0.9) {$\eta$};

\node[anchor=north west, font=\small] at (axis description cs:-0.25,1.05) {(a)};

\end{axis}

\begin{axis}[
  at={(left.east)},
  anchor=west,
  xshift=0.1\columnwidth,
  width=0.55\columnwidth,
  height=0.5\columnwidth, 
  axis lines=left,
  axis line style={-Latex, line width=0.4pt},
  every axis plot/.append style={line cap=round},
  xlabel={$x$},
  ylabel={},
  xtick=\empty, ytick=\empty,
  enlargelimits=false,
  clip=false,
  xmin=0, xmax=1,
  ymin=0.0, ymax=1.0
]

\addplot[gray!55, line width=0.4pt, dashed] coordinates {(0,0.25) (1,0.25)};
\addplot[gray!55, line width=0.4pt, dashed] coordinates {(0,0.75) (1,0.75)};
\node[anchor=west, font=\small] at (axis cs:-0.23,0.25) {$m_-$};
\node[anchor=west, font=\small] at (axis cs:-0.23,0.75) {$m_+$};

\addplot[
  line width=0.7pt,
  smooth
] coordinates {
  (0.00,0.25)
  (0.22,0.25)
  (0.36,0.25)
  (0.43,0.28)
  (0.47,0.36)
  (0.50,0.50)
  (0.53,0.64)
  (0.57,0.72)
  (0.64,0.75)
  (0.80,0.75)
  (1.00,0.75)
};
\node[anchor=south, font=\small] at (axis cs:0.70,0.80) {$m_\mathrm{stat}(x)$};

\node[anchor=south, font=\small] at (axis description cs:+0.1,0.9) {$m$};

\node[anchor=north west, font=\small] at (axis description cs:-0.25,1.05) {(b)};

\end{axis}

\end{tikzpicture}}

\caption{\textbf{Schematic construction of a stationary mesa.}
(a) In $(\eta,m)$ phase space, the flux-balance subspace $\eta=\eta_\mathrm{stat}$ intersects the N-shaped nullcline $\hat A (\eta,m)=0$ at the plateau concentrations $m_\pm$; the middle intersection $m_\mathrm{u}$ lies on the unstable branch. Shaded regions indicate the signed contributions along $\eta=\eta_\mathrm{stat}$ entering the turnover-balance construction.
(b) Corresponding stationary profile $m_\mathrm{stat}(x)$ connecting $m_-$ (left) to $m_+$ (right) under no-flux boundary conditions.}
\label{fig:fbs}
\end{figure}

Setting $\kappa=0$ in the normal-form
equations~\eqref{eq2:normal-form} gives
\begin{subequations}
\label{eq2:dual-kappa-free}
\begin{align}
    \partial_t c
      &= D_c\,\nabla^2 c - \hat{A}(c,m),
      \label{eq2:dual-kappa-free_c} \\
    \partial_t m
      &= D_m\,\nabla^2 m + \hat{A}(c,m).
      \label{eq2:dual-kappa-free_m}
\end{align}
\end{subequations}
These are now standard mass-conserving reaction--diffusion equations
with purely diffusive transport.  Summing the two equations yields the
evolution of the total density $\phi=c+m$,
\begin{equation}
    \partial_t\phi = D_c\,\nabla^2 c + D_m\,\nabla^2 m.
    \label{eq2:phi-dyn-dual}
\end{equation}
Introducing the mass-redistribution potential
\begin{equation}
    \eta \equiv c + d\,m,
    \qquad \mathrm{with} \quad d\equiv \frac{D_m}{D_c},
\end{equation}
this becomes $\partial_t\phi=D_c\,\nabla^2\eta$.  For time-independent
steady states with no-flux boundaries, $\eta$ is therefore constant on
each connected component, and steady states lie on the \emph{flux-balance
subspace} (FBS) $\eta=\eta_{\mathrm{stat}}$; see Fig.~\ref{fig:fbs}.

The connection between this system and the original Cahn--Hilliard model becomes transparent upon rewriting  reaction term in terms of $\eta$ and $\phi$.  
Using $c=(\eta-d\,\phi)/(1-d)$ one finds
\begin{equation}
    \hat{A}
    = \frac{D_c}{\tau M}\,
      \bigl[\eta - \eta^*(\phi)\bigr],
    \label{eq:Ahat_eta}
\end{equation}
where
\begin{equation}
    \eta^*(\phi)
    = d\,\phi_{\mathrm{eq}} + (1-d)\,\mu(\phi)
\end{equation}
is the nullcline value of $\eta$ at given $\phi$.  
The reaction therefore drives $\eta$ toward $\eta^*(\phi)$: on the
nullcline $\hat{A}=0$, i.e., $\eta=\eta^*(\phi)$, local chemical
equilibrium holds; on the flux-balance subspace
$\eta=\eta_{\mathrm{stat}}$, by contrast, the diffusive fluxes of $c$ and
$m$ balance, and $\hat{A}$ does not vanish locally.
A related analysis was given in Ref.~\citep{Robinson_Speck:2025}, who
linearize about the intersection of $\eta$ with the nullcline and obtain
a description local to the binodal densities. That treatment does not
identify the flux-balance subspace $\eta=\eta_{\mathrm{stat}}$ as a
global organizing manifold. The global constraint the flux-balance subspace provides on the densities is central to the construction here: every
steady state satisfies $\eta=\eta_{\mathrm{stat}}$ exactly, and the full
nullcline $\eta^*(\phi)=d\,\phi_{\mathrm{eq}}+(1-d)\,\mu(\phi)$ encodes
$\mu(\phi)$ across the entire state space---including the spinodal
interior---which is what makes the dual interfacial profile coincide
with the Cahn--Hilliard one by construction.

\smallskip

\paragraph*{Steady-state matching.---}
We now show that eliminating $m$ from the steady-state equations using
the flux-balance constraint generates an effective fourth-order term
whose coefficient can be matched to the Cahn--Hilliard stiffness.
Substituting $c=\phi-m$ into Eq.~\eqref{eq2:phi-dyn-dual} and imposing
$\partial_t\phi=0$ gives
\begin{equation}
    \nabla^2 m = \frac{D_c}{M }\,\nabla^2\phi.
    \label{eq:lapm-fb}
\end{equation}
Applying $\tau\,\nabla^2$ to the steady-state form of the
$m$-equation~\eqref{eq2:dual-kappa-free_m} and using the relation
between $\hat{\mu}$ and $\mu$, Eq.~\eqref{eq:def-of-mu-hat}, yields
\begin{equation}
    0 = D_m\tau\,\nabla^4 m
      + \frac{D_c}{M }\,\nabla^2\phi
      - \nabla^2 m
      - \nabla^2\mu(\phi).
    \label{eq:lap-profile-shifted}
\end{equation}
The second and third terms cancel by the flux-balance
identity~\eqref{eq:lapm-fb}, and the same identity converts $\nabla^4 m$
into a fourth-order gradient in $\phi$, leaving
\begin{equation}
    0 = M \,\nabla^2\mu(\phi)
      - \tau D_m D_c\,\nabla^4\phi.
    \label{eq:phi-eff}
\end{equation}
Comparing with the steady-state Cahn--Hilliard equation,
\begin{equation}
    0 = M \,\nabla^2\mu(\phi) - M \kappa\,\nabla^4\phi,
    \label{eq:modelB-steady}
\end{equation}
identifies the \emph{exact matching condition}
\begin{equation}
    \kappa
    = \tau\,\frac{D_m\,D_c}{M }
    = \frac{D_m\,\tau}{1-d}.
    \label{eq:kappa-match}
\end{equation}

\paragraph*{Scaling in the fast relaxation limit.---}
Since both $\kappa$ and $M$ must remain finite as $\tau\to0$, the
matching condition~\eqref{eq:kappa-match} requires
\begin{equation}
    D_m \sim D_c \sim \tau^{-1/2},
    \qquad
    1-d \sim \sqrt{\tau}.
\end{equation}
Reactive turnover is therefore faster than diffusion by a factor $\tau^{-1/2}$,
while $D_m$ approaches $D_c$ from below,
${D_m/D_c = 1-\mathcal{O}(\sqrt{\tau})}$.
In this regime, the reaction rate on the quasi-steady manifold is
large: from the $m$-equation on time scales much longer than $\tau$
one has $\hat{A}\approx -D_m\nabla^2 m = O(\tau^{-1/2})$, so that
reaction and diffusion remain in active, spatially varying balance
across the interface.  It is precisely this balance---the local
deviation of $\eta$ from the nullcline $\eta^*(\phi)$ in
Eq.~\eqref{eq:Ahat_eta}---that generates the emergent $\nabla^4\phi$
term regularising the interface.

\smallskip

\paragraph*{Full dynamics.---}
The steady-state matching carries over to time-dependent states.  In the
fast-reaction regime the auxiliary field $m$ rapidly approaches a
quasi-steady manifold determined by the slowly evolving conserved
density~$\phi$, and eliminating $m$ generates the same effective
fourth-order regularisation in the $\phi$-equation.  More precisely,
adiabatic elimination of $m$ (see App.~\ref{app:duality_McRD_CH})
shows that the reduced $\phi$-dynamics takes the implicit form
\begin{equation}
    \bigl(1-\ell^2\,\nabla^2\bigr)\,\partial_t\phi
    = \mathcal{L}[\phi] + O(\tau),
    \label{eq:dyn_reduction_maintext}
\end{equation}
where
$\mathcal{L}[\phi]\equiv M\,\nabla^2\mu(\phi)-M\kappa\,\nabla^4\phi$
is the Cahn--Hilliard operator and
\begin{equation}
    \ell^2 \equiv (D_m + D_c)\,\tau\,.
    \label{eq:ell_def_maintext}
\end{equation}
Expanding $(1-\ell^2\,\nabla^2)^{-1}$ for small $\ell$ gives
\begin{equation}
    \partial_t\phi
    = \mathcal{L}[\phi]
      + \ell^2\,\nabla^2\mathcal{L}[\phi]
      + O(\tau),
    \label{eq:dyn_correction_maintext}
\end{equation}
showing that the $\kappa$-free dual reproduces Cahn--Hilliard
dynamics up to a controlled $O(\sqrt{\tau})$ correction.
\medskip

The $\kappa$-free system~\eqref{eq2:dual-kappa-free} is therefore a
standard mass-conserving reaction--diffusion model that reproduces the
full Cahn--Hilliard dynamics up to controlled corrections vanishing
with~$\tau \to 0$.  Unlike the $\kappa\neq0$ normal form, where the slow
manifold is the reactive nullcline $\hat{A}=0$, the $\kappa=0$ system
operates in a regime of active reaction--diffusion balance; its steady
states are selected by the flux-balance subspace
$\eta=\eta_{\mathrm{stat}}$, and the effective surface tension emerges
from the interplay of fast diffusion and interconversion in the
auxiliary field.

\subsection{Numerical comparison: Cahn--Hilliard vs.\ $\kappa$-free dual McRD dynamics}\label{sec:num:passive-coarsen}

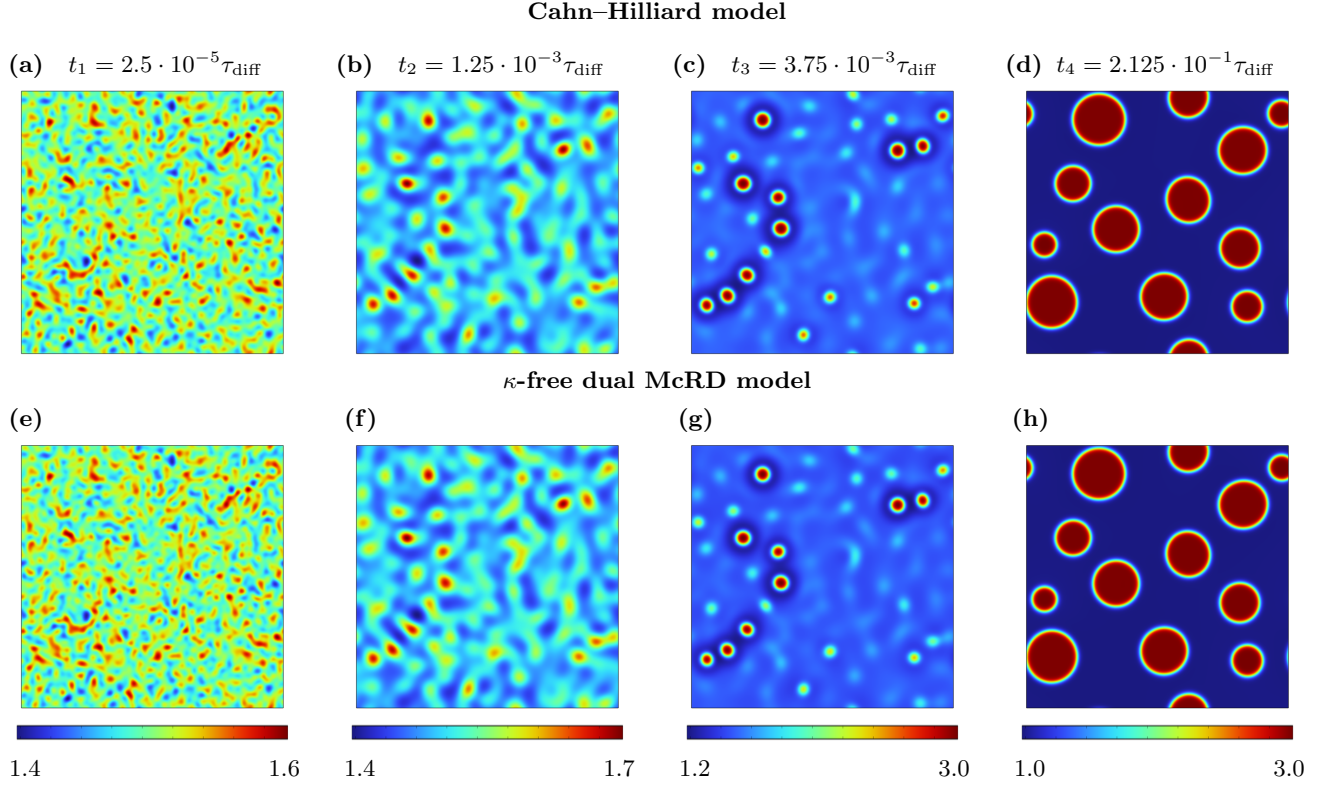
\begin{figure*}[!t]
    \centering
    \begin{tikzpicture}
        \node[anchor=south west, inner sep=0] (img) at (0,0) {\includegraphics[width=0.95\textwidth]{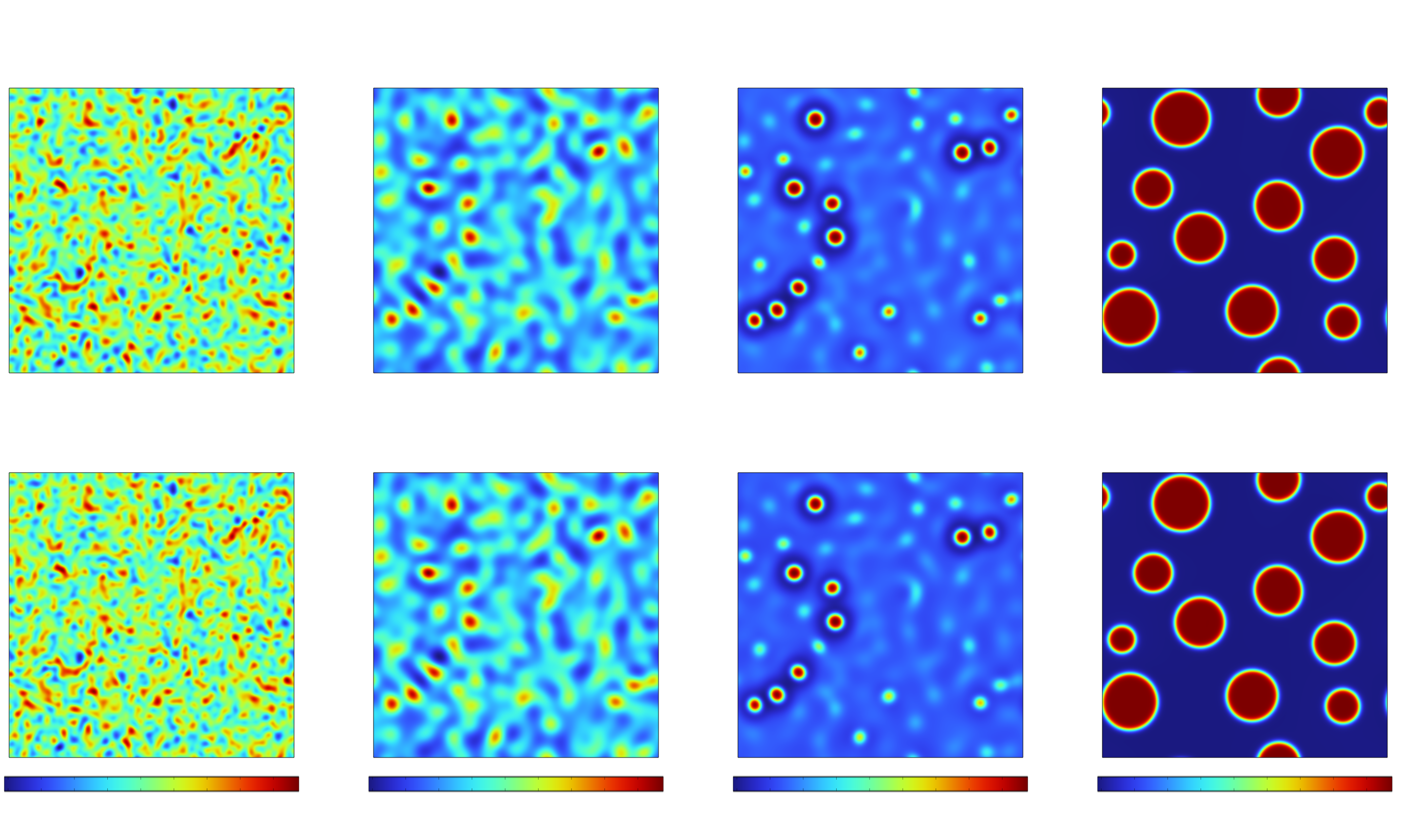}};

        \node[x={(img.south east)},y={(img.north west)}] at (0.5,1) {\textbf{Cahn--Hilliard model}};

        \node[x={(img.south east)},y={(img.north west)}] at (0.5, 0.522) {\textbf{$\kappa$-free dual McRD model}};

        \node[x={(img.south east)},y={(img.north west)}] at (0.095, 0.93){\textbf{(a)}$\quad t_1 = 2.5\cdot 10^{-5} \tau_\mathrm{diff}$};

        \node[x={(img.south east)},y={(img.north west)}] at (0.095+0.26, 0.93){\textbf{(b)}$\quad t_2 =  1.25\cdot 10^{-3} \tau_\mathrm{diff}$};

        \node[x={(img.south east)},y={(img.north west)}] at (0.095+0.26*2, 0.93) {\textbf{(c)}$\quad t_3 = 3.75\cdot 10^{-3} \tau_\mathrm{diff}$};

        \node[x={(img.south east)},y={(img.north west)}] at (0.095+0.26*3, 0.93) {\textbf{(d)}$\;\;t_4 = 2.125\cdot 10^{-1} \tau_\mathrm{diff}$};

        \node[x={(img.south east)},y={(img.north west)}] at (0.01, 0.47) {\textbf{(e)}};

        \node[x={(img.south east)},y={(img.north west)}] at (0.01+0.26, 0.47) {\textbf{(f)}};

        \node[x={(img.south east)},y={(img.north west)}] at (0.01+0.26*2, 0.47) {\textbf{(g)}};

        \node[x={(img.south east)},y={(img.north west)}] at (0.01+0.26*3, 0.47) {\textbf{(h)}};

        \node[x={(img.south east)},y={(img.north west)}] at (0.11+0.26*0, 0.02) {$1.4\qquad\qquad\qquad\qquad\quad\; 1.6$};

        \node[x={(img.south east)},y={(img.north west)}] at (0.11+0.26, 0.02) {$1.4\qquad\qquad\qquad\qquad\quad\; 1.7$};

        \node[x={(img.south east)},y={(img.north west)}] at (0.11+0.26*2, 0.02) {$1.2\qquad\qquad\qquad\qquad\quad\; 3.0$};

        \node[x={(img.south east)},y={(img.north west)}] at (0.11+0.26*3, 0.02) {$1.0\qquad\qquad\qquad\qquad\quad\; 3.0$};
    \end{tikzpicture}

    \vspace{4pt}
\caption{\textbf{Spinodal decomposition and coarsening: Cahn--Hilliard
vs.\ $\kappa$-free dual McRD dynamics.}
Two-dimensional simulations with periodic boundary conditions,
starting from a homogeneous state
$\phi(\boldsymbol{x},0)=\bar\phi=1.5\,\phi_0$ perturbed by uniform
random noise of amplitude $0.5\,\bar\phi$.
(a)--(d)~Snapshots of the conserved field $\phi(\boldsymbol{x},t)$ (in
units of $\phi_0$) for Cahn--Hilliard dynamics at times
$t_1<t_2<t_3<t_4$ (in units of $\tau_\mathrm{ diff}$), from early
spinodal amplification to late-stage coarsening.
(e)--(h)~Corresponding snapshots of the dual McRD system with
identical initial conditions, shown at the same times.
Dual parameters satisfy the matching condition
[Eq.~\eqref{eq:kappa-match}], so that stationary profiles of the
dual model obey the Cahn--Hilliard steady-state equation
[Eq.~\eqref{eq:modelB-steady}] with the same mobility $M$ and
chemical potential $\mu(\phi)$.
All columns share the time and color scale shown at the bottom.
Simulation parameters are given in
Appendix~\ref{app:all_numerics}; here
$\tilde{\kappa}=5\times 10^{-5}$, corresponding to a dimensionless
interface width
$\tilde{\ell}_\kappa \equiv \ell_\kappa/L =
\sqrt{\tilde{\kappa}} \approx 7\times 10^{-3}$.
See Supplemental Material~\cite{supplemental} for movies.}
\label{fig:modelB_vs_dual_spinodal_squarepanels}
\end{figure*}

To demonstrate the duality between Cahn--Hilliard dynamics and the corresponding $\kappa$-free dual McRD system, we numerically solved the Cahn--Hilliard equation [Eqs.~\eqref{eq:CH_model}] and the dual McRD dynamics [Eqs.~\eqref{eq2:dual-kappa-free}] with finite-element methods (FEM), using identical initial conditions and parameter values in the spinodal regime; details of the numerical implementation are provided in App.~\ref{app:all_numerics}.

\smallskip

\paragraph*{Nondimensionalization.---} 
For the Cahn--Hilliard model we use the standard symmetric Landau bulk free-energy density
\begin{equation}
    f(\phi)
    = -\frac{r}{2}\,
    \bigl(\phi-\phi_c\bigr)^{2}
    + \frac{u}{4}\,\bigl(\phi-\phi_c\bigr)^{4},
\end{equation}
corresponding to a Ginzburg--Landau expansion of the mixing free energy about the critical composition~$\phi_c$ of the demixing transition.
Retaining the physical density~$\phi$ rather than working with
the deviation $\phi-\phi_c$ ensures that both binodal
values $\phi_c\pm\phi_0$, with $\phi_0\equiv\sqrt{r/u}$, remain strictly positive, which is essential once reactive terms that depend on
the absolute density are included.

We nondimensionalize lengths by the system size~$L$, time by the
diffusive scale
\begin{equation}
    \tau_{\mathrm{diff}} \equiv \frac{L^{2}}{M\,r}\,,
\end{equation}
and the order parameter by the binodal amplitude~$\phi_0$. In these
rescaled variables the Cahn--Hilliard dynamics becomes
\begin{equation}
    \partial_t \phi
    =\nabla^{2}\!\left[
      \bigl(\phi-\tilde\phi_c\bigr)^{3}
      -\bigl(\phi-\tilde\phi_c\bigr)
      -\tilde\kappa\,\nabla^{2}\phi
    \right],
    \label{eq:CH_general_phic}
\end{equation}
where $\tilde\phi_c\equiv\phi_c/\phi_0$ and $\tilde\kappa\equiv(\ell_\kappa/L)^{2}$ with the interface width $\ell_\kappa\equiv\sqrt{\kappa/r}$.  
In the following we set
$\tilde\phi_c=2$, a convenient choice in dimensionless units that places the binodals at~$1$ and~$3$, so that
\begin{equation}
    \partial_t \phi
    =\nabla^{2}\!\left[
      (\phi-2)^{3}-(\phi-2)
      -\tilde\kappa\,\nabla^{2}\phi
    \right].
    \label{eq:CH_dimless}
\end{equation}
With this, only one free parameter remains in the Cahn--Hilliard system, namely~$\tilde\kappa$.

Using the same rescaling of length and time, the rescaled parameters
in the dual McRD equation read
\begin{equation}
    \tilde D_{c,m}=\frac{D_{c,m}}{M\,r}\,,
    \qquad
    \tilde\tau=\frac{\tau}{\tau_{\mathrm{diff}}}\,.
\end{equation}
The matching conditions are
\begin{equation}
    \tilde\kappa
    =\frac{\kappa}{r\,L^{2}}
    =\tilde\tau\;\tilde D_m\;\tilde D_c\,,
    \qquad
    \frac{1}{r}=\tilde D_c-\tilde D_m\,.
    \label{eq:matching}
\end{equation}
Since $\tilde\kappa$ is considered fixed, the free parameters are~$r$ and~$\tilde\tau$, which in turn determine the rescaled diffusion coefficients~$\tilde D_{c,m}$.
\smallskip

\paragraph*{Spinodal decomposition trajectories.---}
Figure~\ref{fig:modelB_vs_dual_spinodal_squarepanels} compares representative two-dimensional trajectories of the Cahn--Hilliard model and its $\kappa$-free dual McRD system for identical initial conditions (small random perturbations about a homogeneous state) and periodic boundary conditions. 
The Cahn--Hilliard snapshots in Figs.~\ref{fig:modelB_vs_dual_spinodal_squarepanels}(a)–(d) and the corresponding dual snapshots in Figs.~\ref{fig:modelB_vs_dual_spinodal_squarepanels}(e)–(h), evaluated at the same time sequence ${t_1<t_2<t_3<t_4}$, show the same qualitative sequence: linear spinodal amplification selects an initial length scale, domains form and reorganize nonlinearly, and the system enters a coarsening regime. 
The dual-model parameters are chosen using the exact matching condition in Eq.~\eqref{eq:kappa-match}, which fixes $(D_c,D_m,\tau)$ such that the dual reduction reproduces the same Cahn--Hilliard dynamics (same mobility and chemical potential) in the $\tau\to0$ limit and thereby provides a like-for-like dynamical comparison. 

Quantitatively, the transient dynamics need not coincide for finite constraint relaxation time $\tau$ (App.~\ref{app:duality_McRD_CH}). 
The dominant effect is a time lag during the rapid early evolution: at the intermediate time $t_3$ [Figs.~\ref{fig:modelB_vs_dual_spinodal_squarepanels}(c) and \ref{fig:modelB_vs_dual_spinodal_squarepanels}(g)] the dual pattern has reduced contrast and is less advanced along the coarsening trajectory. 
In the next paragraph we make this precise by tracking the characteristic coarsening length $\Lambda(t)$ extracted from the structure factor, Eq.~\eqref{eq:char-length}, showing that the late-stage coarsening law agrees once this preasymptotic time shift is accounted for.

\smallskip

\paragraph*{Quantitative coarsening comparison.---}
Beyond visual agreement, we quantify coarsening by a characteristic length scale $\Lambda(t)$ extracted from the (radially averaged) structure factor $S(k,t)$,
\begin{equation}
\Lambda(t) \equiv \frac{\int \mathrm{d}k\, S(k,t)}{\int \mathrm{d}k\, k\, S(k,t)}.
\label{eq:char-length}
\end{equation}
Here $S(k,t)$ is obtained by angular averaging of the squared Fourier amplitudes, ${S(k,t)  \equiv  \int_{\Omega}\mathrm{d}\Omega\,|\hat\phi(\boldsymbol{k},t)|^{2}}$ with ${\Omega\in[0,2\pi)}$ in two dimensions, where the Fourier transform is defined as ${\hat\phi(\boldsymbol{k},t)  \equiv \int \mathrm{d}\boldsymbol{x}\;\phi(\boldsymbol{x},t)\,e^{-i\,\boldsymbol{k}\cdot\boldsymbol{x}}}$.

\begin{figure}[!t]
    \centering
    \includegraphics[width=\columnwidth]{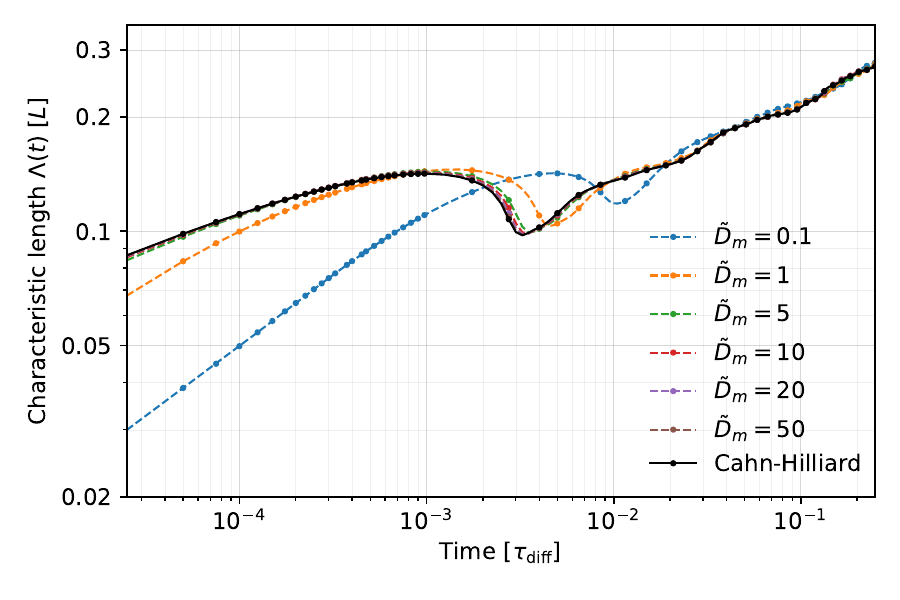}
\caption{Characteristic coarsening length $\Lambda(t)$, defined from the radially averaged structure factor via Eq.~\eqref{eq:char-length} (log--log axes). 
The black curve shows the Cahn--Hilliard reference.
Colored curves show the $\kappa$-free dual dynamics for the indicated auxiliary diffusivities $\tilde D_m$ (legend); for each $\tilde D_m$, the remaining dual parameters are chosen to satisfy the exact matching condition in Eq.~\eqref{eq:kappa-match}. 
Decreasing $\tilde D_m$ leads to a pronounced preasymptotic delay and enhances the transient nonmonotonic feature around $t/\tau_\mathrm{ diff}\sim 10^{-3}$--$10^{-2}$, while increasing $\tilde D_m$ progressively suppresses these transients and yields rapid collapse onto the Cahn--Hilliard coarsening curve at late times. 
Lengths are given in units of the system size $L$ (i.e., $\Lambda/L$), and time in units of the diffusive scale $\tau_\mathrm{ diff}$. 
Simulation parameters are given in App.~\ref{app:all_numerics}.
}
\label{fig:char-length}
\end{figure}

Figure~\ref{fig:char-length} shows $\Lambda(t)$ for the dual dynamics along a sweep in $\tilde D_m$, with the corresponding constraint-relaxation time $\tilde \tau$ fixed by the exact matching condition in Eq.~\eqref{eq:kappa-match}.
For small $\tilde D_m$ the dual dynamics exhibits a pronounced preasymptotic delay: $\Lambda(t)$ grows more slowly at early times, consistent with the time lag visible in Fig.~\ref{fig:modelB_vs_dual_spinodal_squarepanels}.
At late times, however, $\Lambda(t)$ converges to the Cahn--Hilliard reference (black) and follows the same coarsening law.
Increasing $\tilde D_m$ progressively reduces the transient delay and brings the dual trajectory into near-quantitative agreement with the Cahn--Hilliard curve over the full time window.
Most importantly, even in the small-$\tilde D_m$ limit the late-stage coarsening exponent agrees with the Cahn--Hilliard value, indicating that the dual model reproduces the same asymptotic coarsening kinetics once the transient time shift is accounted for. Though the measured power is slightly lower than the one predicted by LSW-theory of $1/3$ \citep{Lifshitz.Slyozov1961, Wagner1961} in both cases.
Overall, under the exact parameter matching in Eq.~\eqref{eq:kappa-match}, deviations between the $\kappa$-free dual formulation and Cahn--Hilliard dynamics are confined to a preasymptotic transient controlled by $\tilde \tau$ and $\tilde D_m$.

\medskip

\paragraph*{Phase-space dynamics.---}

The real-space comparison in Figs.~\ref{fig:modelB_vs_dual_spinodal_squarepanels}
and~\ref{fig:char-length} demonstrates that the $\kappa$-free dual
reproduces Cahn--Hilliard coarsening at late times, with deviations
confined to a preasymptotic transient.  To understand the origin of this
transient, it is instructive to view the dynamics in the
$(\phi,\eta)$ phase plane, where $\eta=c+d\,m$ is the
mass-redistribution potential and $d=D_m/D_c$.
 
\begin{figure*}[!t]
  \centering
 
  \newlength{\phasePanel}
  \setlength{\phasePanel}{0.24\textwidth}
  \newlength{\phaseGap}
  \setlength{\phaseGap}{0.001\textwidth}
 
  \begin{minipage}[t]{\phasePanel}
    \centering
    \textbf{(a)} $t=2.5\cdot 10^{-6}\tau_\mathrm{ diff}$\\[4pt]
    \includegraphics[width=\phasePanel,height=\phasePanel]{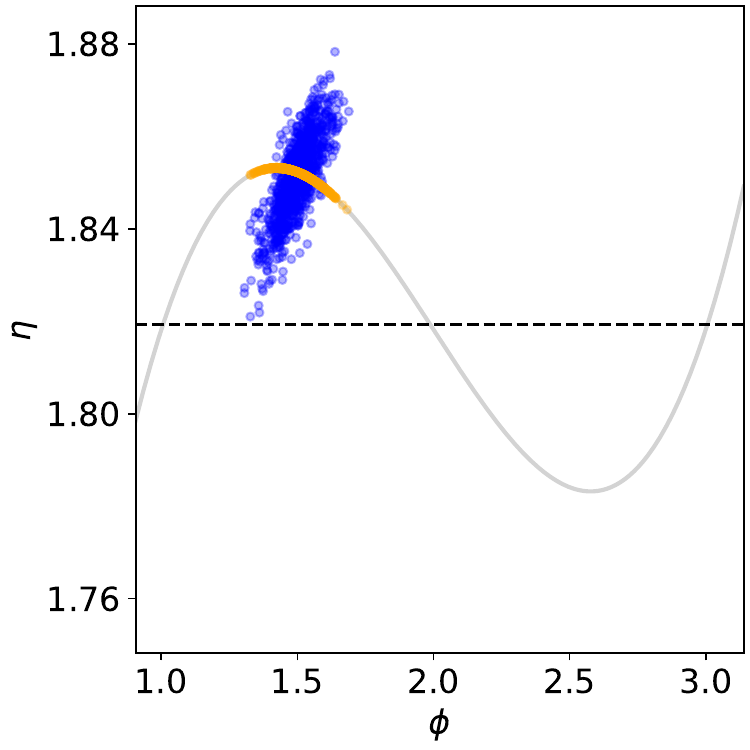}
  \end{minipage}\hspace{\phaseGap}
  \begin{minipage}[t]{\phasePanel}
    \centering
    \textbf{(b)} $t=4.0\cdot 10^{-3}\tau_\mathrm{ diff}$\\[4pt]
    \includegraphics[width=\phasePanel,height=\phasePanel]{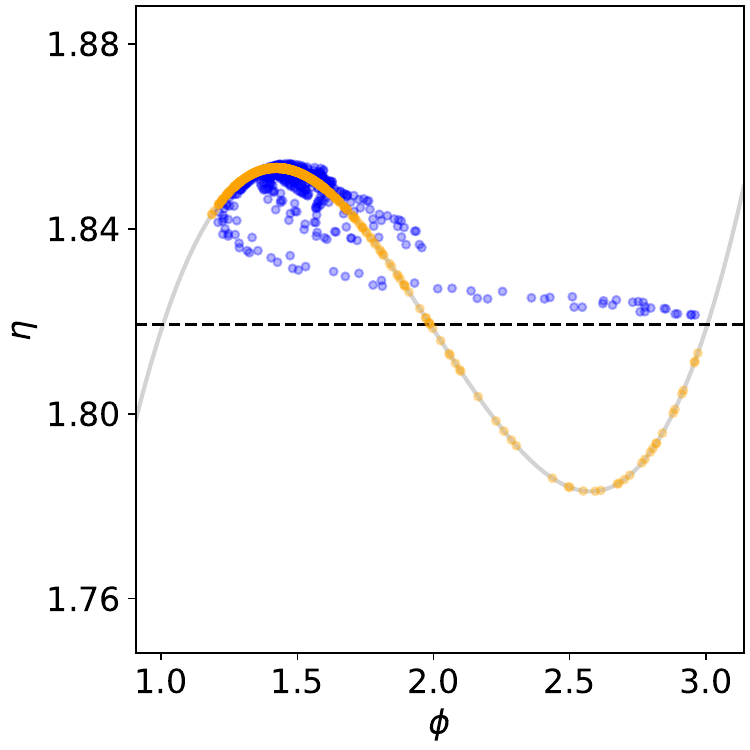}
  \end{minipage}\hspace{\phaseGap}
  \begin{minipage}[t]{\phasePanel}
    \centering
    \textbf{(c)} $t=1.5\cdot 10^{-2}\tau_\mathrm{ diff}$\\[4pt]
    \includegraphics[width=\phasePanel,height=\phasePanel]{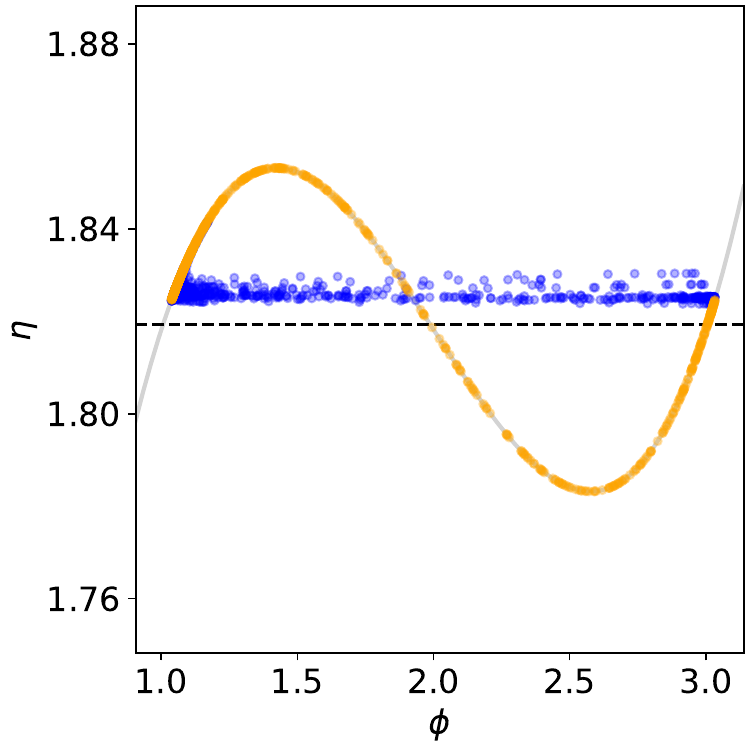}
  \end{minipage}\hspace{\phaseGap}
  \begin{minipage}[t]{\phasePanel}
    \centering
    \textbf{(d)} $t=12.53\,\tau_\mathrm{ diff}$\\[4pt]
    \includegraphics[width=\phasePanel,height=\phasePanel]{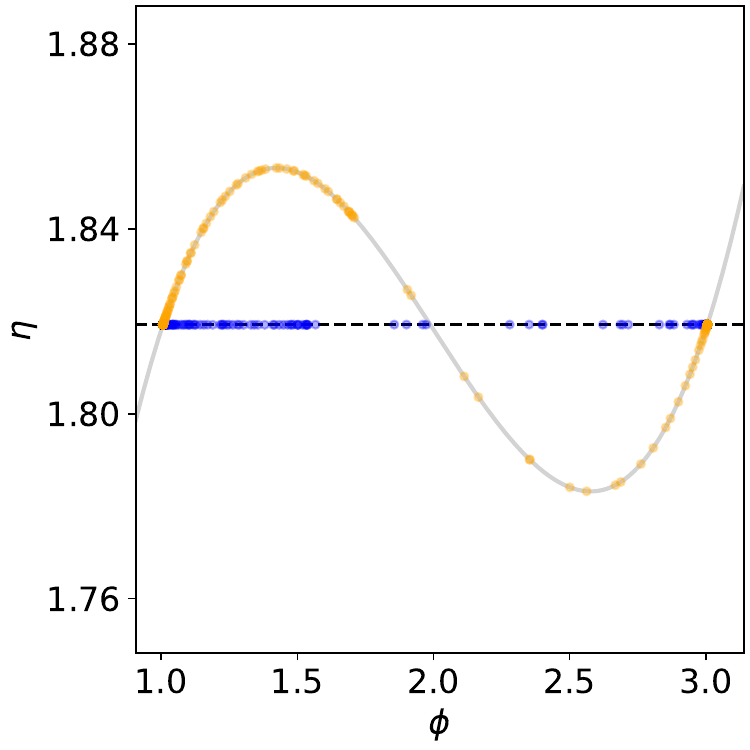}
  \end{minipage}
\caption{
\textbf{Phase-space view of Cahn--Hilliard vs.\ dual McRD dynamics.}
Scatter plots in the $(\phi,\eta)$ plane at four successive times
$t_1<t_2<t_3<t_4$ (indicated above each panel), where each dot
represents one spatial grid point $(x_i,y_j)$.
Orange (light grey) dots: Cahn--Hilliard; blue (dark grey) dots:
$\kappa$-free dual McRD.  Point opacity encodes the local density of
overlapping grid points but saturates well before all points coincide.
The solid grey curve is the reactive nullcline
$\eta=\eta^*(\phi)$; the dashed horizontal line marks the
flux-balance subspace $\eta=\eta_{\mathrm{stat}}$.
(a)~Both clouds coincide near the homogeneous state.
(b)~Spinodal decomposition spreads the CH cloud along the nullcline,
while the dual cloud begins to deviate toward the FBS.
(c)~After well-defined droplets have formed, the two clouds occupy
clearly distinct manifolds.
(d)~At late times both clouds cluster near $\phi_\pm$, where
nullcline and FBS intersect, consistent with the identical plateau
values and coarsening exponents; the apparent absence of blue points
is because most are hidden beneath the orange ones at the
intersection points (the total number of plotted grid points is
the same in every panel).
Parameters: $\tilde{D}_m=10$; axes in dimensionless field units as in
Fig.~\ref{fig:modelB_vs_dual_spinodal_squarepanels}. See Supplemental Material~\cite{supplemental} for movies.
}
\label{fig:phase_space_dynamics}
\end{figure*}

In Cahn--Hilliard dynamics, $c$ and $m$ are not independent degrees of
freedom: the constitutive relation $c=\hat{\mu}(\phi)$ is satisfied at
all times, and hence the system evolves along the reactive nullcline
$\eta=\eta^*(\phi)$ (cf.\ Eq.~\eqref{eq:Ahat_eta}).
Every spatial point $(x,t)$ maps to a location on this curve, and the
entire dynamics is confined to it.
 
The $\kappa$-free dual system, by contrast, does \emph{not} evolve on the
nullcline.  As established in Sec.~\ref{sec:kappa_removal}, its
stationary states are selected by the flux-balance subspace (FBS),
$\eta=\eta_{\mathrm{stat}}$, on which the diffusive fluxes of $c$ and
$m$ balance globally but the reaction $\hat{A}$ remains locally nonzero.
It is precisely this sustained local departure from the nullcline that
generates the effective fourth-order term
$\kappa_{\mathrm{eff}}\,\nabla^4\phi$ responsible for interfacial
regularisation [Eq.~\eqref{eq:kappa-match}].  
In other words, the slow
manifold of the dual model is the FBS, not the nullcline; the distance
between the two in $(\phi,\eta)$ space encodes the surface tension.

The distinction becomes clear during the time-dependent evolution.
Starting from a homogeneous initial condition
$(\phi,\eta)=(\bar\phi,\eta^*(\bar\phi))$---which lies on average
on the nullcline---both models undergo spinodal
decomposition as the homogeneous state becomes unstable.  As the
amplitude of the compositional fluctuations grows, the two
trajectories separate in phase space.

In the Cahn--Hilliard model, every spatial point remains on the
  nullcline $\eta=\eta^*(\phi)$ at all times.  As domains form and
  coarsen, the distribution of points in $(\phi,\eta)$ spreads along
  this N-shaped curve and progressively concentrates near its two
  stable branches, corresponding to the binodal plateau values
  $\phi_\pm$.

In the dual McRD model, the initial perturbation triggers a fast
  transient (duration $\sim\tau$) during which each spatial point
  relaxes toward a quasi-steady state at finite distance from the
  nullcline.  The resulting cloud of points in $(\phi,\eta)$ space
  does not trace the nullcline but instead collapses onto a
  neighbourhood of the FBS, $\eta\approx\eta_{\mathrm{stat}}$.  On
  this manifold, the dynamics is governed by the balance of diffusive
  fluxes, and coarsening proceeds as in the Cahn--Hilliard case, albeit
  with a temporal offset set by the initial relaxation time.

Figure~\ref{fig:phase_space_dynamics} illustrates this separation
by plotting the field configuration as a cloud of points in
the $(\phi,\eta)$ plane at successive times.  At the earliest time
$t_1$, both models are near the homogeneous state and the two clouds
occupy the same region in phase space, with the RD-dual being a smeared out version.  
By $t_2$, the Cahn--Hilliard cloud has
spread along the nullcline, populating the spinodal region between
its two stable branches.  
The dual cloud, by contrast, has begun to flatten toward the FBS: individual spatial points have moved off the nullcline and toward $\eta\approx\eta_{\mathrm{stat}}$.  
At the intermediate time $t_3$---where spinodal decomposition has already taken place---the two clouds occupy clearly distinct regions of phase space: the Cahn--Hilliard distribution is concentrated near the extrema of the nullcline, while the dual distribution lies near the FBS. 
At late times~$t_4$, both models have reached the coarsening regime and the phase-space distributions are qualitatively similar: in both cases, the points cluster near $\phi_\pm$, reflecting well-formed bulk plateaus separated by narrow interfaces. 
The residual difference at $t_4$ is
that the dual cloud lies on the FBS while the Cahn--Hilliard cloud lies
on the nullcline; these two manifolds share the same intersection
points at $\phi_\pm$, consistent with the identical plateau values.

\medskip
 
This phase-space view provides a unifying perspective on the results
of the preceding paragraphs.  The preasymptotic time lag in the
coarsening length (Fig.~\ref{fig:char-length}) reflects the time
needed for the dual system to redistribute its phase-space weight from
the nullcline to the FBS---a process controlled by $\tau$ and $D_m$.
The fact that the late-stage coarsening exponents agree follows from
the fact that, once the plateau values $\phi_\pm$ are reached, both
manifolds coincide at the coexistence points, and the slow dynamics
is governed by the same mass-redistribution mechanism.  The surface
tension, which in the Cahn--Hilliard model is built in through the
$\kappa\,\nabla^2\phi$ term, is generated in the dual model by the
sustained offset between the FBS and the nullcline across the
interfacial region---precisely the mechanism quantified by the
matching condition~\eqref{eq:kappa-match}.

\section{Maxwell Construction and Coexistence in the Dual Description}
\label{sec:maxwell}

We now use the reaction--diffusion embedding developed in Sec.~\ref{sec:duality-transformation_CH} to recast coexistence and interface selection in dual variables. 
The central result is that the Maxwell construction --- the thermodynamic rule that selects coexistence densities and chemical potential --- is precisely equivalent to the reactive turnover-balance condition that selects the mass-redistribution potential level in mass-conserving reaction--diffusion (McRD) systems.
We develop this equivalence in three steps:
we first recall the Maxwell construction in Cahn--Hilliard language,
then formulate the turnover-balance condition in McRD variables,
and finally show that the two are related by an exact change of variables along the flux-balance subspace.
As corollaries, we identify the McRD analogues of the osmotic pressure and the Laplace pressure.

\subsection{Coexistence in the Cahn--Hilliard description}
\label{sec:maxwell}

For a thermodynamic Cahn--Hilliard system, a stationary planar interface requires
\emph{chemical} and \emph{mechanical} equilibrium between the coexisting bulk
plateaus~$\phi_\pm$.
Vanishing diffusive flux $\boldsymbol{J}=-M \,\nabla\mu$ enforces a spatially uniform chemical
potential,
\begin{equation}
    \mu(\phi_-) = \mu(\phi_+) = \mu_\mathrm{coex}\,.
    \label{eq:chem_eq_CH}
\end{equation}
Mechanical equilibrium amounts to equality of the osmotic pressure,
\begin{equation}
    \Pi(\phi) \equiv \phi\,\mu(\phi) - f(\phi)\,,
    \label{eq:osmotic_pressure_def_CH}
\end{equation}
i.e., the Legendre transform of the free-energy density~$f(\phi)$, across the two phases:
\begin{equation}
    \Pi(\phi_-) = \Pi(\phi_+)\,.
    \label{eq:mech_eq_CH}
\end{equation}
Equations~\eqref{eq:chem_eq_CH} and \eqref{eq:mech_eq_CH} together determine $(\phi_-,\phi_+)$
and~$\mu_\mathrm{coex}$; geometrically, $\mu_\mathrm{coex}$ is the slope of the
common tangent to~$f(\phi)$ at~$\phi_\pm$.
Eliminating~$\Pi$ yields the Maxwell equal-area construction,
\begin{equation}
    \int_{\phi_-}^{\phi_+}\!\d\phi \, \bigl[\mu(\phi)-\mu_\mathrm{coex}\bigr] = 0\,,
    \qquad
    \mu(\phi_\pm)=\mu_\mathrm{coex}\,.
    \label{eq:maxwell_eq_area_CH}
\end{equation}
In a closed system the conserved mean density~$\bar\phi$ does not affect
$(\phi_\pm,\mu_\mathrm{coex})$; it only determines the phase fractions via the lever rule,
provided $\phi_- < \bar\phi < \phi_+$.
The equal-area construction will reappear in McRD variables in Fig.~\ref{fig:fbs}(a).

\subsection{Coexistence in the McRD description: reactive turnover balance}

For the dual two-component mass-conserving reaction--diffusion system [Eqs.~\eqref{eq2:dual-kappa-free}],
the conserved total density $\phi = c + m$ obeys $\partial_t\phi = D_c\,\nabla^2\eta$, where
$\eta \equiv c + d\,m$ is the mass-redistribution potential (recall $d \equiv D_m/D_c$ from Sec.~\ref{sec:kappa_removal}).
Stationarity therefore requires $\nabla^2\eta = 0$ and, for periodic or no-flux
boundary conditions on a connected domain, the mass-redistribution potential is
spatially uniform,
\begin{equation}
    \eta(\boldsymbol{x}) \equiv \eta_\mathrm{stat}\,.
    \label{eq:McRD_flux_balance}
\end{equation}
Stationary states are thus confined to the \emph{flux-balance subspace}
$\eta = \eta_\mathrm{stat}$ in $(\eta,m)$ phase space [Fig.~\ref{fig:fbs}(a)].
This is the direct analogue of
$\mu(\boldsymbol{x}) \equiv \mu_\mathrm{coex}$ for a stationary Cahn--Hilliard interface.

Restricting to $\eta = \eta_\mathrm{stat}$ reduces the stationary interface problem
(in the normal coordinate~$x$) to the single profile equation
\begin{equation}
    0 = D_m\,\partial_x^2 m + \hat{A}(m,\eta_\mathrm{stat})\,,
    \label{eq:McRD_profile_eq}
\end{equation}
where $\hat{A}(m,\eta)$ is the local interconversion rate [Eq.~\eqref{eq:hat_A}].
The plateau values~$m_\pm$ are the stable zeros of~$\hat{A}$ on the line
$\eta = \eta_\mathrm{stat}$, i.e., the outer intersections of the flux-balance subspace with the nullcline $\hat{A}(m,\eta) = 0$ [Fig.~\ref{fig:fbs}(a)].

\medskip

\paragraph*{Selection of $\eta_\mathrm{stat}$ by turnover balance.}
The coexistence condition that selects~$\eta_\mathrm{stat}$ follows from a first
integral of Eq.~\eqref{eq:McRD_profile_eq}.
Multiplying by~$\partial_x m$ and integrating once gives
\begin{equation}
    \frac{D_m}{2}\,(\partial_x m)^2 + V(m;\eta_\mathrm{stat})
    = \mathrm{const}\,,
    \label{eq:McRD_first_integral}
\end{equation}
with the effective potential
\begin{equation}
    V(m;\eta)
    \equiv \int_{m_0}^{m}\!\d m'\,\hat{A}(m',\eta)\,.
    \label{eq:McRD_V_def}
\end{equation}
For a heteroclinic orbit (interface) connecting~$m_-$ to~$m_+$, ${\partial_x m \to 0}$ as ${x \to \pm\infty}$, so
the constant in Eq.~\eqref{eq:McRD_first_integral} equals~$V(m_\pm;\eta_\mathrm{stat})$ and therefore
\begin{equation}
    V(m_-;\eta_\mathrm{stat}) = V(m_+;\eta_\mathrm{stat})\,.
    \label{eq:McRD_equal_potential}
\end{equation}
Equivalently, this \emph{turnover-balance} condition reads
\begin{equation}
    \int_{m_-}^{m_+}\!\d m\, \hat{A}(m,\eta_\mathrm{stat}) = 0\,,
    \label{eq:McRD_turnover_balance}
\end{equation}
i.e., the signed area under the reaction rate~$\hat{A}(m,\eta_\mathrm{stat})$ between~$m_-$ and~$m_+$ vanishes [shaded regions in Fig.~\ref{fig:fbs}(a)].
This selects the unique~$\eta_\mathrm{stat}$ for which a stationary
interface exists, in direct analogy to the Maxwell common-tangent construction
that fixes~$\mu_\mathrm{coex}$.
As in the Cahn--Hilliard case, the conserved mean density enters only through the lever rule once $(m_\pm,\eta_\mathrm{stat})$ are fixed.
The turnover-balance condition is well known in the McRD literature~\citep{Brauns_Frey:2020}; the new contribution below is the proof of its exact equivalence to the Maxwell construction.

\subsection{Equivalence of the Maxwell construction and turnover balance}

We now prove that the turnover-balance condition, Eq.~\eqref{eq:McRD_turnover_balance}, in McRD variables is exactly the Maxwell equal-area rule, Eq.~\eqref{eq:maxwell_eq_area_CH}, expressed in the dual description.
The bridge is the affine relation between $(\eta,m)$ and~$\phi$ on the flux-balance subspace.

Along $\eta = \eta_\mathrm{stat}$, the identity $\phi = \eta + (1-d)\,m$ gives a one-to-one mapping between~$m$ and~$\phi$:
\begin{equation}
    \d\phi = (1-d)\,\d m\,,
    \qquad
    \phi_\pm = \eta_\mathrm{stat} + (1-d)\,m_\pm\,.
    \label{eq:phi_m_eta_affine}
\end{equation}
On the same subspace, evaluating the reaction term [Eq.~\eqref{eq:hat_A}] at $\eta = \eta_\mathrm{stat}$ and using $c = (\eta_\mathrm{stat} - d\,\phi)/(1-d)$ together with the definition of~$\hat{\mu}$ [Eq.~\eqref{eq:def-of-mu-hat}] gives
\begin{equation}
    \hat{A}\!\bigl(m(\phi,\eta_\mathrm{stat}),\,\eta_\mathrm{stat}\bigr)
    =
    \frac{1}{\tau}\bigl[\mu_\mathrm{coex} - \mu(\phi)\bigr],
    \label{eq:A_mu_coex_reduced}
\end{equation}
with
\begin{equation}
    \mu_\mathrm{coex}
    \equiv \frac{\eta_\mathrm{stat} - d\,\phi_\mathrm{eq}}{1-d}\,.
    \label{eq:mucoex_def}
\end{equation}
The plateau condition $\hat{A}(m_\pm,\eta_\mathrm{stat}) = 0$ then immediately yields chemical equilibrium,
\begin{equation}
    \mu(\phi_-) = \mu(\phi_+) = \mu_\mathrm{coex}\,.
    \label{eq:mu_equal_from_nullcline}
\end{equation}
The remaining selection condition follows by substituting Eq.~\eqref{eq:A_mu_coex_reduced} into the turnover balance, Eq.~\eqref{eq:McRD_turnover_balance}, and changing variables via $\d m = \d\phi/(1-d)$:
\begin{align}
    0
    &= \int_{m_-}^{m_+}\!\hat{A}(m,\eta_\mathrm{stat})\,\d m
    \nonumber\\
    &= \frac{1}{(1-d)\,\tau}
       \int_{\phi_-}^{\phi_+}\!\d\phi \, \bigl[\mu_\mathrm{coex} - \mu(\phi)\bigr]\,.
    \label{eq:turnover_to_mu}
\end{align}
Since $(1-d)\,\tau \neq 0$ for any finite $\tau > 0$, this is equivalent to the Maxwell equal-area rule,
\begin{equation}
    \int_{\phi_-}^{\phi_+}\!\d\phi \, \bigl[\mu(\phi) - \mu_\mathrm{coex}\bigr] = 0\,,
    \label{eq:maxwell}
\end{equation}
together with $\mu(\phi_\pm) = \mu_\mathrm{coex}$.
Note that the equivalence holds for any finite $\tau > 0$ and $d < 1$; the fast-relaxation limit $\tau \to 0$, in which $(1-d) \sim \sqrt{\tau} \to 0$ (Sec.~\ref{sec:kappa_removal}), does not affect the selection rule but only the rate at which the interface profile approaches the stationary heteroclinic orbit.

Thus, the turnover balance in $(m,\eta)$ space is exactly the Maxwell construction expressed in reaction--diffusion variables: it selects~$\eta_\mathrm{stat}$ and thereby fixes the coexistence chemical potential~$\mu_\mathrm{coex}$.
Physically, this is a \emph{mass-budget} statement: the rate at which the interconversion~$\hat{A}$ deposits mass into the high-density phase must exactly equal the rate at which it removes mass from the low-density phase, so that neither phase grows at the expense of the other.
It is this balance of reactive fluxes, rather than a free-energy minimization, that pins the interface.

\subsection{Osmotic pressure and Laplace pressure in McRD variables}

\paragraph*{Osmotic pressure.}
The first integral, Eq.~\eqref{eq:McRD_first_integral}, introduces the reaction potential $V(m;\eta)$ [Eq.~\eqref{eq:McRD_V_def}].
On the flux-balance subspace, substituting Eq.~\eqref{eq:A_mu_coex_reduced} and changing variables via Eq.~\eqref{eq:phi_m_eta_affine} gives
\begin{align}
    V(m;\eta_\mathrm{stat})
    &= \frac{1}{(1-d)\,\tau}
       \int^{\phi}\!\d\varphi\, \bigl[\mu_\mathrm{coex} - \mu(\varphi)\bigr]
    \nonumber\\
    &= \frac{1}{(1-d)\,\tau}
       \bigl[\phi\,\mu_\mathrm{coex} - f(\phi)\bigr]
       + \mathrm{const}\,,
    \label{eq:V_grand_potential}
\end{align}
where we used $\mu(\phi) = f'(\phi)$.
Up to the prefactor $(1-d)\,\tau$, the reaction potential~$V$ thus coincides with the Legendre transform of~$f$, i.e., the osmotic pressure $\Pi(\phi) = \phi\,\mu(\phi) - f(\phi)$.
The interface condition $V(m_-) = V(m_+)$ [Eq.~\eqref{eq:McRD_equal_potential}] is therefore equivalent to pressure balance, $\Pi(\phi_-) = \Pi(\phi_+)$, across the two phases.
\medskip

\begin{table*}[ht]
\caption{\textbf{Coexistence dictionary.}
Correspondence between the Cahn--Hilliard (CH) phase-field description and the dual mass-conserving reaction--diffusion (McRD) formulation established in this section.}
\label{tab:dictionary}
\begin{ruledtabular}
\begin{tabular}{lll}
 & Cahn--Hilliard & McRD dual \\
\hline
Dynamical variable
  & density $\phi$
  & $c$, $m$ \quad ($\phi = c + m$) \\[4pt]
Transport potential
  & chemical potential $\mu$
  & mass-redistribution potential $\eta$ \\[4pt]
Stationarity
  & $\mu(\boldsymbol{x}) = \mu_\mathrm{coex}$
  & $\eta(\boldsymbol{x}) = \eta_\mathrm{stat}$ \\[4pt]
Coexistence selection
  & Maxwell equal-area rule
  & turnover balance \\
  & $\displaystyle\int_{\phi_-}^{\phi_+}\!\d\phi \, \bigl[\mu - \mu_\mathrm{coex}\bigr] = 0$
  & $\displaystyle\int_{m_-}^{m_+}\!\d m\, \hat{A}(m,\eta_\mathrm{stat}) = 0$ \\[10pt]
Osmotic pressure
  & $\Pi(\phi) = \phi\,\mu - f$
  & $(1\!-\!d)\,\tau\,V(m;\eta)$ \\[4pt]
Mechanical equil.
  & $\Pi(\phi_-) = \Pi(\phi_+)$
  & $V(m_-;\eta_\mathrm{stat}) = V(m_+;\eta_\mathrm{stat})$ \\[4pt]
Laplace pressure
  & $\Delta\Pi = -(d_\mathrm{sp}\!-\!1)\,\sigma/R$
  & same, with $\sigma = \kappa\!\displaystyle\int\!\d x \, (\partial_x\phi)^2$ \\[4pt]
Interfacial stiffness
  & $\kappa$ (explicit parameter)
  & $\kappa = \tau\,D_m/(1\!-\!d)$ (emergent) \\
\end{tabular}
\end{ruledtabular}
\end{table*}

\paragraph*{Laplace pressure.}
The duality predicts a Young--Laplace law for the McRD system, despite the absence of any free-energy functional with an interface penalizing term in its formulation.
We show that this law emerges from the flux-balance constraint alone, with an effective interfacial tension built directly from the dual variables.
On the flux-balance subspace, the auxiliary field $m$ is slaved to the order parameter through the affine relation $\phi = \eta_\mathrm{stat} + (1-d)\,m$, so that $\nabla^2 m = (1-d)^{-1}\,\nabla^2\phi$ and the stationary profile equation [Eq.~\eqref{eq:McRD_profile_eq}] closes on $\phi$:
\begin{equation}
    0
    = \frac{D_m}{1-d}\,\nabla^2\phi
      + \hat{A} \bigl(m(\phi,\eta_\mathrm{stat}),\,\eta_\mathrm{stat}\bigr).
    \label{eq:McRD_profile_phi}
\end{equation}
Substituting Eq.~\eqref{eq:A_mu_coex_reduced} gives
\begin{equation}
    0
    = \kappa\,\nabla^2\phi + \mu_\mathrm{coex} - \mu(\phi)\,,
    \qquad
    \kappa \equiv \frac{\tau\,D_m}{1-d}\,,
    \label{eq:McRD_profile_phi_kappa}
\end{equation}
which is the steady-state Allen--Cahn equation with the same effective stiffness~$\kappa$ identified in the steady-state matching of Sec.~\ref{sec:kappa_removal} [Eq.~\eqref{eq:kappa-match}].

For a radially symmetric droplet $\phi = \phi(r)$ in spatial dimension~$d_\mathrm{sp}$,
\begin{equation}
    \nabla^2\phi
    = \partial_r^2\phi + \frac{d_\mathrm{sp}-1}{r}\,\partial_r\phi\,.
    \label{eq:Laplace_radial_Laplacian}
\end{equation}
Multiplying Eq.~\eqref{eq:McRD_profile_phi_kappa} by~$\partial_r\phi$ and using $\mu(\phi) = f'(\phi)$ yields
\begin{equation}
    0 = \partial_r \biggl[\frac{\kappa}{2}\,(\partial_r\phi)^2 
        + \mu_\mathrm{coex}\,\phi - f(\phi)\biggr]
        + \kappa\,\frac{d_\mathrm{sp}-1}{r}\,(\partial_r\phi)^2\,.
    \label{eq:Laplace_first_integral_radial}
\end{equation}
Integrating across the interface and using $\partial_r\phi \to 0$ far from the interface gives
\begin{equation}
    \Pi(\phi_+) - \Pi(\phi_-)
    = -\kappa\,(d_\mathrm{sp}-1)
      \int_\mathrm{interface}\!\!\d r \, \frac{(\partial_r\phi)^2}{r}\,,
    \label{eq:Laplace_pressure_exact}
\end{equation}
where $\Pi(\phi) = \phi\,\mu(\phi) - f(\phi)$ is the osmotic pressure [Eq.~\eqref{eq:osmotic_pressure_def_CH}].
For a large droplet of radius $R \gg \ell_\mathrm{int}$, replacing $1/r \simeq 1/R$ across the narrow interface yields the Laplace form
\begin{equation}
    \Pi(\phi_+) - \Pi(\phi_-)
    \simeq -\frac{(d_\mathrm{sp}-1)\,\sigma}{R}\,,
    \label{eq:Laplace_pressure_asymptotic}
\end{equation}
with the effective interfacial tension
\begin{equation}
    \sigma
    \equiv \kappa\!\int_{-\infty}^{\infty}\!\d x \, (\partial_x\phi)^2\,,
    \label{eq:sigma_def}
\end{equation}
where $x$ denotes the coordinate normal to the interface.
Equation~\eqref{eq:sigma_def} has the same form as the Cahn--Hilliard interfacial tension, but the construction makes its origin transparent: in the McRD picture the tension is generated entirely by the flux-balance subspace, not by any underlying free energy.
This recovers the effective interfacial tension derived directly in Ref.~\citep{Weyer_Frey:2026}, and identifies the duality with Cahn--Hilliard dynamics as its thermodynamic shadow.

The result connects to a broader question in active matter: how the equilibrium notions of interfacial tension, Maxwell construction, and Laplace pressure generalise once detailed balance is broken.
Scalar active field theories address this top-down.
Active Model~B introduces a leading-order time-reversal-symmetry-breaking term in the chemical potential that renders the pressure profile-dependent across interfaces~\citep{Wittkowski.etal2014}; Active Model~B$^+$ adds a non-gradient TRS-breaking current that can reverse Ostwald ripening and stabilise microphase separation~\citep{Tjhung_Cates:2018}; we refer the reader to~\citep{Cates_Nardini:2025} for a comprehensive overview.
A parallel bottom-up question is whether mass-conserving reaction--diffusion dynamics, which possess no free-energy functional, nonetheless inherit a thermodynamic-like vocabulary.
They do: an effective interfacial tension, Laplace-pressure-driven coarsening, and curvature-dependent domain dynamics all emerge from the reaction kinetics alone~\citep{Weyer_Frey:2026}.
The duality makes this parallel exact: the surface tension [Eq.~\eqref{eq:sigma_def}], the Laplace relation [Eq.~\eqref{eq:Laplace_pressure_asymptotic}], and the osmotic pressure of the Cahn--Hilliard description reappear in the McRD language as geometric objects on the flux-balance subspace, with $\eta$ taking the structural role of $\mu$ and the turnover-balance condition that of the Maxwell construction (Table~\ref{tab:dictionary}).
These features of mass-conserving reaction--diffusion dynamics are therefore not coincidental but a direct consequence of the dual chemical-potential representation in the fast-interconversion limit.
Beyond the strict dual form considered here, one can ask which effective gradient terms arise from more general reaction networks.
This question is taken up in Ref.~\citep{Toffenetti_Nettuno_Frey:2026}, where a three-component mass-conserving network is reduced to a single conserved scalar field whose square-gradient coefficient depends on density and can change sign, opening a finite-wavelength instability that lies outside the correspondence developed here.

\section{Duality of systems with broken mass-conservation}
\label{sec:broken_mass-conservation}

Weak breaking of global mass conservation is a generic route to wavelength selection far from equilibrium~\citep{Zwicker:2022,Frey_Weyer:2026}.
It arises both in phase-separating mixtures with driven chemical turnover~\citep{Weber_Lee:2019,Zwicker:2022} and in (nearly) mass-conserving reaction--diffusion systems with weak production and degradation~\citep{Frey_Weyer:2026}.
At the molecular level these systems are governed by very different physics---chemical interconversion in a phase-separating mixture versus state-switching and transport in a multi-component reaction--diffusion network---yet the mesoscale phenomenology is strikingly similar.
After an initial coarsening stage driven by mass redistribution, bulk turnover interrupts coarsening and yields stable domain sizes~\citep{Glotzer_Jan:1994, Puri_Frisch:1994, Christensen_Fogedby:1996, Carati_Lefever:1997,
Brauns_Frey:2021, Weyer_Frey:2023}.  
For sufficiently large domains, a distinct instability can set in and drive domain splitting~\citep{Zwicker_Juelicher:2017, Brauns_Frey:2021, Weyer_Frey:2026}.
In this section we show that this similarity is structural.
Cahn--Hilliard dynamics with reactive turnover admits a dual representation within a broader class of reaction--diffusion models, making the shared mechanism explicit.

\smallskip

\paragraph*{Nonequilibrium phase separation.---}
For phase-separating mixtures with chemical interconversion, a standard coarse-grained description augments Cahn--Hilliard transport by local reactive turnover~\cite{Huberman:1976,Puri_Frisch:1994,Glotzer_Jan:1994,Glotzer_Muthukumar:1995}
\begin{align}
\partial_t \phi
= \nabla\!\cdot\!\bigl(M(\phi)\,\nabla \mu_\mathrm{tot}[\phi]\bigr) + s(\phi),
\label{eq:CHR}
\end{align}
where $\phi$ is a composition field, $\mu_\mathrm{tot}[\phi]$ is the Cahn--Hilliard chemical potential [Eq.~\eqref{eq:chemical_potential_CH_model}], and $s(\phi)$ represents energy-fueled local interconversion kinetics.
A formally related Landau--Ginzburg-based ``generalized diffusion'' term also appears in population dispersal models, where $\phi$ denotes a population density and $s(\phi)$ represents local growth~\cite{Cohen_Murray:1981}.

A canonical example is a ternary mixture comprising two species $A$
and $B$ that undergo chemical interconversion,
$A \rightleftharpoons B$, dissolved in an inert solvent~$C$; here
$\phi(\boldsymbol{x},t)$ denotes the local concentration of species~$A$.
Self-attractive interactions of~$A$ relative to the solvent drive
phase separation into $A$-rich and $A$-poor domains, as encoded in
the Cahn--Hilliard free energy~$f(\phi)$.
Mass-action kinetics yield a net production rate $s(\phi) = k_{\leftarrow}[B] - k_{\to}\phi$.
If species~$B$ diffuses much faster than the reaction timescale, its concentration remains spatially uniform and can be absorbed into a redefined rate constant, yielding
\begin{align}
s(\phi)=k_{+}-k_{-}\phi,
\label{eq:linear_source}
\end{align}
where $k_+ \equiv k_{\leftarrow}[B]$ represents constant production of~$A$ from the uniform background of~$B$, and $k_- \equiv k_{\to}$ governs consumption of~$A$ proportional to its local concentration. Such Cahn--Hilliard dynamics with local turnover, when driven out of equilibrium by an external fuel source, can arrest coarsening and select a finite length scale~\citep{Glotzer_Jan:1994, Christensen_Fogedby:1996, Carati_Lefever:1997, Cates_Tailleur:2010}.
Formally, the reactive term can be absorbed into an effective free energy containing a nonlocal long-range repulsive interaction mediated by a Coulomb-like kernel; wavelength selection then arises from the competition between short-range thermodynamic attraction and this effective long-range repulsion~\citep{Christensen_Fogedby:1996}.
An analogous mechanism operates for phase separation on fluctuating membranes, where elastic deformations mediate long-range interactions that arrest coarsening and select a characteristic domain size~\citep{Winter_Frey:2025}.
A general field-theoretic framework for phase separation with non-local interactions has recently been put forward, who find that long-range interactions generically suppress coarsening, while non-local short-range interactions can additionally drive a continuous transition into patterned states~\citep{Thewes_Zwicker:2025}.

\smallskip

\paragraph*{Pattern formation in reaction--diffusion systems with weakly broken mass conservation.}
A different route to a similar phenomenology is provided by (nearly) mass-conserving reaction--diffusion (McRD) systems \citep{Brauns_Frey:2020,Brauns_Frey:2021,Weyer_Frey:2023}.
In their simplest form, two-component McRD models describe diffusion-coupled interconversion between two internal states, $m$ and $c$,
\begin{subequations}
\label{eq:McRD_weak_sources}
\begin{align}
\partial_t m
&= D_m \nabla^2 m + f(m,c) + s_m(m,c),
\\
\partial_t c
&= D_c \nabla^2 c - f(m,c) + s_c(m,c),
\end{align}
\end{subequations}
where $f(m,c)$ redistributes mass between the two states, while the weak terms $s_{m,c}$ represent production and degradation and thereby break strict conservation of the total density $\phi\equiv m+c$.

For strictly mass-conserving two-component dynamics, it was shown that patterns coarsen through mass competition between domains and, generically, do not select an intrinsic wavelength \citep{Brauns_Frey:2020}.
When mass conservation is weakly broken, this mass-competition dynamics is interrupted and a band of stable wavelengths emerges after an initial coarsening stage, while sufficiently wide plateaus can become unstable and split \citep{Brauns_Frey:2021}. 
A singular-perturbation framework unifies these regimes and quantifies how the crossover from coarsening to stable patterns, as well as the onset of plateau instabilities and splitting, depend on the separation of timescales between interconversion and diffusive redistribution \citep{Weyer_Frey:2023}.

\begin{figure*}[!t]
    \centering
    \begin{tikzpicture}
        \node[anchor=south west, inner sep=0] (img) at (0,0) {\includegraphics[width=0.95\textwidth]{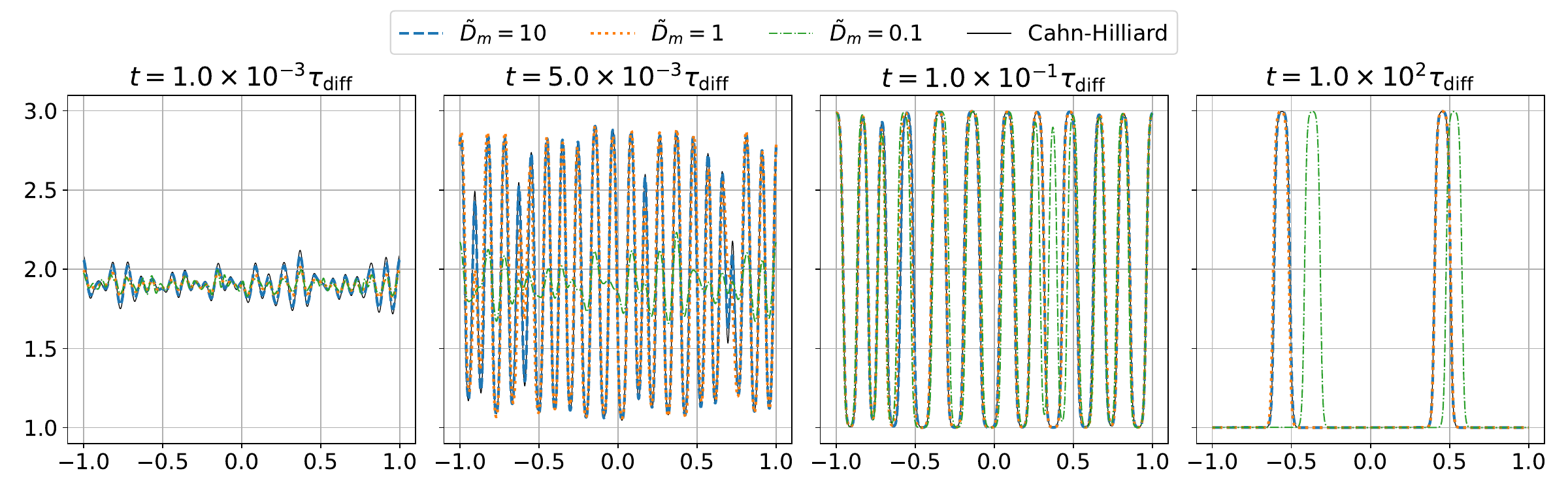}};

        \node[x={(img.south east)},y={(img.north west)}] at (-0.008,0.45) {$\phi$};

        \node[x={(img.south east)},y={(img.north west)}] at (0.5,-0.02) {$x\,[L]$};
    \end{tikzpicture}

    \vspace{4pt}
\caption{Convergence of the dual two-field dynamics to Cahn–Hilliard with reactive turnover in the stiff limit.
Snapshots of the order-parameter profile $\phi(x,t)$ at the indicated times for Cahn–Hilliard with reactive turnover [Eq.~\eqref{eq:CH_model}] (solid black) and for the dual two-field system [Eq.~\eqref{eq:dual_RD}] at different $\tilde{D}_m$ (dashed blue/dark gray: $\tilde{D}_m = 10$; dotted orange/medium gray: $\tilde{D}_m = 1$; dash-dotted green/light gray: $\tilde{D}_m = 0.1$).
As $\tilde{D}_m$ increases, the dual system approaches the one-field Cahn--Hilliard limit; deviations are visible only for $\tilde{D}_m = 0.1$ at late times.
One-dimensional simulations on a periodic domain of length $L = 2$ with initial condition $\phi(x,0) = \bar{\phi} + \delta\phi(x)$, where $\bar{\phi} = 2$ and $\delta\phi(x)$ is small-amplitude random noise.
Time is measured in units of the diffusive timescale $\tau_\mathrm{diff} = L^2 / D_c$.
Remaining parameters are listed in Appendix~\ref{app:all_numerics}.}
    \label{fig:active_time_slides}
\end{figure*}

\subsection{Dual reaction-diffusion model of Cahn–Hilliard with reactive turnover dynamics}\label{sec:num:nmcRD}
Having introduced the one-field Cahn–Hilliard with reactive turnover description \eqref{eq:CHR} and the two-field McRD setting
\eqref{eq:McRD_weak_sources}, we now make the duality explicit.
As in the conservative case discussed in Sec.~\ref{sec:duality-transformation_CH}, the key step is to resolve the dynamics into a fast interconversion between
two internal states and a slow diffusive redistribution of total mass.
This leads to a two-field formulation whose slow-manifold limit reproduces the Cahn--Hilliard equation with reactive turnover.

Writing the state densities as $c$ and $m$, and denoting the interconversion kinetics by an exchange term $\hat A(c,m)$,
we consider
\begin{subequations}
\label{eq:dual_RD}
\begin{align}
\partial_t c
&= D_c\,\nabla^2 c - \hat A(c,m)
   + s \big(\phi(c,m)\big),
\label{eq:dual_RD_c}\\
\partial_t m
&= D_m\,\nabla^2 m + \hat A(c,m).
\label{eq:dual_RD_m}
\end{align}
\end{subequations}
The source term is written in the same functional form as in Eq.~\eqref{eq:CHR}, emphasizing that it breaks
conservation of the $(c,m)$ subsystem in the same controlled way.
The construction and the subsequent reduction proceed exactly as without a source term: the source affects only the
$\phi$-balance, while the fast interconversion encoded in $\hat A$ still enforces local quasi-equilibrium and thereby
defines the same slow manifold.
We relegate the explicit reduction and the mapping back to Eq.~\eqref{eq:CHR} to Appendix~\ref{app:duality_McRD_CH}.

\smallskip

\paragraph*{Numerical analysis of duality.---} 
The dual reduction is exact only in the limit ${\tau \to 0}$, where the exchange term $\hat A(c,m)$ rapidly relaxes the system to the reaction manifold and the remaining slow dynamics is carried by the total field $\phi=c+m$. 
As for the Cahn-Hilliard model [Sec.~\ref{sec:duality-transformation_CH}], the question is therefore how far one can move away from this limit while retaining the same pattern-forming phenomenology.
To test this, we compare Cahn–Hilliard with reactive turnover with the dual two-field system for identical initial conditions while increasing $\tau$ (equivalently, decreasing $D_m$ in our parameterization). Figure~\ref{fig:active_time_slides} shows that slower conversion systematically prolongs the transient: coarsening is delayed and, for sufficiently large $\tau$, the competition between peaks can select different survivors, leading to a shifted late-time configuration. In contrast, for small $\tau$ (or moderate-to-large $D_m$) the dual system closely tracks the one-field dynamics throughout the evolution and converges to the same stationary pattern. 

\subsection{Duality of mechanisms}

The duality developed above is not merely a change of variables: it identifies a common mechanism underlying reactive phase separation and pattern-forming reaction--diffusion dynamics. In Cahn–Hilliard with reactive turnover, diffusive transport is driven by the chemical potential $\mu(\phi)$, whereas in mass-conserving reaction--diffusion systems it is organized by the mass-redistribution potential $\eta$; in both cases, diffusion redistributes the conserved field $\phi$ and competes with local source--sink terms that weakly break global conservation. This section makes that correspondence explicit by translating the droplet-level intuition for Cahn–Hilliard with reactive turnover into the $(\phi,\eta)$ description used for McRD systems.

\medskip

\subsubsection*{Weak sources interrupt conservative coarsening}

In both Cahn--Hilliard dynamics with reactive turnover and McRD systems with weakly broken mass-conservation, the strictly conservative limit produces phase-separated plateaus~\footnote{We restrict ourselves to the case where there is enough mass for each high density domain such that well defined mesas form, and not peaks.} at coexistence values $\phi_-<\phi_+$ connected by narrow interfaces, and the ensuing dynamics is conservative mass competition (coarsening) mediated by gradients of the relevant redistribution potential (the chemical potential $\mu$ for Cahn--Hilliard, and the mass-redistribution potential $\eta$ for McRD).
Weak sources interrupt this conservative coarsening process by selecting a preferred far-field state.
A necessary condition for stationary domains to persist is therefore that the net source admits a \emph{stable} homogeneous fixed point and that this fixed point lies \emph{between} the low- and high-density plateau values selected by the conservative dynamics.
For Cahn–Hilliard with reactive turnover, the fixed point condition is simply
\begin{equation}
s(\phi_\ast)=0,
\qquad
\partial_\phi s(\phi_\ast)<0,
\label{eq:CH_fixed_point}
\end{equation}
while for McRD systems the corresponding condition is evaluated on the manifold of local equilibria 
\begin{equation}
s \bigl(m(\phi_\ast,\eta_\ast),c(\phi_\ast,\eta_\ast)\bigr)=0,
\qquad
\partial_\phi s \bigl(\phi_\ast,\eta_\ast\bigr)<0.
\label{eq:McRD_fixed_point}
\end{equation}
In either case, placing the stable fixed point $\phi_*$ between $\phi_-$ and $\phi_+$ implies opposite signs of the net source on the two plateaus,
\begin{equation}
s(\phi_-)>0,
\qquad
s(\phi_+)<0,
\label{eq:CH_signs}
\end{equation}
and, respectively,
\begin{equation}
s (\phi_-,\eta_-)>0,
\qquad
s (\phi_+,\eta_+)<0,
\label{eq:McRD_signs}
\end{equation}
where $(\phi_\pm,\eta_\pm)$ denote the plateau values on the flux-balance subspace.
Thus, in both settings, production occurs in the dilute phase while degradation occurs in the dense phase, and the resulting steady exchange is compensated by conservative transport across interfaces.

\subsubsection*{Equivalence of the dual models in the sharp-interface limit}

We consider the strong-segregation regime, where a droplet (a ``mesa'' in one dimension) is well approximated by two nearly uniform plateaus, a dense interior at $\phi_\mathrm{ in}$ and a dilute exterior at $\phi_\mathrm{ out}$, separated by a narrow interface.
In the Cahn--Hilliard formulation the microscopic interface width is set by the square-gradient term and scales as $\ell_\kappa\sim\sqrt{\kappa}$.
In the dual reaction--diffusion formulation the corresponding microscopic width is controlled by the internal conversion length of the fast interconversion mode.
Throughout, we assume weak conservation breaking, so that reactions perturb the conservative plateau structure only on mesoscopic scales large compared to the microscopic interface width.
The plateau-scale analysis below is independent of dimension; curvature enters only through interfacial matching (Gibbs--Thomson shift of the interfacial potential) and is not needed here.

We now take the sharp-interface limit.
On the Cahn--Hilliard side this corresponds to $\kappa\to0$.
By the matching relation Eq.~\eqref{eq:kappa-match}, $\kappa=0$ implies $D_m=0$ for finite $M$, $\tau$, and $D_c$, hence $D_c=M$ and the dual mass-redistribution potential reduces to $\eta=c$.

In a stationary state, the Cahn-Hilliard equation with reactive turnover obeys
\begin{equation}
0 = M\,\nabla^2 \mu(\phi) + s(\phi).
\end{equation}
On a plateau $\alpha\in\{\mathrm{in},\mathrm{out}\}$ we write $\phi=\phi_\alpha+\delta\phi$ and linearize
\begin{equation}
\mu(\phi)=\mu(\phi_\alpha)+\chi_\alpha\,\delta\phi,
\qquad
s(\phi)=s_\alpha+s'_\alpha\,\delta\phi,
\end{equation}
with $\chi_\alpha\equiv \mu'(\phi_\alpha)$, $s_\alpha\equiv s(\phi_\alpha)$, and $s'_\alpha\equiv s'(\phi_\alpha)<0$.
In terms of $\delta\mu\equiv \mu-\mu(\phi_\alpha)=\chi_\alpha\,\delta\phi$, one obtains
\begin{equation}
\nabla^2 \delta\mu - \frac{1}{\ell_\alpha^2}\,\delta\mu = -\Gamma_\alpha,
\label{eq:SI_CHR_Helmholtz}
\end{equation}
with screeding length $\ell_\alpha$ and basal source strength $\Gamma_\alpha$:
\begin{equation}
\ell_\alpha^2=\frac{M\chi_\alpha}{|s'_\alpha|},
\qquad
\Gamma_\alpha=\frac{s_\alpha}{M}.
\label{eq:SI_CHR_defs}
\end{equation}

For $D_m=0$, the stationary dual equations reduce to
\begin{subequations}
\begin{align}
0 &= M\nabla^2 c + s(\phi),
\label{eq:SI_dual_stat_c}\\
0 &= A(c,m).
\label{eq:SI_dual_stat_A}
\end{align}
\label{eq:SI_dual_stationary_balance}
\end{subequations}
The nullcline condition \eqref{eq:SI_dual_stat_A} implies, to leading order $0\simeq A_c\,\delta c + A_m\,\delta m$ with $A_c\equiv \partial_c A\big|_\alpha$ and $A_m\equiv \partial_m A\big|_\alpha$ so that ${\delta\phi= (1-A_c/A_m) \, \delta c}$.
Substituting into Eq.~\eqref{eq:SI_dual_stationary_balance} yields
\begin{equation}
\nabla^2\delta c-\frac{1}{\tilde\ell_\alpha^2}\,\delta c=-\Gamma_\alpha,
\qquad
\tilde\ell_\alpha^2=\frac{M}{|s'_\alpha|\,(1-A_c/A_m)}.
\label{eq:SI_dual_Helmholtz}
\end{equation}
Along the nullcline the slow manifold can be parameterized as $c=c_\star(\phi)$, and differentiating $A(c,m)=0$ gives
$\mathrm{d}c_\star/\mathrm{d}\phi=1/(1-A_c/A_m)$.
The duality construction enforces $c_\star(\phi)=\mu(\phi)$, so $\mathrm{d}c_\star/\mathrm{d}\phi=\mu'(\phi)$ and therefore $1-A_c/A_m=1/\chi_\alpha$ on plateau $\alpha$.
Consequently $\tilde\ell_\alpha=\ell_\alpha$ and $\delta c\equiv\delta\mu$, so Eq.~\eqref{eq:SI_dual_Helmholtz} is identical to Eq.~\eqref{eq:SI_CHR_Helmholtz}.
In this sharp-interface and fast-relaxation limit one may thus identify $c=\eta=\mu$ at the level of the plateau-scale theory.
All algebraic steps are given in Appendix~\ref{app:interface_theory_non-conserved}.

\smallskip

The sharp-interface equivalence, $c=\eta=\mu$, implies that the Cahn--Hilliard equation with reactive turnover and its dual McRD formulation share the same mesoscopic description of how domains interact: in both cases, plateau-scale transport and inter-domain coupling are mediated by the same screened-Poisson problem for the redistribution potential, with identical screening lengths and source strengths on each plateau.
This immediately unifies the mesoscale physics: coarsening arrest (interrupted mass competition) and the existence of chemically maintained stationary droplet sizes \citep{Zwicker_Juelicher:2015,Brauns_Frey:2021,Weyer_Frey:2023}, wavelength selection in extended systems as an interrupted-coarsening mechanism that terminates at the stability threshold \citep{Brauns_Frey:2020,Brauns_Frey:2021,Frey_Weyer:2026,Zwicker:2022}, and splitting instabilities of sufficiently large domains, which appear as mesa splitting in 1D and as droplet-division shape instabilities in higher dimensions \citep{Zwicker_Juelicher:2017,Brauns_Frey:2021,Weyer_Frey:2026}.
Thus, once the sharp-interface mapping is established, the Cahn--Hilliard equation with reactive turnover and the dual McRD model become interchangeable representations of the same mesoscopic instability landscape; differences are confined to microscopic interface structure and to regimes where fast interconversion no longer holds.

\subsection{Numerical test of the duality in the multi-mesa regime: phase diagram}
\label{subsec:numerical_tests}

The sharp-interface theory developed above rationalizes stationary mesa widths in the mesa-number-conserving regime, but it does not resolve how \emph{multiple} mesas interact over long times, nor when topology-changing events such as mesa disappearance or mesa splitting occur.  
Numerics therefore serve two purposes: (i) to verify that the dual two-field formulation [Eq.~\eqref{eq:dual_RD}] reproduces the mesoscale dynamics of Cahn–Hilliard with reactive turnover evolution [Eq,~\eqref{eq:CHR}] beyond the single-mesa setting, and (ii) to map, in a controlled way, the dynamical phase diagram separating coarsening, mesa splitting, and a mesa-number-conserving regime.

\smallskip

\paragraph*{Model comparison and choice of dual parameters.---}
Throughout this section we compare direct simulations of Eq.~\eqref{eq:CHR} to simulations of the dual system, Eq.~\eqref{eq:dual_RD}, for identical initial conditions and the same nondimensionalization.  
Guided by the time-dependent benchmark in Fig.~\ref{fig:active_time_slides}, we fix the dual relaxation parameters at values for which the dual dynamics tracks the one-field evolution quantitatively over the full transient, so that deviations from the stiff limit do not affect the outcome classification.
Specifically, in the simulation, we used $\tilde D_m = 10$, which gives the relaxation time scale using the exact matching condition [\eqref{eq:matching}]. 
As seen in Fig.~\ref{fig:active_time_slides}, an auxiliary diffusion of this size will be sufficient to obtain the same profile, even during the fast dynamics after onset of instability.

\smallskip

\begin{figure}[!t]
    \centering
    \includegraphics[width=\columnwidth]{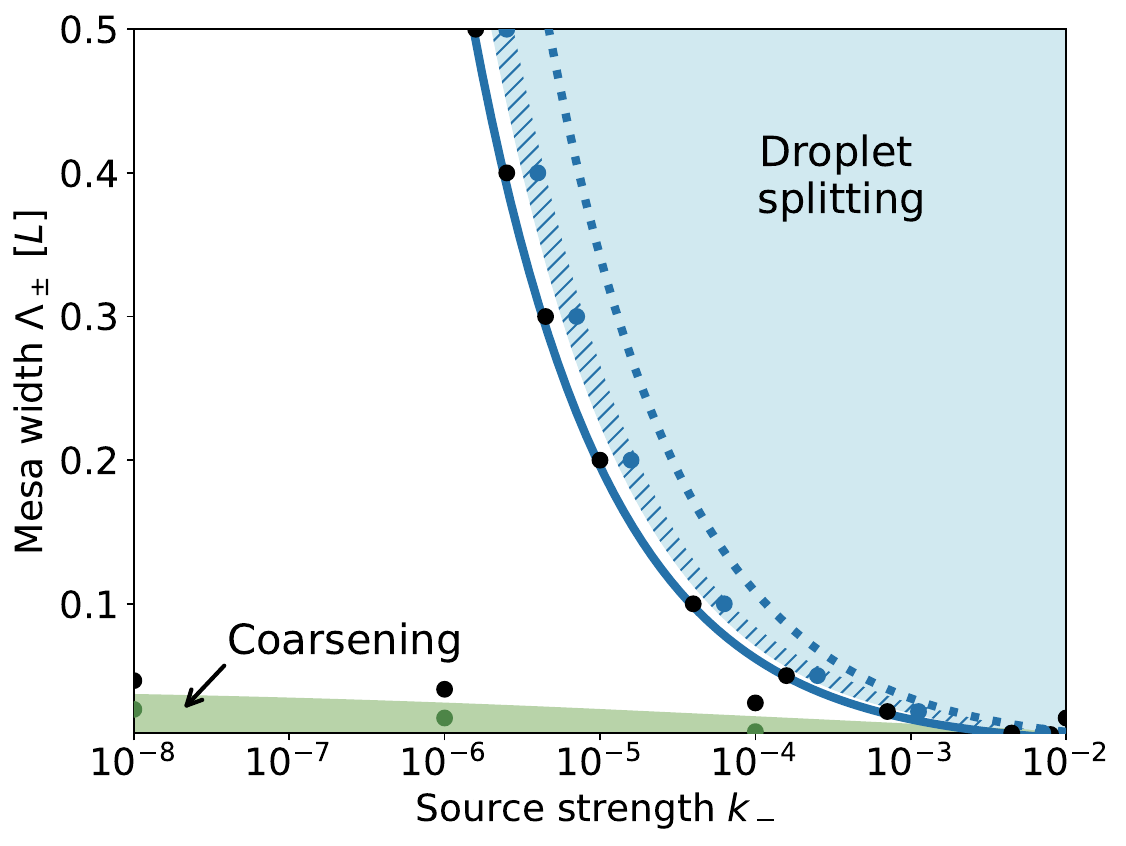}
\caption{
\textbf{Phase diagram: long-time fate of mesa patterns under
weakly broken mass conservation.}
Long-time outcome of a periodic one-dimensional mesa train
for the Cahn--Hilliard model with reactions (CHR) and its dual
McRD system, as a function of two control parameters:
the decay rate $k_-$ at fixed ratio $k_+/k_-$ (horizontal
axis, logarithmic), which sets the overall source strength
without shifting the homogeneous fixed point $\phi^*$; and the
initial mesa width $\Lambda_\pm/L$ (vertical axis), where
$\Lambda_+=\Lambda_-$ by construction of the initial condition
(see text).
Shaded regions indicate coarsening (green/light gray) and mesa splitting (blue/gray); in the intermediate regime the mesa
number is conserved and mesas relax to stationary widths. 
The mesa splitting region exhibits a regime (hatched area), which represents the
increased splitting domain in the alternate simulation protocol (see main text).
Simulations were performed for both models; phase boundaries
and late-time patterns agree within numerical accuracy, so a
single set of symbols is shown.
Symbols mark simulated parameter points at the phase boundary
(black: intermediate regime; green/light gray: coarsening;
blue/dark gray: splitting). (Not plotted are the simulated points at the non-hatched blue/gray to hatched phase boundary as the data points overlap.)
The thick blue (dark gray) curve is the
sharp-interface prediction for the splitting threshold at quadratic order, dotted line of same color is the first order calculation; see
Appendix~\ref{app:interface_theory_non-conserved}.
Remaining parameters and nondimensionalization are given in
Appendix~\ref{app:all_numerics}.
}
\label{fig3:cont-multi-stab-PD}
\end{figure}

\paragraph*{Control parameters.---}
For the canonical linear turnover ${s(\phi)=k_{+}-k_{-}\phi}$,
we parametrize the overall strength of mass nonconservation,
which we refer to as `source strength' in
Fig.~\ref{fig3:cont-multi-stab-PD},
by rescaling $k_+$ and $k_-$ simultaneously at fixed ratio
$k_{+}/k_{-}=\phi^*$, so that the homogeneous fixed point
$\phi^*$ remains unchanged.  We use the decay rate $k_-$ as the
control parameter (horizontal axis in
Fig.~\ref{fig3:cont-multi-stab-PD}); it sets the reaction
timescale $k_-^{-1}$ without shifting the reaction nullcline.

\smallskip

\paragraph*{Initial conditions (mesa train).---}
Simulations are performed in one spatial dimension on a periodic domain of length $2L$.  
Initial conditions are periodic arrays of alternating high- and low-density plateaus of equal width $\Lambda_\pm$, i.e., a mesa train of spatial period $\Lambda = 2 \Lambda_\pm$; accordingly, the vertical axis in Fig.~\ref{fig3:cont-multi-stab-PD} is the normalized initial mesa width $\Lambda_\pm/L$.
Plateau values are chosen close to the two bulk densities selected by phase separation in the conservative limit of Eq.~\eqref{eq:CHR}.
To disentangle droplet shrinking from droplet splitting, we employ a distinct simulation protocol to obtain the additional hatched region of the droplet-splitting regime (Fig.~\ref{fig3:cont-multi-stab-PD}). Rather than initializing mesas of wavelength $\Lambda_\pm$, we simulate a system of fixed size $\Lambda_+$ and impose Dirichlet boundary conditions that pin the field to its high-density value. This setup mimics the upper plateau of a mesa while holding its width fixed, allowing us to isolate the splitting instability from any size evolution. For details, see App.~\ref{app:all_numerics}.

\smallskip

\paragraph*{Numerics and outcome classification.---}
We integrate the Cahn–Hilliard with reactive turnover equation
[Eq.~\eqref{eq:CHR}] and its dual [Eq.~\eqref{eq:dual_RD}]
using a finite-element method with an implicit time integrator
until the pattern reaches a stationary state; discretization
parameters and stopping criteria are given in
Appendix~\ref{app:all_numerics}.

We classify long-time outcomes by comparing the number of high-density mesas in
the final stationary profile to the initial count.
Coarsening corresponds to a net decrease in mesa
number through merging or
disappearance of mesas.
Mesa splitting corresponds to a net increase via nucleation of a
low-density gap within a high-density plateau.
The intermediate regime is defined by conserved mesa number and
relaxation to a stationary multi-mesa pattern.
Individual mesa widths may still evolve within the
intermediate (white/unshaded) region of
Fig.~\ref{fig3:cont-multi-stab-PD}; the classification refers
only to the number of mesas, not their widths.
\smallskip

\paragraph*{Equivalence of the two formulations.---}
For every parameter point, both the
Cahn–Hilliard with reactive turnover equation [Eq.~\eqref{eq:CHR}] and the
dual system [Eq.~\eqref{eq:dual_RD}] were integrated with
identical initial conditions.
Within numerical accuracy, the phase boundaries and late-time
patterns are indistinguishable; accordingly,
Fig.~\ref{fig3:cont-multi-stab-PD} shows a single set of
symbols.
To quantify the agreement, we define the normalized $L^2$
deviation
\begin{equation}
e = \frac{\int_\Omega \d x\, (c+m-\phi_\mathrm{CH})^2}
         {\int_\Omega \d x\, (\phi_\mathrm{CH})^2},
\label{eq:L2_error}
\end{equation}
where $\Omega$ denotes the spatial domain.  For
$\tilde{D}_m=10$, one obtains $e\sim 10^{-12}$ across all
probed parameter points, confirming that the two formulations
yield pointwise identical profiles to within numerical
precision.

\smallskip

\paragraph*{Regime structure in the phase diagram.---}
Figure~\ref{fig3:cont-multi-stab-PD} organizes the
long-time dynamics as a function of two control parameters: the
decay rate $k_-$ (at fixed ratio $k_+/k_-$) and the initial
mesa width $\Lambda_\pm$.
At fixed $k_-$, increasing $\Lambda_\pm$
drives a sequence of outcomes with two sharp transitions:
narrow mesas coarsen (green/light gray
region), intermediate widths evolve at fixed mesa number and
relax to stationary profiles, and sufficiently
wide mesas undergo splitting
(blue/dark gray region).
Both transition lines shift systematically with
$k_-$, reflecting the growing influence of local
reactive turnover on the mass-redistribution
dynamics.\footnote{The choice of initial mesa width as a control parameter is reasonable as the relaxation of the profile from a step function to its quasi steady state happens much quicker than dynamics such as coarsening, shrinkage or mesa splitting.}

\smallskip

\paragraph*{Interrupted coarsening.---}
In the coarsening regime (small initial mesa width), the dynamics proceeds as in mass-conserving phase separation: mesas merge by mass competition and the typical mesa width grows. 
The reason is that, for narrow mesas, the net source--sink imbalance generated in the bulk is weak on the coarsening time scale. 
More precisely, the integrated reaction imbalance inside a mesa scales with its bulk volume ($\sim \Lambda$), whereas the compensating exchange is mediated through the interfaces. 
As mesas widen, the bulk imbalance increases until the interfacial diffusive exchange required to sustain further growth becomes rate limiting~\citep{Brauns_Frey:2021,Weber_Lee:2019}.
The dynamics then crosses over from coarsening to relaxation toward a stationary mesa width $\Lambda^\ast(k)$.

\smallskip

\paragraph*{Mesa splitting.---}
For large initial mesa widths, the plateau regions are no longer flat on the slow turnover time scale: weak production and degradation generate a shallow \emph{curvature} of the profile along each plateau. 
In particular, the high-density plateau is depressed by net degradation and is replenished by inflow through the interfaces, while the dilute plateau is elevated by net production.  
As a result, the plateau values bend toward the homogeneous source-balance composition $\phi_*$ defined by $s(\phi_*)=0$, and the deviation from a constant density increases with plateau length and source strength $k$. 
If $\phi_{-}<\phi_*<\phi_{+}$, the center of a sufficiently long dense plateau is driven far enough away from the coexistence value that it eventually enters the interval of \emph{lateral instability} (spinodal-type regime).  
At that point the plateau cannot remain laterally uniform: an internal trough nucleates inside the mesa and develops into a new low-density region, thereby splitting the original mesa into two or more, depending on how far away from the phase boundary one is.  
Accordingly, increasing $k$ shifts the splitting threshold to smaller $\Lambda$, because stronger turnover increases the plateau curvature and drives the interior more rapidly toward the unstable regime.

The thick and dotted blue curves in Fig.~\ref{fig3:cont-multi-stab-PD} are analytic estimates for the mesa-splitting boundary obtained from the sharp-interface theory in Appendix~\ref{app:interface_theory_non-conserved}. Both are derived by perturbing the upper plateau value of the Cahn--Hilliard equation with reactive turnover, Eq.~\eqref{eq:CHR}, and asking when that plateau becomes unstable to the growth of a central trough. The dotted curve keeps only the leading (linear) term in the perturbation amplitude; the solid curve retains terms through second order. The dual-system formulation makes this expansion straightforward and, in principle, extendable to any order in the perturbation amplitude.

Both curves describe the splitting boundary under a \emph{no-shrinkage assumption}: the mesa width is held fixed while the internal trough develops. This assumption is realized numerically by the alternate protocol, which pins the density to $\phi_+$ on the domain boundary and thereby decouples trough formation from plateau shrinkage; the resulting additional splitting region is the \emph{hatched} area in Fig.~\ref{fig3:cont-multi-stab-PD}. With a free droplet in a periodic domain, net degradation instead shrinks the dense plateau in parallel with trough formation: near the analytic boundary, the mesa can shrink into the stable-width regime before the central minimum reaches the instability threshold, so splitting is averted. This is why the \emph{solid} (non-hatched) splitting region is the smaller of the two. The dotted (linear) curve already locates the hatched boundary qualitatively, but the profile just before splitting deviates appreciably from the $\cosh$-shape predicted at linear order; the second-order (solid) curve corrects for this and gives the quantitative match.

\smallskip

\paragraph*{Crossover and regime of validity.---}
Close to the junction where the coarsening and splitting boundaries meet, the morphology crosses over from strongly segregated mesas to weakly modulated profiles. 
In this region, sufficiently strong sources can dissolve mesas altogether, or leave only small-amplitude peaks rather than well-developed plateaus. 
This is precisely the parameter range where the sharp-interface assumptions underlying the width-selection and splitting theories (flat bulk plateaus separated by narrow interfaces) cease to apply. 
Importantly, despite this crossover, the Cahn–Hilliard with reactive turnover \eqref{eq:CHR} and the dual two-field system \eqref{eq:dual_RD} continue to yield the same regime structure and multistability within numerical accuracy. 
Together with the sharp-interface calculations, these simulations establish that reaction-interrupted coarsening and large-mesa splitting in \eqref{eq:CHR} are mirrored one-to-one, under the duality, by the corresponding weakly nonconservative mechanism in the reaction--diffusion formulation \eqref{eq:dual_RD}.

\section{Multi-component systems}
\label{sec:multi-component_systems}

We now extend the duality developed in Sec.~\ref{sec:duality-transformation_CH} to a set of $N$ conserved fields $\boldsymbol{\phi}(\boldsymbol{x},t)=(\phi_1,\ldots,\phi_N)$, each obeying a continuity equation
\begin{equation}
    \partial_t\phi_i(\boldsymbol{x},t)
    = -\nabla\cdot\boldsymbol{J}_i(\boldsymbol{x},t),
    \qquad i=1,\ldots,N.
    \label{eq:mu_n_continuity}
\end{equation}
We restrict throughout to the \emph{potential-current class}, in which each flux is driven by gradients of chemical potentials,
\begin{equation}
    \boldsymbol{J}_i
    = -\sum_{j=1}^{N}
      M_{ij}(\boldsymbol{\phi})\,
      \nabla\mu_j[\boldsymbol{\phi}],
    \label{eq:mu_n_potential_current}
\end{equation}
where $M_{ij}(\boldsymbol{\phi})$ is a mobility matrix that may depend on the local composition, such that the dynamics takes the form
\begin{equation}
    \partial_t\phi_i
    = \nabla\cdot\sum_{j=1}^{N}
      M_{ij}(\boldsymbol{\phi})\,
      \nabla\mu_j [\boldsymbol{\phi}].
    \label{eq:mu_n_potential_form}
\end{equation}
Equation~\eqref{eq:mu_n_potential_form} reduces to equilibrium multi-component Cahn--Hilliard dynamics when two independent structural conditions are satisfied: the mobility matrix is symmetric, $M_{ij}=M_{ji}$ (Onsager reciprocity), and positive semidefinite; and the chemical potentials are variational, $\mu_i = \delta\mathcal{F}/\delta\phi_i$, for a single free-energy functional $\mathcal{F}[\boldsymbol{\phi}]$.
When both hold, $\mathcal{F}$ is a Lyapunov functional and the dynamics is a gradient flow.
Either condition can fail independently, giving rise to two distinct classes of nonequilibrium behavior.

\smallskip

\paragraph*{Violation of Onsager reciprocity.---}
The chemical potentials may remain variational while the mobility matrix loses its symmetry, breaking Onsager reciprocity~\citep{Onsager:1931a,Casimir:1945,DeGroot_Mazur:1962}.
Decomposing $M = M^{\mathrm{S}} + M^{\mathrm{A}}$ with $M^{\mathrm{S}} = \tfrac{1}{2}(M + M^{\mathsf{T}})$ and $M^{\mathrm{A}} = \tfrac{1}{2}(M - M^{\mathsf{T}})$, the antisymmetric part $M^{\mathrm{A}}$ generates currents that do not contribute to free-energy dissipation while $M^{\mathrm{S}}$ controls dissipation.
The free energy $\mathcal{F}$ therefore remains a Lyapunov functional, but the dynamics is no longer a gradient flow.
\smallskip

\paragraph*{Failure of integrability.---}
Conversely, the mobility matrix may be symmetric while the chemical potentials fail to derive from a single free-energy functional.
For local constitutive laws $\mu_i = \mu_i(\boldsymbol{\phi})$, this requires Maxwell symmetry,
\begin{equation}
    \frac{\partial\mu_i}{\partial\phi_j}
    = \frac{\partial\mu_j}{\partial\phi_i},
    \label{eq:mu_n_maxwell_local}
\end{equation}
i.e., the integrability condition for a free-energy density $f(\boldsymbol{\phi})$ with $\mu_i = \partial f/\partial\phi_i$.
For gradient-dependent theories, integrability furthermore requires self-adjointness of the linearised response operator $L_{ij} = \delta\mu_i/\delta\phi_j$, equivalently the vanishing of the functional curl in field space~\citep{Graham_Haken:1971,Callen:1985}; a canonical violation is provided by Active Model~B-type gradient nonlinearities~\citep{Nardini_Cates:2017,Tjhung_Cates:2018}.

\smallskip

\paragraph*{Equivalent formulations and the role of the mobility matrix.---}
For constant mobilities $M_{ij}$, off-diagonal entries can always be absorbed into a redefinition of the chemical potentials.
Choosing any set of positive constants $M_i > 0$ and defining
\begin{equation}
    \tilde\mu_i(\boldsymbol{\phi})
    \equiv \frac{1}{M_i}\sum_{j} M_{ij}\,\mu_j(\boldsymbol{\phi}),
    \label{eq:mu_tilde_def}
\end{equation}
the dynamics [Eq.~\eqref{eq:mu_n_potential_form}] takes the diagonal form
\begin{equation}
    \partial_t\phi_i
    = M_i\,\nabla^2\tilde\mu_i(\boldsymbol{\phi}).
    \label{eq:diagonal_form}
\end{equation}
No symmetry or diagonalizability of $M_{ij}$ is required; the entire multicomponent coupling now resides in $\tilde\mu_i(\boldsymbol{\phi})$.
If the original $\mu_i$s satisfy Maxwell symmetry [Eq.~\eqref{eq:mu_n_maxwell_local}], the $\tilde\mu_i$ generically do not, and vice versa: the integrability classification depends on the choice of representation.

For composition-dependent mobilities $M_{ij}(\boldsymbol{\phi})$ the situation is more subtle.
The dynamics can be cast in the form $\partial_t\phi_i = \nabla\cdot M_i(\boldsymbol{\phi})\,\nabla\tilde\mu_i$ only if the flux $\sum_j M_{ij}(\boldsymbol{\phi})\,\nabla\mu_j$ is proportional to a gradient for each $i$, which imposes an additional integrability condition on the product $M_{ij}(\boldsymbol{\phi})\,\partial\mu_j/\partial\phi_k$.
For a single conserved field this is automatic: $M(\phi)\mu'(\phi)$ can always be integrated to a function $G(\phi)$ with $G' = M\mu'$, so the flux $M\nabla\mu = \nabla G$ is a pure gradient.
For multiple fields this is not generically satisfied; we leave a systematic treatment to future work.

In the following, we restrict to constant diagonal mobilities $M_{ij} = M_i\,\delta_{ij}$, which by Eqs.~\eqref{eq:mu_tilde_def}--\eqref{eq:diagonal_form} entails no loss of generality for systems with constant mobility matrix.

\subsection{Multi-component Cahn--Hilliard dynamics}
\label{sec:CH_multicomponent}
With constant diagonal mobilities $M_{ij} = M_i\,\delta_{ij}$, we now include the standard square-gradient regularization by taking chemical potentials of the form
\begin{equation}
    \mu_{i,\mathrm{tot}}[\boldsymbol{\phi}]
    = \mu_i(\boldsymbol{\phi})
      - \kappa_i\,\nabla^2\phi_i,
    \label{eq:mu_tot_multicomponent}
\end{equation}
where $\mu_i(\boldsymbol{\phi})$ is a local constitutive relation that may couple all components and need not satisfy Maxwell symmetry [Eq.~\eqref{eq:mu_n_maxwell_local}], and $\kappa_i > 0$ controls the interfacial stiffness for species $i$.
The multi-component Cahn--Hilliard dynamics then reads
\begin{equation}
    \partial_t\phi_i
    = M_i\,\nabla^2\bigl[
        \mu_i(\boldsymbol{\phi})
        - \kappa_i\,\nabla^2\phi_i\bigr].
    \label{eq:CH_multicomponent}
\end{equation}

\smallskip

\paragraph*{Dual reaction--diffusion system.---}
Following the scalar construction (Sec.~\ref{sec:dual_system}), we introduce for each species $i$ a pair of dual fields $c_i$ and $m_i$ with $\phi_i = c_i + m_i$, coupled by a local interconversion reaction,
\begin{subequations}
\label{eq:dual_multicomponent}
\begin{align}
    \partial_t c_i
      &= D_i^c\,\nabla^2 c_i
         - \hat{A}_i(\mathbf{c},\mathbf{m}),
    \label{eq:dual_c_multi} \\
    \partial_t m_i
      &= D_i^m\,\nabla^2 m_i
         + \hat{A}_i(\mathbf{c},\mathbf{m}),
    \label{eq:dual_m_multi}
\end{align}
\end{subequations}
with the interconversion flux
\begin{equation}
    \hat{A}_i(\mathbf{c},\mathbf{m})
    = \frac{1}{\tau_i}\,\bigl[
        c_i - \hat{\mu}_i(\boldsymbol{\phi})\bigr],
    \label{eq:Ahat_multi}
\end{equation}
where $\tau_i > 0$ is the interconversion time scale for species $i$.
The resulting embedding $\boldsymbol{\gamma}(\boldsymbol{\phi})$ is injective and regular, as required by the scalar construction: injectivity follows because $\hat{\mu}_i(\boldsymbol{\phi})=\hat{\mu}_i(\boldsymbol{\phi'})$ for all $i$ together with $m_i = \phi_i - \mu_i(\boldsymbol{\phi})$ implies $\boldsymbol{\phi}=\boldsymbol{\phi'}$; regularity holds by direct differentiation.\footnote{The argument generalizes the scalar injectivity condition of Sec.~\ref{sec:dual_system}: the multicomponent curve $\boldsymbol{\gamma}$ has components $(\hat{\mu}_i, m_i)$ for each species, and the proof proceeds componentwise.}
Each species diffuses independently and interconverts solely between its own $c_i$ and $m_i$ forms, so that no mass is transferred between different species.
The multicomponent coupling enters only through the target function $\hat{\mu}_i(\boldsymbol{\phi})$, which determines the \emph{rate} of the $c_i\leftrightarrow m_i$ interconversion as a function of all species densities via the constitutive relation $\mu_i(\boldsymbol{\phi})$.

Substituting $c_i = (\eta_i - d_i\,\phi_i)/(1-d_i)$ [cf.\ Eq.~\eqref{eq:Ahat_eta}], the interconversion flux takes the form
\begin{equation}
    \hat{A}_i
    = \frac{D_i^c}{\tau_i\,M_i}\,
      \bigl[\eta_i - \eta_i^*(\boldsymbol{\phi})\bigr],
    \label{eq:Ahat_eta_multi}
\end{equation}
where $\eta_i \equiv c_i + d_i\,m_i$ is the componentwise mass-redistribution potential ($d_i \equiv D_i^m/D_i^c$) and
\begin{equation}
    \eta_i^*(\boldsymbol{\phi})
    = d_i\,\phi_{i,\mathrm{eq}}
      + (1-d_i)\,\mu_i(\boldsymbol{\phi})
    \label{eq:eta_star_multi}
\end{equation}
is the reactive nullcline for species $i$ in the $(\phi_i,\eta_i)$-plane; here $\phi_{i,\mathrm{eq}}$ is an arbitrary reference density for each species [cf.\ Eq.~\eqref{eq:def-of-mu-hat}].
As in the scalar case, the reaction drives $\eta_i$ toward $\eta_i^*(\boldsymbol{\phi})$: on the nullcline $\hat{A}_i = 0$ and local chemical equilibrium holds; on the flux-balance subspace $\eta_i = \eta_{i,\mathrm{stat}}$, by contrast, the diffusive fluxes of $c_i$ and $m_i$ balance globally, and $\hat{A}_i$ does not vanish locally.

\smallskip

\paragraph*{Matching conditions.---}
Since the interconversion acts independently on each species, the adiabatic reduction to the $\phi_i$-dynamics proceeds for each component $i$ by the same argument as in the scalar case  (Sec.~\ref{sec:kappa_removal}).
The modified chemical potential
\begin{equation}
    \hat{\mu}_i(\boldsymbol{\phi})
    = \mu_i(\boldsymbol{\phi})
      - \frac{D_i^m}{M_i}\,
        (\phi_i - \phi_{i,\mathrm{eq}})
    \label{eq:muhat_multi}
\end{equation}
absorbs the Fickian diffusion of $m_i$, and the diffusivity contrast fixes the mobility,
\begin{equation}
    D_i^c - D_i^m = M_i.
    \label{eq:Dc_match_multi}
\end{equation}
Eliminating $m_i$ from the steady-state equations using the componentwise flux-balance constraint $\nabla^2\eta_i = 0$ generates an effective fourth-order term whose coefficient matches the Cahn--Hilliard stiffness when
\begin{equation}
    \kappa_i \, M_i
    =  \tau_i\,D_i^m\,D_i^c \, ,
    \label{eq:kappa_match_multi}
\end{equation}
the componentwise generalization of Eq.~\eqref{eq:kappa-match}.
The detailed calculation is given in Appendix~\ref{app:duality_McRD_CH}.

\smallskip

\paragraph*{Full dynamics.---}
In the fast-reaction limit $\tau_i \to 0$ at fixed $M_i$ and $\kappa_i$, the adiabatic elimination of $m_i$ yields the implicit form (see Appendix~\ref{app:duality_McRD_CH}; cf.\ Eq.~\eqref{eq:dyn_reduction_maintext} for the scalar case)
\begin{equation}
    \bigl(1 - \ell_i^2\,\nabla^2\bigr)\,
    \partial_t\phi_i
    = M_i\,\nabla^2\bigl[
        \mu_i(\boldsymbol{\phi})
        - \kappa_i\,\nabla^2\phi_i\bigr]
      + O(\tau_i),
    \label{eq:implicit_multi}
\end{equation}
with the reaction--diffusion length
\begin{equation}
    \ell_i^2
    \equiv (D_i^m + D_i^c)\,\tau_i\,.
    \label{eq:ell_multi}
\end{equation}
Inverting $(1-\ell_i^2\nabla^2)$ for small $\ell_i$ confirms that each component recovers its Cahn--Hilliard dynamics [Eq.~\eqref{eq:CH_multicomponent}] up to a controlled $O(\sqrt{\tau_i}\,)$ correction,
\begin{equation}
    \partial_t\phi_i
    = M_i\,\nabla^2\bigl[
        \mu_i(\boldsymbol{\phi})
        - \kappa_i\,\nabla^2\phi_i\bigr]
      + O(\sqrt{\tau_i}).
    \label{eq:CH_recovered_multi}
\end{equation}
\smallskip

\paragraph*{Scaling regime.---}
The matching condition [Eq.~\eqref{eq:kappa_match_multi}] requires both diffusivities to diverge in the fast-reaction limit, $D_i^m \sim D_i^c \sim \tau_i^{-1/2}$ and $1 - d_i \sim \sqrt{\tau_i}$, while their difference $D_i^c - D_i^m = M_i$ stays finite.
As in the scalar case, the reaction rate on the quasi-steady manifold scales as $\hat{A}_i \approx -D_i^m\,\nabla^2 m_i = O(\tau_i^{-1/2})$: reaction and diffusion remain in active, spatially varying balance across the interface for each species independently.

\medskip

To summarize: since any constant mobility matrix can be diagonalized by absorbing off-diagonal entries into the chemical potentials [Eqs.~\eqref{eq:mu_tilde_def}--\eqref{eq:diagonal_form}], the dual construction applies componentwise and covers the full class of constant-mobility multicomponent models.
The resulting dual system is species-diagonal: diffusion and interconversion act independently on each $(c_i, m_i)$ pair, with no cross-species transport or reaction terms.
All multicomponent structure, including interspecies coupling, the topology of the phase diagram, and possible violations of Maxwell symmetry or Onsager reciprocity, is encoded in the constitutive relations $\mu_i(\boldsymbol{\phi})$; the dual requires no integrability or specific functional form of these relations.
In the following subsection, we exploit this generality to analyze traveling-wave solutions in a two-component system with broken Maxwell symmetry.

\subsection{Sharp-interface theory of nonreciprocal traveling waves via the duality}
\label{subsec:tw_velocity}

We now specialize to two interacting conserved fields $\phi_1(\boldsymbol{x},t)$ and $\phi_2(\boldsymbol{x},t)$ evolving as
\begin{subequations}\label{eq:tw:orig}
\begin{align}
  \partial_t \phi_1
    &= D_1\nabla^2\bigl(\mu_1(\phi_1,\phi_2)
       - \kappa\,\nabla^2\phi_1\bigr),
  \label{eq:tw:orig_phi1}\\
  \partial_t \phi_2
    &= D_2\nabla^2\mu_2(\phi_1,\phi_2),
  \label{eq:tw:orig_phi2}
\end{align}
\end{subequations}
with local chemical potentials
\begin{subequations}\label{eq:tw:pot}
\begin{align}
  \mu_1(\phi_1,\phi_2)
    &= -r\,\phi_1 + \phi_1^3 + \alpha_{12}\,\phi_2,
  \label{eq:tw:mu1}\\
  \mu_2(\phi_1,\phi_2)
    &= \phi_2 + \alpha_{21}\,\phi_1.
  \label{eq:tw:mu2}
\end{align}
\end{subequations}
Here $D_1$ and $D_2$ play the role of the mobilities $M_i$ in Eq.~\eqref{eq:CH_multicomponent}.
The parameter $r>0$ places the $\phi_1$-sector in the phase-separating regime (symmetric $\phi^4$ theory) with interfacial stiffness $\kappa$, while $\phi_2$ is purely diffusive ($\kappa_2=0$).
Nonreciprocity enters through the cross-couplings $\alpha_{12}\neq\alpha_{21}$, which violate Maxwell symmetry [Eq.~\eqref{eq:mu_n_maxwell_local}].
Equations~\eqref{eq:tw:orig} and~\eqref{eq:tw:pot} define the \emph{nonreciprocal Cahn--Hilliard} (NRCH) model~\citep{Saha_Golestanian:2020,You_Marchetti:2020} as a minimal continuum description of pattern formation in active mixtures with broken Maxwell symmetry.

\begin{figure}[!t]
  \centering
  \includegraphics[width=\columnwidth]{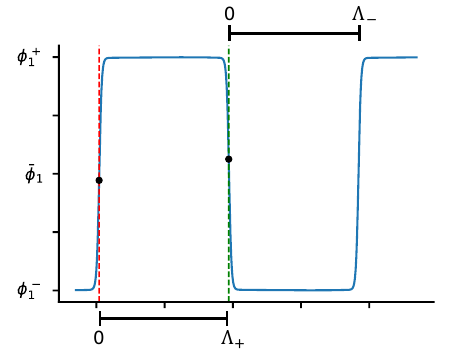}
\caption{%
    \textbf{Geometry of the sharp-interface traveling-wave ansatz.}
    Comoving-frame profile of $\phi_1(z)$ from a numerical
    simulation (blue/dark gray solid curve), showing alternating
    high- and low-density plateaus $\phi_1^\pm$ separated by
    narrow interfaces.
    Plateau endpoints are defined by the inflection points of
    each interface ($\phi_1''=0$): the left inflection point
    (red/medium gray dashed line) marks the start, and the right
    inflection point (green/light gray dashed line) marks the end
    of the adjacent plateau.
    This convention defines the mesa widths $\Lambda_{+}$
    (high-density) and $\Lambda_{-}$ (low-density).
    Each plateau is parameterized by a local coordinate
    $z\in[0,\Lambda_{\pm}]$; black dots mark the periodic
    matching points where adjacent plateaus connect.
    The example shown has a moderate interface width chosen for
    visual clarity; see Fig.~\ref{fig:TW-ansatz-breakdown} for
    additional cases.
    Simulation parameters are listed in
    Appendix~\ref{app:all_numerics}.
}
\label{fig:TW-phi-profile}
\end{figure}

In one spatial dimension, the NRCH model exhibits two regimes~\cite{Saha_Golestanian:2020, You_Marchetti:2020,Brauns_Marchetti:2024, Frohoff_Thiele:2021, Rana_Golestanian:2024}.
For weak nonreciprocity the system coarsens to complete phase separation, as in the conventional Cahn--Hilliard equation.
Beyond a critical coupling strength, $\alpha_{12}\alpha_{21}<0$ with $|{\alpha_{12}\alpha_{21}}|$ sufficiently large, stationary states cease to exist and are replaced by \emph{traveling mesa structures} (Fig.~\ref{fig:TW-phi-profile}), extended plateaus at the coexistence densities~$\phi_1^\pm$, separated by narrow interfaces, that propagate at a constant velocity.
The onset of propagation is an oscillatory (conserved-Hopf) instability of the interfacial region, where the two hydrodynamic modes associated with the conserved densities coalesce at an exceptional point~\cite{Fruchart_Vitelli:2021,Brauns_Marchetti:2024}.

A striking feature of the traveling-wave regime is \emph{interrupted coarsening without wavelength selection}: in the initial transient, domains travel at different speeds and coarsen via chase-and-run collisions until all domains achieve the same velocity, at which point coarsening halts.
The resulting steady state is a periodic wave train whose wavelength (equivalently, the mesa width~$\Lambda_+$) is not uniquely selected by the dynamics but depends sensitively on initial conditions~\cite{Brauns_Marchetti:2024,Saha_Golestanian:2020}.
Stable wave trains exist for a finite range of mesa widths, from the interface width up to an upper bound.  Where exactly this upper bound is, remains an open question at present.
The propagation velocity decreases with increasing mesa width, so that both wavelength and speed are set by the preparation of the system rather than by the model parameters alone.

\medskip

\paragraph*{Sharp-interface theory via the duality.}
Despite the detailed numerical and linear-stability characterization of the NRCH traveling waves, an analytical sharp-interface theory for their velocity has remained out of reach.
The fundamental obstacle is that the sharp-interface limit $\kappa\to 0$ in the Cahn--Hilliard formulation [Eq.~\eqref{eq:tw:orig}] is a singular limit: the $\kappa$-term, which regularizes the interface, generates distributional contributions that must be tracked through a formal matched-asymptotic expansion, and the absence of a variational structure in the nonreciprocal case prevents the use of standard free-energy-based sharp-interface reductions.

The duality developed in the preceding sections resolves this difficulty.
In the dual McRD representation, the sharp-interface limit corresponds not to a singular limit but to the regular limit $D^m_i\to 0$ in which the auxiliary pool becomes immobile.
In this limit, only the diffusive variables~$c_i$ transport mass, and the problem acquires a clean structure: bulk plateaus are characterized by \emph{local reactive equilibrium} ($A_i=0$), the spatially constant redistribution potential is fixed by an \emph{interfacial turnover balance} (the McRD counterpart of a Maxwell construction), and traveling-wave profiles can be constructed by elementary matching of piecewise-analytic solutions.
The simple calculation below, which yields a closed, implicit formula for the wave velocity, would not be possible in the original Cahn--Hilliard variables, and thus illustrates the analytical power of the duality in a concrete, physically relevant setting.

\medskip

\paragraph*{Dual McRD model.}
We now carry out this program.
Setting the interface stiffness ${\kappa=0}$ and taking the sharp-interface limit ${D^m_i\to 0}$, we decompose ${\phi_i=c_i+m_i}$ and write the dual McRD system as
\begin{subequations}\label{eq:tw:dual}
\begin{align}
  \partial_t c_i &= D_i\nabla^2 c_i - A_i,
  \label{eq:tw:dual_c}\\
  \partial_t m_i &= A_i,
  \label{eq:tw:dual_m}
\end{align}
\end{subequations}
where $D_i \equiv D_i^c = M_i$ in the limit $D_i^m=0$ [Eq.~\eqref{eq:Dc_match_multi}].
Summing these two equations eliminates the exchange terms and gives the conservation law
\begin{equation}
  \partial_t \phi_i = D_i\nabla^2 c_i,
  \qquad
  \boldsymbol{J}_i = -D_i\nabla c_i,
  \label{eq:tw:mredistribution}
\end{equation}
identifying $c_i$ as the mass-redistribution potential whose gradients drive the flux of the total density~$\phi_i$ (cf.\ Sec.~\ref{sec:kappa_removal}).
The exchange terms $A_i$ encode local interconversion between $m_i$ and~$c_i$.
Their form is chosen so that local reactive equilibrium, $A_i=0$, enforces the constitutive relation $c_i=\mu_i(\phi_1,\phi_2)$:
\begin{subequations}\label{eq:tw:Adefs}
\begin{align}
  A_1 &\equiv c_1 - \mu_1(\phi_1,\phi_2)
       = c_1 + r\,\phi_1 - \phi_1^3 - \alpha_{12}\,\phi_2,
  \label{eq:tw:A1}\\
  A_2 &\equiv c_2 - \mu_2(\phi_1,\phi_2)
       = -m_2 - \alpha_{21}\,\phi_1,
  \label{eq:tw:A2}
\end{align}
\end{subequations}
where in the second line we used $c_2-\phi_2=-m_2$ and $\mu_2=\phi_2+\alpha_{21}\phi_1$.
Setting $A_i=0$ and substituting the resulting $c_i=\mu_i$ into Eq.~\eqref{eq:tw:mredistribution} recovers the original conserved dynamics at $\kappa=0$.
The dual formulation thus cleanly separates \emph{transport} (diffusion of the redistribution potentials~$c_i$) from \emph{local interconversion} (exchange via~$A_i$), a structure we exploit throughout the remainder of this section.

\medskip

\paragraph*{Stationary sharp-interface states.}
Before constructing traveling waves, we determine the stationary plateau values.
In the dual McRD picture, bulk \emph{densities} follow from local reactive equilibrium ($A_i=0$), while the uniform level of the redistribution potential is selected by the interfacial turnover balance (Sec.~\ref{sec:maxwell}).

\emph{Bulk plateaus from local reactive equilibrium.}  In a sharp-interface stationary pattern the bulk domains are nearly homogeneous, so $A_i\simeq 0$.
From $A_2=0$ one immediately obtains
\begin{equation}
  m_2 = -\alpha_{21}\,\phi_1.
  \label{eq:tw:elim2}
\end{equation}
Substituting this result into $A_1=0$ yields
\begin{equation}
  0 = \tilde c_1 + \tilde r\,\phi_1 - \phi_1^3,
  \label{eq:tw:A1_reduced}
\end{equation}
with the effective parameters~\cite{Brauns_Marchetti:2024}
\begin{equation}
  \tilde r \equiv r + \alpha_{12}\alpha_{21},
  \qquad
  \tilde c_1 \equiv c_1 - \alpha_{12}\,c_2.
  \label{eq:tw:tilde}
\end{equation}
Bulk reactive equilibrium therefore constrains $(\phi_1,\tilde c_1)$ to the cubic nullcline
\begin{equation}
  \tilde c_1 = -\tilde r\,\phi_1 + \phi_1^3.
  \label{eq:tw:nullcline}
\end{equation}
For a given uniform level~$\tilde c_1$, the admissible plateau values $\phi_1^\pm$ are the roots of Eq.~\eqref{eq:tw:nullcline}; the value of $\tilde c_1$ itself is not fixed by bulk equilibrium alone.

\emph{Potential levels from turnover balance.}  Stationarity requires that the mass-redistribution potentials $c_i$ are spatially uniform ($\nabla^2 c_i=0$).
Their values are selected by the reactive turnover balance, which in the sharp-interface regime reduces to the solvability condition [Eq.~\eqref{eq:McRD_equal_potential}], the McRD counterpart of Maxwell's equal-area construction (Sec.~\ref{sec:maxwell}).
For symmetric mean density, ${\bar\phi_1=0}$, the ${\phi_1\to -\phi_1}$ symmetry of the nullcline [Eq.~\eqref{eq:tw:nullcline}] forces ${\tilde c_1=0}$, and Eq.~\eqref{eq:tw:nullcline} then gives the binodal densities
\begin{equation}
  \phi_1^\pm = \pm\sqrt{\tilde r}
             = \pm\sqrt{r + \alpha_{12}\alpha_{21}}\,.
  \label{eq:tw:binodals}
\end{equation}

\medskip

\paragraph*{Comoving-frame formulation.}
To analyze traveling structures we pass to the comoving coordinate $z \equiv x - vt$ and seek time-independent profiles.
The substitutions $\partial_t\to -v\,\partial_z$ and $\nabla\to\partial_z$ (restricting to one-dimensional flat interfaces) cast the dual system [Eqs.~\eqref{eq:tw:dual}] into
\begin{subequations}\label{eq:tw:comoving}
\begin{align}
  -v\,c_i' &= D_i\,c_i'' - A_i,
  \label{eq:tw:comoving_c}\\
  -v\,m_i' &= A_i,
  \label{eq:tw:comoving_m}
\end{align}
\end{subequations}
where primes denote $\partial_z$.
Summing Eqs.~\eqref{eq:tw:comoving_c} and~\eqref{eq:tw:comoving_m} eliminates the exchange terms and yields an exact conservation law for the total density,
\begin{equation}
  -v\,\phi_i' = D_i\,c_i''.
  \label{eq:tw:phi_cons_comoving}
\end{equation}
This is the key structural simplification: Eq.~\eqref{eq:tw:phi_cons_comoving} involves only the redistribution potential~$c_i$ and the conserved field~$\phi_i$, irrespective of the form of the reaction term~$A_i$.
Integrating once gives the spatially constant mass flux
\begin{equation}
  J_i = -D_i\,c_i' - v\,\phi_i
  \label{eq:tw:fluxlaw}
\end{equation}
as required by stationarity in the comoving frame, and provides a useful identity for the immobile pool,
\begin{equation}
  m_i = \phi_i - c_i
      = -\frac{D_i}{v}\,c_i' - c_i - \frac{J_i}{v}\,,
  \label{eq:tw:m_from_c}
\end{equation}
which allows us to eliminate $m_i$ in favor of $c_i$ and a few integration constants.
Importantly, Eq.~\eqref{eq:tw:m_from_c} is exact in the comoving steady state; no sharp-interface approximation has been invoked so far.

\medskip

\paragraph*{Sharp-interface plateau ansatz for $\phi_1$.}
As illustrated in Fig.~\ref{fig:TW-phi-profile}, the phase-separating field~$\phi_1$ in the sharp-interface regime consists of extended bulk plateaus separated by narrow interfacial layers.
In the bulk,
\begin{equation}
  \phi_1(z) \simeq \phi_1^\pm,
  \qquad
  \phi_1'(z) \simeq 0,
  \label{eq:tw:plateau_phi1}
\end{equation}
with all rapid variation of $\phi_1$ confined to the interfaces, and $\phi_1^\pm$ the coexistence values from Eq.~\eqref{eq:tw:binodals}.
To make the sharp-interface construction precise at finite interface width, we define plateau \emph{endpoints} operationally as the inflection points $\phi_1''(z)=0$ bracketing each interface (Fig.~\ref{fig:TW-phi-profile}).
The distances between successive endpoints define the mesa widths $\Lambda_{+}$ and $\Lambda_{-}$.
Each plateau is parameterized on its own coordinate interval, $z\in[0,\Lambda_{+}]$ for the $(+)$ domain and $z\in[0,\Lambda_{-}]$ for the $(-)$ domain, with periodicity imposed by identifying the interval endpoints,
\begin{equation}
\begin{aligned}
  z=0\ \text{in $(+)$}
    &\equiv z=\Lambda_{-}\ \text{in $(-)$},\\
  z=\Lambda_{+}\ \text{in $(+)$}
    &\equiv z=0\ \text{in $(-)$}.
\end{aligned}
\label{eq:tw:periodic_id}
\end{equation}

\medskip

\begin{figure}[!t]
  \centering
  \includegraphics[width=\columnwidth]{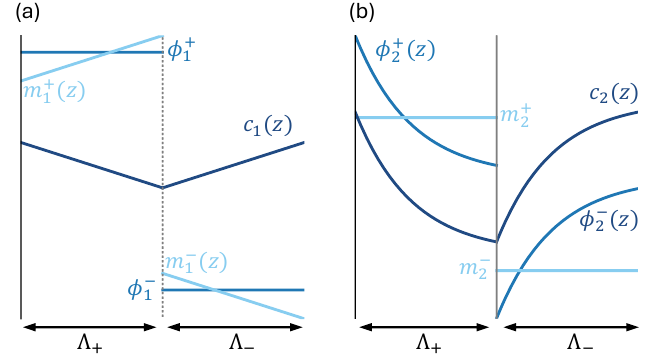}
\caption{%
    \textbf{Sharp-interface traveling-wave profiles from the analytical construction.}
    Analytical profiles in the comoving coordinate ${z=x-vt}$ over one period ${\Lambda=\Lambda_++\Lambda_-}$ (cf.\ Fig.~\ref{fig:TW-phi-profile}).
    (a)~Phase-separating component: ${\phi_1(z)}$ is approximated by flat coexistence plateaus ${\phi_1^\pm}$ of widths ${\Lambda_\pm}$ (blue/gray); the redistribution potential is piecewise affine, ${c_1^\pm(z)=s_\pm z+b_\pm}$ [Eq.~\eqref{eq:tw:c1_linear}] (dark blue/dark gray), with slopes fixed by enforcing endpoint reactive equilibrium, ${\tilde c_1 \equiv c_1-\alpha_{12}c_2\simeq 0}$ [Eq.~\eqref{eq:tw:c1_alpha12c2_endpoints}].
    (b)~Coupled component: on each $\phi_1$-plateau, bulk equilibrium implies a piecewise constant immobile pool ${m_2^\pm=-\alpha_{21}\phi_1^\pm}$ [Eq.~\eqref{eq:tw:elim2}] (light blue/light gray), while the total density relaxes exponentially, ${\phi_2^\pm(z)=C+B_\pm e^{-vz/D_2}}$ [Eq.~\eqref{eq:tw:phi2_sol}] (blue/gray).
    Continuity of ${c_2}$ across the sharp interfaces determines~${B_\pm}$ [Eq.~\eqref{eq:tw:B_solutions}] (dark blue/dark gray).
    The resulting slope mismatch ${s_- - s_+}$ selects the propagation speed via the kinematic jump condition ${v\,\Delta\phi=D_1(s_- - s_+)}$.
    }
  \label{fig:TW-analytical-profiles}
\end{figure}

\paragraph*{Linear profiles of the dual fields.}
On each plateau, $\phi_1(z)\simeq\phi_{1,0}$ implies $c_1''\simeq 0$ via Eq.~\eqref{eq:tw:phi_cons_comoving}, so the redistribution potential~$c_1(z)$ is piecewise affine [Fig.~\ref{fig:TW-analytical-profiles}(a)],
\begin{equation}
  c_1^{\pm}(z) = s_{\pm}\,z + b_{\pm}\,.
  \label{eq:tw:c1_linear}
\end{equation}
The slopes are determined by the constant comoving flux [Eq.~\eqref{eq:tw:fluxlaw}],
\begin{equation}
  s_\pm = -\frac{J_1 + v\,\phi_1^\pm}{D_1}\,,
  \label{eq:tw:c1_slope}
\end{equation}
and $\phi_1=c_1+m_1$ gives $m_1^{\pm}(z)=\phi_1^\pm-c_1^{\pm}(z)$.
The traveling-wave problem thus reduces to a finite set of unknowns, the slopes~$s_{\pm}$, offsets~$b_{\pm}$, flux constants~$J_i$, and the speed~$v$, constrained by matching conditions across the interfaces.

\medskip

\paragraph*{Solving the $\phi_2$-sector on $\phi_1$ plateaus.}
On a $\phi_1$-plateau, bulk local equilibrium in the $\phi_2$-sector ($A_2\simeq 0$) fixes the immobile pool algebraically via Eq.~\eqref{eq:tw:A2},
\begin{equation}
  m_2^\pm = -\alpha_{21}\,\phi_1^\pm,
  \label{eq:tw:m2_pm}
\end{equation}
so that $m_2$ jumps together with $\phi_1$ across each interface [Fig.~\ref{fig:TW-analytical-profiles}(b)].
Since $m_2$ is constant on each plateau, $\phi_2' = c_2'$, and the comoving flux condition [Eq.~\eqref{eq:tw:fluxlaw}] reduces to a first-order ordinary differential equation,
\begin{equation}
  D_2\,\phi_2' + v\,\phi_2 = -J_2\,.
  \label{eq:tw:phi2_ode}
\end{equation}
The total density~$\phi_2$ therefore relaxes exponentially on each plateau,
\begin{equation}
  \phi_2^\pm(z) = C + B_\pm\,e^{-zv/D_2},
  \qquad
  C \equiv -\frac{J_2}{v}\,,
  \label{eq:tw:phi2_sol}
\end{equation}
and the corresponding mass redistribution potential reads
\begin{equation}
  c_2^\pm(z) = \phi_2^\pm(z) - m_2^\pm
             = C - m_2^\pm + B_\pm\,e^{-zv/D_2}.
  \label{eq:tw:c2_sol}
\end{equation}
Because $c_2$ is the redistribution potential for $\phi_2$ [Eq.~\eqref{eq:tw:mredistribution}], it must be continuous across the sharp interfaces [Fig.~\ref{fig:TW-analytical-profiles}(b)].
With the plateau parameterization of Fig.~\ref{fig:TW-phi-profile}, this continuity requirement reads
\begin{equation}
  c_2^{+}(0) = c_2^{-}(\Lambda_-),
  \qquad
  c_2^{+}(\Lambda_+) = c_2^{-}(0).
  \label{eq:tw:c2_cont}
\end{equation}
Inserting Eq.~\eqref{eq:tw:c2_sol} and writing $\Delta\phi\equiv\phi_1^+-\phi_1^-$ gives
\begin{subequations}\label{eq:tw:B_relations}
\begin{align}
  B_{+} + \alpha_{21}\,\Delta\phi
    &= B_{-}\,e^{-\Lambda_- v/D_2},\\
  B_{+}\,e^{-\Lambda_+ v/D_2} + \alpha_{21}\,\Delta\phi
    &= B_{-}\,,
\end{align}
\end{subequations}
whose solution is
\begin{subequations}\label{eq:tw:B_solutions}
\begin{align}
  B_{+}
    &= -\alpha_{21}\,\Delta\phi\;
       \frac{1 - e^{-\Lambda_- v/D_2}}{1 - e^{-\Lambda v/D_2}}\,,\\
  B_{-}
    &= +\alpha_{21}\,\Delta\phi\;
       \frac{1 - e^{-\Lambda_+ v/D_2}}{1 - e^{-\Lambda v/D_2}}\,,
\end{align}
\end{subequations}
with $\Lambda\equiv \Lambda_+ + \Lambda_-$ the total period.

The remaining constant $C=-J_2/v$ is fixed not by redistribution potential continuity but by the conserved spatial mean~$\bar{\phi}_2$.
Integrating $\phi_2$ over one period,
\begin{equation}
  \int_{0}^{\Lambda_+}\!\d z\,\phi_2^{+}(z)
  + \int_{0}^{\Lambda_-}\!\d z\,\phi_2^{-}(z)
  = \Lambda\,\bar{\phi}_2\,,
  \label{eq:tw:phi2_mass}
\end{equation}
and using Eqs.~\eqref{eq:tw:phi2_sol} and~\eqref{eq:tw:B_solutions}, one finds
\begin{equation}
  C = \bar{\phi}_2\,.
  \label{eq:tw:C_from_mass}
\end{equation}
Crucially, $C$ cancels from the matching conditions [Eq.~\eqref{eq:tw:c2_cont}] and from every redistribution-potential difference that enters below; it is needed to reconstruct the full $\phi_2$-profile but does not affect the velocity relation derived next.

\medskip

\paragraph*{Fixing the redistribution potential $c_1$ from local reactive equilibrium.}
We next determine the piecewise-affine mass-redistribution potential $c_1(z)$ on each $\phi_1$-plateau.
Away from the narrow interfaces the bulk dynamics relaxes close to local reactive equilibrium, while transport is encoded by mass-redistribution potential gradients.
Within the mesa ansatz (flat $\phi_1$ and affine $c_1$), the exchange constraints cannot be enforced pointwise throughout the plateau; instead, we impose $A_1\simeq 0$ at the plateau endpoints (matching points to the interfacial inner layers), which fixes the boundary values of $c_1$ and thereby the affine plateau profile.

Using $\phi_1=\phi_1^\pm$ [Eq.~\eqref{eq:tw:binodals}] and the $\phi_2$-sector equilibrium result $m_2^\pm=-\alpha_{21}\phi_1^\pm$ [Eq.~\eqref{eq:tw:m2_pm}], imposing $A_1=0$ at a plateau endpoint gives
\begin{equation}
  c_1 = -r\,\phi_1 + \phi_1^3 + \alpha_{12}(c_2 + m_2).
  \label{eq:tw:c1_mu1_bulk}
\end{equation}
On the coexistence plateaus, $(\phi_1^\pm)^2 = r + \alpha_{12} \, \alpha_{21}$, so the polynomial contribution cancels and we obtain the endpoint relation $c_1(z_\mathrm{ end}) \simeq \alpha_{12}\,c_2(z_\mathrm{ end})$, i.e.,
\begin{equation}
  \tilde c_1(z_\mathrm{ end})
    \equiv c_1 - \alpha_{12}\,c_2
    \simeq 0.
  \label{eq:tw:c1_alpha12c2_endpoints}
\end{equation}
Equation~\eqref{eq:tw:c1_alpha12c2_endpoints} is used as an endpoint matching condition to fix the affine $c_1$ profiles; it need not hold pointwise in the plateau interior because $c_2$ relaxes exponentially on scale $\ell_v \equiv D_2/|v|$ [Eq.~\eqref{eq:tw:phi2_sol}] whereas $c_1$ is affine.
Since $c_2$ is continuous across sharp interfaces [Eq.~\eqref{eq:tw:c2_cont}], Eq.~\eqref{eq:tw:c1_alpha12c2_endpoints} implies continuity of $c_1$ at the same order without imposing it as an independent matching condition.

On each plateau $c_1$ is affine [Eq.~\eqref{eq:tw:c1_linear}] and the endpoint condition [Eq.~\eqref{eq:tw:c1_alpha12c2_endpoints}] applies at $z=0$ and $z=\Lambda_\pm$.
Subtracting the two endpoint values eliminates the offsets $b_\pm$ and yields
\begin{equation}
  s_\pm
    = \frac{c_1^\pm(\Lambda_\pm) - c_1^\pm(0)}{\Lambda_\pm}
    \simeq \alpha_{12}\,
      \frac{c_2^\pm(\Lambda_\pm) - c_2^\pm(0)}{\Lambda_\pm}\,.
  \label{eq:tw:slope_from_c2}
\end{equation}
Since $m_2^\pm$ is constant on a plateau, $c_2^\pm = \phi_2^\pm - m_2^\pm$ differs from $\phi_2^\pm$ only by an additive constant, so
\begin{equation}
  s_\pm
    \simeq \alpha_{12}\,
      \frac{\phi_2^\pm(\Lambda_\pm) - \phi_2^\pm(0)}{\Lambda_\pm}\,.
  \label{eq:tw:slope_from_phi2}
\end{equation}
Using the plateau solution [Eq.~\eqref{eq:tw:phi2_sol}], we obtain
\begin{equation}
  s_\pm
    = -\alpha_{12}\,\frac{B_\pm}{\Lambda_\pm}
      \bigl(1 - e^{-v\Lambda_\pm/D_2}\bigr),
  \label{eq:tw:slope_from_B}
\end{equation}
where the constant $C$ drops out.
Inserting $B_\pm$ from Eqs.~\eqref{eq:tw:B_solutions} gives the explicit plateau slopes
\begin{equation}
  s_\pm
    = \pm\,\alpha_{12}\alpha_{21}\,\Delta\phi\;
      \frac{\Phi(v;\,\Lambda_+,\Lambda_-)}{\Lambda_\pm}\,,
  \label{eq:tw:slopes}
\end{equation}
with $\Delta\phi \equiv \phi_1^+ - \phi_1^-$ and
\begin{equation}
  \Phi(v;\,\Lambda_+,\Lambda_-)
    \equiv
    \frac{\bigl(1 - e^{-v\Lambda_+/D_2}\bigr)
          \bigl(1 - e^{-v\Lambda_-/D_2}\bigr)}
         {1 - e^{-v\Lambda/D_2}}\,.
  \label{eq:tw:F_def}
\end{equation}
Thus, endpoint reactive equilibrium reduces the $c_1$-slopes to explicit functions of the mesa widths and the speed~$v$.

\medskip

\paragraph*{Kinematic jump condition and velocity selection.}
The plateau slopes are related to the propagation speed by the exact comoving-frame continuity equation $-v\,\phi_1' = D_1\,c_1''$ [Eq.~\eqref{eq:tw:phi_cons_comoving}].
Integrating across a sharp interface gives the jump condition
\begin{equation}
  -v\,[\phi_1] = D_1\,[c_1'],
  \label{eq:tw:jump_general}
\end{equation}
where $[f]\equiv f_\mathrm{ right}-f_\mathrm{ left}$ denotes the jump across the interface.
Consider the interface from the $(-)$ to the $(+)$ plateau: here $[\phi_1]=\phi_1^+-\phi_1^-\equiv\Delta\phi>0$ and $[c_1']=s_+-s_-$, so Eq.~\eqref{eq:tw:jump_general} becomes
\begin{equation}
  v\,\Delta\phi = D_1\,(s_- - s_+).
  \label{eq:tw:jump_phi1}
\end{equation}
(The opposite interface, $(+)$ to $(-)$, gives $[\phi_1]=-\Delta\phi$ and $[c_1']=s_--s_+$, yielding the same relation.)
Physically, the moving interface converts material between the two coexistence densities at rate~$v\,\Delta\phi$, balanced by the mismatch of redistribution fluxes~$-D_1 s_\pm$ on the two sides.

Substituting the slopes [Eq.~\eqref{eq:tw:slopes}] into Eq.~\eqref{eq:tw:jump_phi1} cancels $\Delta\phi$ and yields the central result of this section, a closed, implicit equation for the traveling-wave speed
\begin{equation}
  \frac{v\Lambda}{D_2}
    = -\frac{D_1}{D_2}\,\alpha_{12}\alpha_{21}\;
      \frac{\Lambda^2}{\Lambda_+\,\Lambda_-}\;
      \Phi(v;\,\Lambda_+,\Lambda_-).
  \label{eq:tw:v_final}
\end{equation}
The mesa widths~$\Lambda_\pm$ are linked by mass conservation of~$\phi_1$,
\begin{equation}
  \Lambda_+\,\phi_1^+ + \Lambda_-\,\phi_1^-
    = (\Lambda_+ + \Lambda_-)\,\bar\phi_1\,.
  \label{eq:tw:mass_phi1}
\end{equation}
Three ingredients enter Eq.~\eqref{eq:tw:v_final}: (i)~local reactive equilibrium in the McRD bulk, imposed as a matching condition at the plateau edges; (ii)~continuity of the redistribution potential in the $\phi_2$-sector, which fixes the exponential amplitudes~$B_\pm$; and (iii)~the exact interfacial jump condition [Eq.~\eqref{eq:tw:jump_general}], a direct consequence of mass conservation in the comoving frame.

For the orientation $z=x-vt$ with $v>0$, the right-hand side of Eq.~\eqref{eq:tw:v_final} is proportional to $-D_1\alpha_{12}\alpha_{21}$, while $\Phi$ is positive for $v>0$ and $\Lambda_\pm>0$ (each factor $(1-e^{-x})$ in Eq.~\eqref{eq:tw:F_def} is positive for $x>0$).
Traveling solutions therefore require
\begin{equation}
  \alpha_{12}\alpha_{21} < 0.
  \label{eq:tw:sign_condition}
\end{equation}
Reversing the sign of $v$ merely reverses the propagation direction ($z\mapsto -z$), so Eq.~\eqref{eq:tw:sign_condition} is the physically relevant existence condition: traveling waves arise only when the two cross-couplings have opposite sign.

To analyze the onset quantitatively, it is convenient to introduce the dimensionless nonreciprocity parameter
\begin{equation}
  \gamma \equiv -\frac{D_1}{D_2}\,\alpha_{12}\alpha_{21}\,.
  \label{eq:tw:gamma_def}
\end{equation}
Expanding $\Phi$ for small $|v|$ to order $\mathcal{O}(v^3)$ gives
\begin{equation}
  \Phi(v;\,\Lambda_+,\Lambda_-)
    = \frac{v}{D_2}\,\frac{\Lambda_+\Lambda_-}{\Lambda_++\Lambda_-}
      \biggl[1 - \frac{\Lambda_+\Lambda_-}{12D_2^2}\,v^2\biggr],
  \label{eq:tw:smallv_ratio_cubic}
\end{equation}
and substituting into Eq.~\eqref{eq:tw:v_final} produces the standard cubic normal form
\begin{equation}
  0 = (\gamma-1)\,v
      - \gamma\,\frac{\Lambda_+\Lambda_-}{12D_2^2}\,v^3.
  \label{eq:tw:amplitude_equation}
\end{equation}
The traveling branch therefore bifurcates at\footnote{For the common parameterization $\alpha_{12}=-\alpha_{21}\equiv\alpha$, the nonreciprocity parameter becomes $\gamma=(D_1/D_2)\,\alpha^2$ and the onset condition [Eq.~\eqref{eq:tw:onset_condition}] reduces to $|\alpha|>\sqrt{D_2/D_1}$.}
\begin{equation}
  \gamma > 1
  \qquad\Leftrightarrow\qquad
  -\frac{D_1\,\alpha_{12}\alpha_{21}}{D_2} > 1,
  \label{eq:tw:onset_condition}
\end{equation}
with a supercritical square-root scaling,
\begin{equation}
  v^2
    = \frac{12D_2^2}{\Lambda_+\Lambda_-}
      \left(1 - \frac{1}{\gamma}\right)
      + \mathcal{O} \bigl((\gamma-1)^2\bigr).
  \label{eq:tw:sqrt_onset}
\end{equation}
This threshold coincides with the onset of the oscillatory (conserved-Hopf) instability obtained via linear stability analysis~\cite{Brauns_Marchetti:2024} of the interfacial region: in both cases the two hydrodynamic modes associated with the conserved densities coalesce at an exceptional point precisely when $\gamma=1$.
The present sharp-interface calculation recovers the same condition from a fully nonlinear construction.

Moreover, $\Phi(v)/v$ is a strictly decreasing function of $v$ for $v>0$.\footnote{Writing $\Phi(v)/v = (\Lambda_+\Lambda_-/\Lambda)\, f(v\Lambda_+/D_2)\,f(v\Lambda_-/D_2)/f(v\Lambda/D_2)$ with $f(x)\equiv (1-e^{-x})/x$, the function $\log f$ is strictly convex for $x>0$ (since $(\log f)''=x^{-2}-e^x/(e^x{-}1)^2>0$).
Setting $a=\Lambda_+/D_2$ and $b=\Lambda_-/D_2$, convexity implies that $(\log f)'$ is increasing.
Since $av<(a{+}b)v$ and $bv<(a{+}b)v$ for $v>0$, it follows that $a(\log f)'(av)+b(\log f)'(bv)\le(a{+}b)(\log f)'((a{+}b)v)$, giving $\frac{d}{dv}\log(\Phi/v)\le 0$.}
Consequently, the slope criterion extracted at $v=0$ is also global: for $\gamma<1$, Eq.~\eqref{eq:tw:v_final} admits no nonzero solution.

Far above onset, in the regime $|v|\Lambda_\pm/D_2\gg 1$, the exponentials in $\Phi$ are negligible, $\Phi\to 1$, and Eq.~\eqref{eq:tw:v_final} simplifies to
\begin{equation}
  v \simeq \gamma\,D_2\,\frac{\Lambda_+ + \Lambda_-}{\Lambda_+\,\Lambda_-}\,.
  \label{eq:tw:largev_asymptotic}
\end{equation}
Introducing the mesa fraction $\lambda\equiv \Lambda_+/\Lambda$ (with $\Lambda=\Lambda_++\Lambda_-$) gives
\begin{equation}
  v \simeq \frac{\gamma\,D_2}{\lambda(1-\lambda)}\,\frac{1}{\Lambda}\,,
  \qquad 0<\lambda<1,
  \label{eq:tw:largev_scaling}
\end{equation}
so that at fixed~$\lambda$ the speed scales as $|v|\propto \Lambda^{-1}$.
Physically, once $\Lambda_\pm$ exceed the relaxation length $\ell_v\equiv D_2/|v|$, the $\phi_2$-profile decays essentially to its baseline within each mesa.
The endpoint matching then fixes the plateau slopes as $s_\pm\propto\pm 1/\Lambda_\pm$, and the jump condition [Eq.~\eqref{eq:tw:jump_phi1}] selects~$v$ in terms of the mass-constrained mesa widths.
In this regime, Eq.~\eqref{eq:tw:largev_scaling} implies $\ell_v\simeq\lambda(1-\lambda)\,\Lambda/\gamma$.

Collecting the two limits, the near-onset expansion [Eq.~\eqref{eq:tw:sqrt_onset}] and the large-speed asymptote [Eq.~\eqref{eq:tw:largev_scaling}] yield
\begin{equation}
  \frac{v\Lambda}{D_2}
    = \begin{cases}
        \displaystyle
        \sqrt{\frac{12}{\lambda(1-\lambda)}
              \left(1-\frac{1}{\gamma}\right)},
        & \gamma\to 1^+,\\[10pt]
        \displaystyle
        \frac{\gamma}{\lambda(1-\lambda)}\,,
        & \gamma\gg 1,
      \end{cases}
  \label{eq:tw:v_limits_onecase}
\end{equation}
with ${\lambda=\Lambda_+/\Lambda}$.
On the rescaled axis $\lambda(1{-}\lambda)\,v\Lambda/D_2$, the far-above-onset scaling collapses onto the universal line $\lambda(1{-}\lambda)\,v\Lambda/D_2=\gamma$, independent of the mesa fraction.

\begin{figure}[!t]
\centering
\includegraphics[width=\columnwidth]{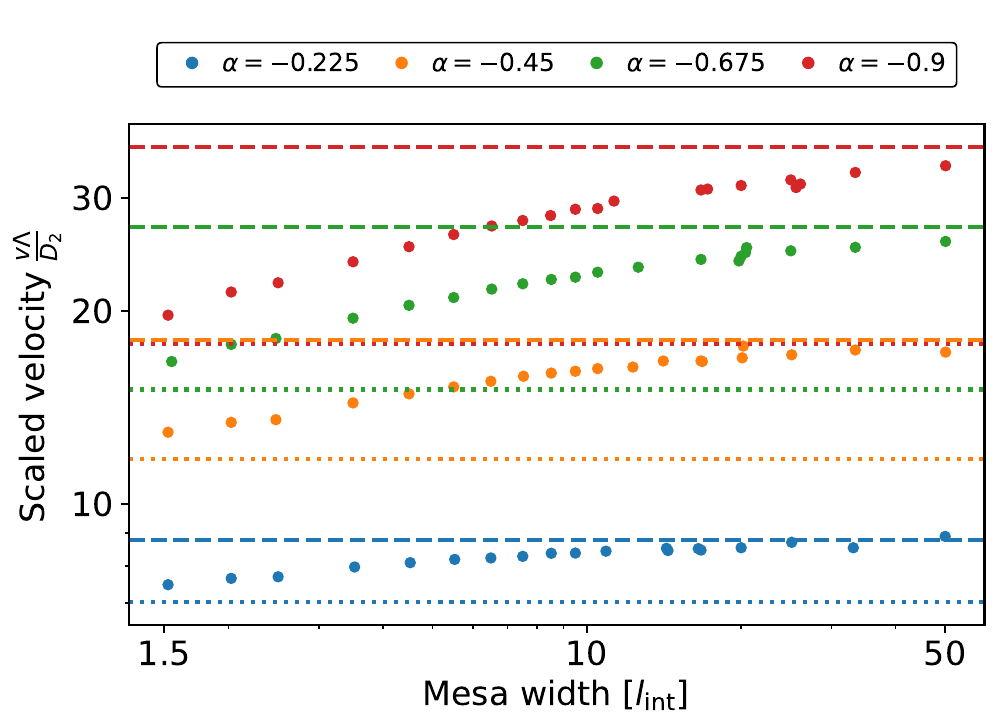}
\caption{\textbf{Speed--width relation for traveling mesas.}
Dimensionless propagation speed $|v|\Lambda/D_2$ of periodic
traveling-wave solutions in 1D as a function of the mesa width
$\Lambda_+/\ell_\mathrm{ int}$ (log scale), extracted from the
$\phi_1$ profile via inflection-point boundaries
(cf.\ Fig.~\ref{fig:TW-phi-profile}).
Filled circles: numerical integration of
Eqs.~\eqref{eq:tw:orig} for four values of the nonreciprocal
coupling $\alpha$ (see legend).
From bottom to top: $\alpha = -0.225$ (blue/light gray),
$-0.450$ (orange/medium gray), $-0.675$ (green/dark gray),
$-0.900$ (red/black).
Dashed lines: large-mesa asymptote of the sharp-interface
prediction [Eq.~\eqref{eq:tw:largev_scaling}]; dotted lines:
LSA-extrapolated prediction [Eq.~\eqref{eq:tw:veloc_LSA}].
For wide mesas ($\Lambda_+ \gtrsim 10\,\ell_\mathrm{ int}$) and
moderate $|\alpha|$, the numerical data approach the
sharp-interface solution [Eq.~\eqref{eq:tw:v_final}] from below.
At stronger driving (e.g.\ $\alpha=-0.900$, red/black), the
data remain systematically below the asymptote, consistent with
the breakdown of the flat-plateau approximation (see text).
For narrow mesas the speed approaches the LSA prediction from
above.
Statistical errors on $v$ are smaller than the symbol size.
Simulation parameters are listed in
Appendix~\ref{app:all_numerics}.
}
\label{fig:TW-velocity}
\end{figure}

\medskip

\paragraph*{Range of validity.}
The derivation of Eq.~\eqref{eq:tw:v_final} rests on three assumptions: (i)~sharp interfaces, (ii)~flat $\phi_1$-plateaus at the stationary coexistence values~$\phi^\pm_{1}$, and (iii)~approximate local reactive equilibrium, $A_1\simeq A_2\simeq 0$, away from interfaces.
All three are self-consistent for sufficiently large $r$ at moderate speeds, provided the mesa widths greatly exceed the interface width.
In the dual McRD system, assumption~(iii) can always be enforced by taking a sufficiently fast relaxation rate in the exchange terms~$A_i$.
Assumptions~(i) and~(ii), however, are controlled by the fixed parameter~$r$ and break down independently: (i)~fails when the mesa width becomes comparable to the interface width, and (ii)~fails when the velocity is large enough that the bulk profiles develop a visible tilt.
In the latter case, corrections to the flat-plateau and piecewise-linear forms become important, and Eq.~\eqref{eq:tw:v_final} overestimates the speed.
For sufficiently narrow mesas the profile approaches a near-harmonic shape, and a Stefan-type law motivated by the dispersion relation at the interface provides a better description~\cite{Brauns_Marchetti:2024}.
We therefore regard Eq.~\eqref{eq:tw:v_final} as a controlled sharp-interface prediction in the regime of moderate speeds and sufficiently large~$r$, and compare it to numerics in this parameter window.

\subsection{Comparison to numerical simulations}
\label{sec:num:TW}

We now test the sharp-interface velocity law [Eq.~\eqref{eq:tw:v_final}] against direct numerical integration of the nonreciprocal Cahn--Hilliard dynamics [Eqs.~\eqref{eq:tw:orig}] in one spatial dimension.
Equations~\eqref{eq:tw:orig} are integrated with a finite-element method using an implicit backward-differentiation time stepper; full details of the discretization, mesh resolution, and convergence criteria are given in Appendix~\ref{app:all_numerics}.
The initial condition is a periodic mesa profile for~$\phi_1$, with plateau values close to~$\phi_1^\pm$ and~$\phi_2$ initialized at the corresponding local reactive equilibrium.
After an initial transient (identified by the adaptive time step converging to a constant value), the system settles into a uniformly translating state.

\medskip

\paragraph*{Measurement protocol.}
All lengths in this section are measured in units of the interface width $\ell_{\rm int} \approx 8/\sqrt{r}$, estimated from the numerical profiles; since $\kappa$ and $r$ are held fixed throughout, this provides a uniform reference scale.
The mesa widths $\Lambda_\pm$ are extracted from the inflection-point boundaries of the $\phi_1$ profile as defined in Fig.~\ref{fig:TW-phi-profile}.
We distinguish a large-mesa regime ($\Lambda_+ \gtrsim 10\,\ell_{\rm int}$) and a small-mesa regime.
The propagation speed is extracted from a post-transient time series of equally spaced snapshots: in the large-mesa regime we track the $\phi_2$ peak position; in the small-mesa regime, where the profile is nearly sinusoidal and the $\phi_2$ peaks are poorly defined, we track the inflection points of $\phi_1$ instead.
In both cases the instantaneous velocity $v_i \equiv (x_i - x_{i-1})/\Delta t$ is computed at each step and the reported speed is the average over all post-transient steps; the choice of $\Delta t$ is specified in Appendix~\ref{app:all_numerics}.

The rationale for using different tracking features is numerical resolution relative to the interface width.
In the large-mesa regime, the mesh ($\Delta x = 0.05\,\ell_{\rm int}$) is chosen coarse enough for feasible run times, so only a handful of grid points span each interface.
The inflection point of $\phi_1$ is then localized only to within a fraction of $\ell_{\rm int}$, which translates into a sizeable relative error on the velocity of slowly drifting wide mesas.
The sharply localized $\phi_2$ peaks provide a more robust positional reference in this regime.
Conversely, for small mesa widths the finer mesh ($\Delta x = 0.025\,\ell_{\rm int}$) resolves the interface accurately and the inflection points of the nearly sinusoidal $\phi_1$ profile are well defined, whereas the $\phi_2$ peaks have essentially vanished.

The results are insensitive to the choice of inflection-point versus half-maximum-crossing boundaries:
in the large-mesa regime the two definitions differ by at most an interface width $\ell_{\rm int}\ll\Lambda_\pm$, and in the small-mesa limit they coincide.
Statistical errors on $v$, estimated from the variance of $\{v_i\}$, are smaller than the symbol size in all figures.

\begin{figure}[!t]
\centering
\includegraphics[width=\columnwidth]{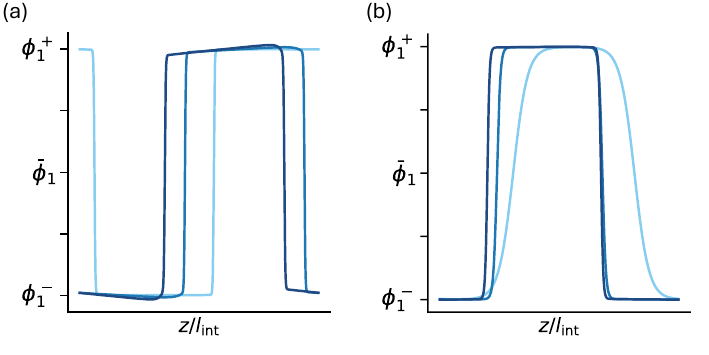}
\caption{\textbf{Breakdown of the sharp-interface ansatz.}
Comoving-frame profiles $\phi_1(z)$ from numerical simulations
of Eqs.~\eqref{eq:tw:orig}, illustrating the two independent
failure modes of the flat-plateau approximation.
(a)~Decreasing mesa width at fixed nonreciprocity: as
$\Lambda_+/\ell_\mathrm{ int}$ decreases (dark to light blue/dark
to light gray), the profile evolves from a well-defined mesa
with flat plateaus to a near-sinusoidal shape, invalidating
the piecewise-constant ansatz for $\phi_1$. We plot the cases ${\Lambda_+/\ell_\mathrm{ int}\in \{1.5,\,5.5,\,9.5\}}$.
(b)~Increasing nonreciprocal driving at fixed mesa width: as
$|\alpha|$ increases (dark to light blue/dark to light gray),
the plateaus develop a visible tilt, so that the
constant-$\phi_1$ approximation underlying the slope
calculation [Eq.~\eqref{eq:tw:slopes}] becomes inaccurate. We plot the cases ${\alpha \in \{-0.2,\,-1,\,-2\}}$.
In both cases, coexistence plateaus $\phi_1^\pm$ and the mean
density $\bar\phi_1$ are indicated on the vertical axis.
Simulation parameters are listed in
Appendix~\ref{app:all_numerics}.
}
\label{fig:TW-ansatz-breakdown}
\end{figure}

\medskip

\paragraph*{Velocity versus mesa width.}

Figure~\ref{fig:TW-velocity} shows the measured dimensionless speed $|v|\Lambda/D_2$ as a function of the mesa width $\Lambda_+/\ell_{\rm int}$ for several values of the nonreciprocal coupling $\alpha$.
The dashed lines show the full solution [Eq.~\eqref{eq:tw:v_final}] and the dotted lines the LSA-extrapolated prediction [Eq.~\eqref{eq:tw:veloc_LSA}], both evaluated from the simulation parameters with no adjustable constants.
Two features of Fig.~\ref{fig:TW-velocity} merit discussion.

\emph{Crossover to the LSA regime at small mesa width.}
For narrow mesas, the dimensionless speed $|v|\Lambda/D_2$ does not follow the $|v|\propto \Lambda^{-1}$ scaling implied by the large-mesa asymptote [Eq.~\eqref{eq:tw:largev_scaling}], but instead approaches from above a lower bound set by the linear-stability analysis (LSA) of the spatially extended traveling-wave branch.
This behavior reflects the profile $\phi_1$ approaching a harmonic (near-sinusoidal) shape [Fig.~\ref{fig:TW-ansatz-breakdown}, left].
The LSA-extrapolated speed, shown as a dotted line, is derived following \citep{Brauns_Marchetti:2024} in Appendix~\ref{app:TW-LSA} and reads
\begin{equation}
    v = D_2\frac{\pi}{2}\frac{\Lambda_+ + \Lambda_-}{\Lambda_+\Lambda_-}\sqrt{\gamma-1}.
\label{eq:tw:veloc_LSA}
\end{equation}
It scales differently with $\gamma$ near onset than the sharp-interface formula [Eq.~\eqref{eq:tw:v_final}].

The coexistence of these two limiting behaviors, harmonic near onset and mesa-like far from onset, implies a nonmonotonic crossover in $|v|(\Lambda_+)$ when the mesa width becomes comparable to the interface width.
A systematic treatment of this crossover, for instance via a matched asymptotic expansion, lies beyond the scope of the present work.

\emph{Persistent deviation at large $\gamma$.}
For sufficiently strong nonreciprocity, even wide mesas do not reach the predicted speed.
The origin is the breakdown of the flat-plateau assumption: increasing $\gamma$ increases $|v|$, which in turn tilts the $\phi_1$-profile [Fig.~\ref{fig:TW-ansatz-breakdown}, right].
Crucially, it is \emph{not} the local reactive equilibrium condition $A_1\simeq 0$ that fails first; this condition continues to hold at the plateau endpoints because the reaction kinetics are fast ($\tau \to 0$).
Rather, what fails is the \emph{global} flat-plateau ansatz: $\phi_1$ varies appreciably across the domain, so the affine parameterization of $c_1$ [Eq.~\eqref{eq:tw:c1_linear}] and the constant-$\phi_1$ approximation used to derive the slopes $s_\pm$ [Eq.~\eqref{eq:tw:slopes}] become inaccurate.
The sharp-interface prediction [Eq.~\eqref{eq:tw:v_final}] then systematically overestimates $|v|$ because it approximates $\phi_1$ as piecewise constant.
In principle, incorporating a linearly tilted plateau into the ansatz should yield improved large-mesa asymptotics and a corrected scaling exponent. This provides an interesting avenue for further exploration.

\begin{figure}[!t]
\centering
\includegraphics[width=\columnwidth]{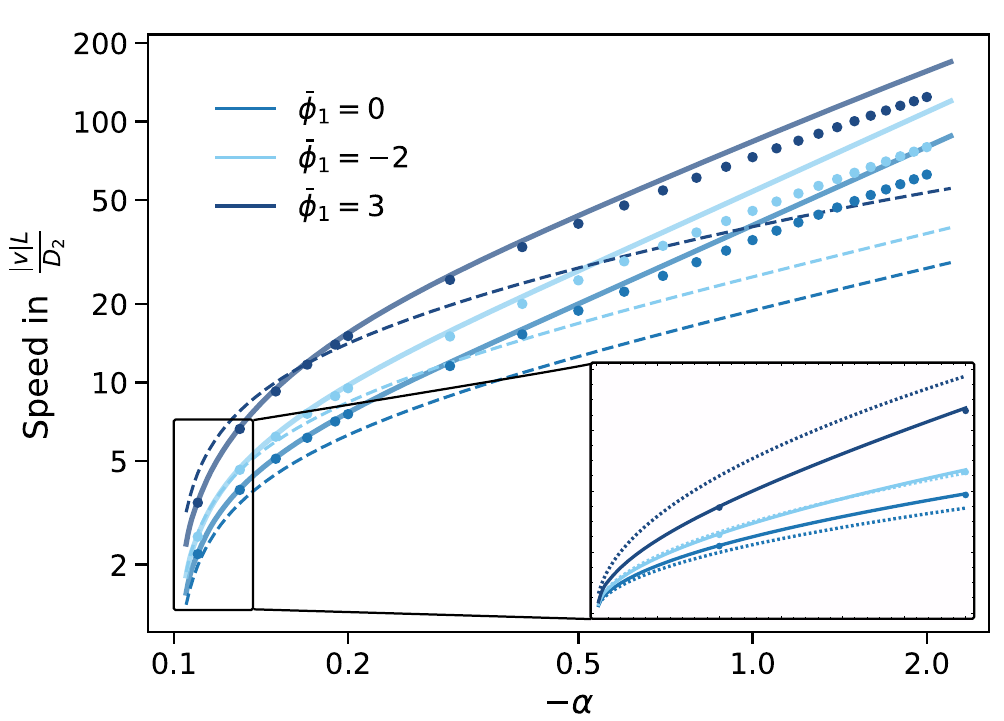}
\caption{\textbf{Speed versus nonreciprocity at fixed mesa geometry.}
Dimensionless propagation speed $|v|\Lambda/D_2$ as a function
of the nonreciprocal coupling $-\alpha$ (log--log scale), for
three values of the mean density: $\bar\phi_1 = 0$ (medium
blue/medium gray), $\bar\phi_1 = -2$ (light blue/light gray),
and $\bar\phi_1 = 3$ (dark blue/dark gray).
Mesa widths are fixed deep in the large-mesa regime:
$\Lambda_+ = 50\,\ell_\mathrm{ int}$ ($\bar\phi_1 = 0$),
$25\,\ell_\mathrm{ int}$ ($\bar\phi_1 = -2$), and
$87.5\,\ell_\mathrm{ int}$ ($\bar\phi_1 = 3$), all at fixed
system size $2L=400$ (four mesa trains of period
$\Lambda = 100$).
Filled circles: numerical integration of
Eqs.~\eqref{eq:tw:orig}.
Solid curves: sharp-interface prediction
[Eq.~\eqref{eq:tw:v_final}], evaluated at the measured
widths with no fit parameters.
Dashed curves: LSA-extrapolated prediction
[Eq.~\eqref{eq:tw:veloc_LSA}].
Near onset the sharp-interface prediction tracks the
numerical data more closely than the LSA approximation.
At large $|\alpha|$, the data fall systematically below the
sharp-interface prediction, consistent with the breakdown of
the flat-plateau approximation (see text). Zooming into the onset (inset bottom right),
it is clear that the points for $\bar \phi = 3,\, 0$ lie on the full solution
rather than the LSA curve.
Simulation parameters are listed in
Appendix~\ref{app:all_numerics}.
}
\label{fig:TW-velocity_vs_alpha}
\end{figure}

\medskip
\paragraph*{Velocity versus nonreciprocity strength.}
To probe the dependence on the nonreciprocal coupling directly, we fix the mesa widths $\Lambda_\pm$ deep in the large-mesa regime and vary $\alpha$ at several values of the mean density $\bar\phi_1$ (see Fig.~\ref{fig:TW-velocity_vs_alpha} caption for the specific initial conditions).
We use the antisymmetric parameterization $\alpha_{12}=-\alpha_{21}\equiv \alpha$, for which $\gamma=(D_1/D_2)\alpha^2$ and the onset condition becomes $|\alpha|>\alpha_c\equiv\sqrt{D_2/D_1}$ [Eq.~\eqref{eq:tw:onset_condition}].

Figure~\ref{fig:TW-velocity_vs_alpha} displays the dimensionless speed $|v|\Lambda/D_2$ as a function of $|\alpha|$ on a logarithmic axis.
The sharp-interface theory makes two distinct predictions for the scaling with $\alpha$.
Near onset, Eq.~\eqref{eq:tw:sqrt_onset} predicts a supercritical square-root growth,
\begin{equation}
|v|\propto \sqrt{|\alpha|^2-\frac{D_2}{D_1}}\,,
\label{eq:tw:v_onset_alpha_disc}
\end{equation}
which is non-analytic at $|\alpha|=\alpha_c$.
Far above onset, $\Phi \to 1$ and the large-$|v|$ asymptote [Eq.~\eqref{eq:tw:largev_asymptotic}] implies
\begin{equation}
|v|\propto \alpha^2
\label{eq:tw:v_alpha_asymptotic_disc}
\end{equation}
at fixed $\Lambda_\pm$.
Thus the sharp-interface prediction traces a crossover from concave, non-analytic onset to convex, quadratic growth, with an intermediate window in which an approximately linear effective dependence on $|\alpha|$ can arise.
This crossover structure goes beyond what is accessible from a linear-stability analysis: Eq.~\eqref{eq:tw:v_final} remains predictive at moderate distances from onset because it incorporates the full nonlinear matching at the interfaces.

Since only $|\alpha_{21}|$ is varied at fixed $\alpha_{12}$, the predicted asymptotic scaling in Fig.~\ref{fig:TW-velocity_vs_alpha} is $|v|\propto|\alpha|$.
The numerical data instead follow an effective power law $|v|\sim|\alpha|^p$ with $p$ systematically below unity: fitting the three mean-density data sets yields $p=0.84,\,0.82,\,0.77$ for $\bar\phi_1=0,\,-2,\,3$ respectively ($95\%$ confidence intervals $\pm 0.005$).
This sub-linear scaling is consistent with the breakdown of the flat-plateau approximation discussed for Fig.~\ref{fig:TW-velocity}: at large $|\alpha|$ the increasing propagation speed tilts the plateaus, and the sharp-interface prediction systematically overestimates $|v|$.

\medskip

\paragraph*{Summary.}
Within its regime of validity, wide mesas ($\Lambda_\pm/\ell_{\rm int}\gg 1$) at moderate nonreciprocal driving, the dual McRD sharp-interface construction yields a closed, implicit velocity law [Eq.~\eqref{eq:tw:v_final}] that quantitatively predicts the traveling-wave speed from the mesa widths, with no adjustable parameters.
The numerical comparison in Figs.~\ref{fig:TW-velocity} and~\ref{fig:TW-velocity_vs_alpha} supports this picture:
the harmonic small-mesa limit, captured by the LSA-extrapolated formula [Eq.~\eqref{eq:tw:veloc_LSA}], crosses over smoothly to the nonlinear sharp-interface law [Eq.~\eqref{eq:tw:v_final}] as the mesa width grows.
Both predictions share the same onset at $\gamma=1$, but the supercritical curves differ: in the large-mesa regime the data follow the sharp-interface law rather than the LSA extrapolation already near onset (cf.\ inset of Fig.~\ref{fig:TW-velocity_vs_alpha}).
Quantitative agreement extends well away from onset, and the systematic deviation at large driving is consistent with the anticipated breakdown of the flat-plateau approximation.
Accordingly, the velocity law [Eq.~\eqref{eq:tw:v_final}] should be regarded as a controlled prediction in the regime of moderate speeds and sharp interfaces, rather than as a uniformly valid formula for arbitrarily strong nonreciprocal driving.

All analysis and simulations in this section are strictly one-dimensional.
In two dimensions, traveling-wave interfaces can become unstable to transverse undulations, giving rise to spatiotemporal chaos~\citep{Brauns_Marchetti:2024}.
Closely related interfacial instabilities, including persistent capillary waves propagating along domain boundaries, have been identified in conserved active emulsions with nonreciprocal chemotactic coupling~\citep{Rasshofer_Frey:2025}.
Whether the dual McRD sharp-interface framework developed here can provide analytical access to such 2D interfacial dynamics remains an open question.

\section{Discussion and outlook}
\label{ref:discussion_outlook}

\subsection*{Summary of the duality}

The construction we have developed answers, at the equation level, the structural question raised in the introduction: the two continuum frameworks introduced there are not parallel descriptions of the same physics but stand in an asymmetric relation.
Reaction--diffusion is the broader class: every Cahn--Hilliard-type field theory with passive square-gradient interfacial terms, strictly mass-conserving or augmented by weak reactive turnover, embeds as the slow dynamics on an attracting manifold of a correspondingly conservative or weakly nonconservative two-component-per-species reaction--diffusion system. Conversely, every such system with attractive nullcline admits, in the fast-interconversion limit, an exact chemical-potential representation whose effective constitutive relation is read off from the reaction nullcline.
The equivalence holds beyond the linear and weakly nonlinear regimes accessible to the precedents reviewed in the introduction, mapping the nonlinear, far-from-equilibrium steady states of the two descriptions onto each other.
From this vantage, the thermodynamic vocabulary of phase-field theory---chemical equilibrium, Maxwell constructions, interfacial tension, Laplace pressure---is not external structure imported into reaction--diffusion by analogy, but the imprint of the slow manifold rendered in the language of the extended two-field dynamics.
The dynamical-geometric vocabulary of McRD---the flux-balance subspace, nullcline geometry, mass-redistribution potential, and turnover-balance condition---is, conversely, what a passive thermodynamic description becomes once the fast modes that pin its slow manifold are resolved.
The phenomenological overlap that motivated this paper does not require a further structural origin: chemical-potential dynamics is the thermodynamic shadow of an underlying mass-conserving reaction--diffusion system, and the dictionary developed in this paper is the explicit form of that projection.

The construction proceeds by \emph{unfolding} the local relation $\mu(\phi)$, whose non-invertibility in phase-separating regimes precludes a single-field reaction--diffusion formulation, into an extended two-field description in which it appears as an attracting manifold enforced by fast local interconversion.
Because the embedding conserves total mass exactly and does not rely on proximity to an instability threshold, the dual inherits the full nonlinear structure of the original dynamics rather than only its near-onset phenomenology.
The non-uniqueness of the off-manifold continuation is a representational freedom that we exploit to arrive at a transparent normal form with diagonal diffusion and local conversion kinetics [Sec.~\ref{sec:normal_form}].

\medskip

\subsection*{Conceptual advantage of the dual system}

A broad class of biological and soft-matter problems involves chemical species that interact both through pairwise potentials, giving rise to phase separation, and through chemical reactions fueled, for example, by ATP hydrolysis.
Such mixed systems have been treated in several ways.
The simplest is to add a non-conserving reactive source term to the Cahn--Hilliard dynamics, which selects a finite domain size~\citep{Glotzer_Coniglio:1994b,Glotzer_Muthukumar:1995}.
A thermodynamically consistent formulation specifies the chemical-potential differences $\Delta\mu$ that drive the reactions, so that the kinetics inherit the constraints of the underlying free energy~\citep{Zwicker:2022}.
In the limit of linear reactions, the effect of the reactive part can be absorbed into the chemical potential as a long-range interaction, recovering an equilibrium problem of the Ohta--Kawasaki type with a Coulomb-like non-local kernel~\citep{Ohta_Kawasaki:1986,Liu_Goldenfeld:1989}.
Beyond this limit, descriptions necessarily become hybrid: a passive thermodynamic part for the interactions and a separate, driven part for the reactions, each carrying its own structure.

The duality removes this dichotomy.
In the dual McRD description, pairwise potential couplings and chemical reactions enter on the same footing as kinetic conversions between species: every interaction, whether inter- or intraspecies and whether of thermodynamic or reactive origin, is encoded in the geometry of nullclines and the flux-balance subspace.
The price is that no free-energy functional is available in general; the gain is a single object---the reaction kinetics, together with its nullcline geometry---that organizes both contributions.
Systems whose original formulations distribute the physics differently between thermodynamic interactions and explicit reactions can therefore be compared directly once dualized.

This common reaction--diffusion structure has, in turn, recently been found to exhibit phenomenology long thought distinctive of phase-field systems: an effective interfacial tension, curvature-dependent domain dynamics, and interrupted coarsening with a selected wavelength, despite the absence of any free-energy functional~\citep{Weyer_Frey:2026}.
These observations, which in the introduction motivated the search for a structural origin of the parallels between the two languages, find that origin in the mass-redistribution potential $\eta$, which takes the role of the chemical potential $\mu$, the turnover-balance condition, which takes that of the Maxwell construction, and the osmotic and Laplace pressures of the Cahn--Hilliard description, which reappear as geometric objects on the flux-balance subspace (Table~\ref{tab:dictionary}).

\subsection*{Analytical payoff: the traveling-wave velocity of nonreciprocal Cahn--Hilliard dynamics}

For nonreciprocal multicomponent theories, the duality not only provides a dictionary but also new analytical tools.
The sharp-interface traveling-wave theory of Sec.~\ref{subsec:tw_velocity} illustrates this concretely: the velocity law Eq.~(\ref{eq:tw:v_final}) was derived entirely within the dual McRD description, where the flux-balance geometry organizes the calculation---the redistribution-potential slopes on the mesa plateaus fix the mass currents, and the velocity is selected by a kinematic jump condition at the interface.
The result is a closed-form expression for the propagation velocity of nonreciprocal traveling waves in the wide-mesa regime, where the bulk plateaus carry the dynamics and the interfaces enter only through matching conditions.

Two earlier results address the same problem.
Linear stability analysis at the interface, far from the homogeneous steady state, fixes the propagation speed and yields a formula~\citep{Brauns_Marchetti:2024} that sets the lower bound on the speed approached at narrow mesas, where the profile becomes near-sinusoidal (Sec.~\ref{sec:num:TW}).
A complementary derivation has very recently been obtained through matched asymptotic expansions in the original Cahn--Hilliard variables: an effective sharp-interface description of the two-dimensional dynamics, in the form of a modified Mullins--Sekerka problem, from which explicit speeds for one-dimensional periodic wave trains can be extracted~\citep{Mullins.Sekerka1963,Gomez_Strikwerda:2025}; the resulting expressions agree with Eq.~(\ref{eq:tw:v_final}) in their scaling with the nonreciprocity, with matching deviations at large nonreciprocity and a smooth crossover into the harmonic regime.
As is often the case for dual formulations, working in the variables of the dual McRD description renders the structure of the calculation more transparent: the bulk plateaus, the redistribution-potential profile, and the interfacial matching condition appear as separate geometric objects on the flux-balance subspace, whereas in the original two-field variables the same physics is reached only through the technically more involved machinery of multiscale asymptotic analysis.

\subsection*{Scope and limitations}

The construction is a \emph{duality}, not a microscopic identification: it does not claim that phase-separating mixtures and intracellular protein-pattern systems share a common molecular description.
The relation is between coarse-grained continuum equations, established by an embedding whose approximation error is controlled by the finite relaxation rate $\tau^{-1}$ onto the slow manifold.
On the slow manifold the dual reaction--diffusion system reproduces the patterns, interface profiles, and phase-space structure of the original chemical-potential dynamics; the two formulations are not equivalent off the manifold.

The error of the dual description scales as $\partial_t \nabla^2 \phi/\sqrt{\tau}$ (Sec.~\ref{sec:num:passive-coarsen}): the construction is accurate at long wavelengths and late times, and degrades during fast transients.
For systems with continuous symmetries---translation invariance in coarsening, phase invariance for periodic patterns---the steady states form symmetry-related families.
Because each transient incurs a finite displacement of order $\sqrt{\tau}$, the dual trajectory generically reaches a different member of the family from the one selected by the original dynamics, and the two descriptions agree only modulo the corresponding symmetry transformation: a rigid spatial shift in coarsening, a phase shift for periodic patterns.
The displacement vanishes as $\tau \to 0$, and pointwise agreement within the symmetry orbit is recovered in this limit.

Two structural restrictions limit the class of field theories to which the construction applies.
The first is on the form of the current: the conserved dynamics must be of potential-current form [Eq.~(\ref{eq:mu_n_potential_form})], with fluxes given by gradients of chemical potentials through a mobility matrix that need not be symmetric.
Non-gradient contributions to the current---self-propulsion terms in active scalar field theories, advective couplings to a velocity field as in Model~H---are not covered.

The second restriction is on the dual network topology. Multicomponent Cahn--Hilliard models with passive interfacial terms, including all nonreciprocal variants, admit dual representations in which each conserved species carries two components coupled by interconversion kinetics. The broken-mass-conservation system of Sec.~\ref{sec:broken_mass-conservation}, a Cahn--Hilliard equation with reactive turnover dual to a non-mass-conserving two-component reaction--diffusion system, can be viewed as a reduction of an underlying conserved system in which additional components have been integrated out. On the Cahn--Hilliard side, the underlying system would involve at least two species coupled both through their chemical potentials and through reactive interconversion, and would dualize to a network of at least four chemically coupled components. Reaction--diffusion networks in which each species carries more than two components, with the Min system and its active and inactive cytosolic pools as the canonical example, lie outside the present construction; their effective long-time behavior has been derived for a three-component case in~\citep{Toffenetti_Nettuno_Frey:2026}, but an exact dualization remains an open problem to which we return in the Outlook.

\subsection*{Outlook}

Three directions extend the construction beyond the cases analyzed here, ordered by increasing scope.

The first direction is to dualize \emph{partially}.
The construction developed here dualizes the entire chemical-potential dynamics, including its passive variational core, and is therefore not always the most economical representation: a variational system is already parsimonious, and dualizing it introduces reactive structure that is not strictly needed.
A more economical alternative is to leave the variational part in its original form and apply the construction only to the contributions that break detailed balance---for example, the nonreciprocal couplings of Sec.~\ref{sec:multi-component_systems}.
The result is a hybrid Cahn--Hilliard backbone with a reactive correction that carries the non-equilibrium content, developed in a forthcoming paper~\citep{Zhou_Frey:2026}.

The second direction is to extend the duality to stochastic dynamics, relevant whenever fluctuations are dynamically important---small particle numbers, mesoscopic biological systems, or dynamics close to a critical point.
The two sides constrain noise differently: on the chemical-potential side, fluctuation--dissipation ties the noise tensor to the mobility matrix and to a temperature; on the McRD side, the noise inherits its structure from the master equation governing the underlying reaction kinetics.
These two structures are not in general compatible, and whether the duality lifts to the stochastic level---and in what regime---requires either decomposing the noise into a duality-respecting part and a residual mismatch, or analyzing the consequences of the mismatch directly.

The third and most consequential direction is to use the duality as an organizing principle.
Multicomponent Cahn--Hilliard systems with reactions, and systems with different splits between thermodynamic and reactive contributions, can dualize to the same reduced McRD network. The duality thereby groups continuum models with disparate microscopic origins by their dual realization, and a result obtained for one representative, whether an interface profile, a coarsening law, a coexistence condition, transfers to every member of the class.
This transfer runs in both directions. On one side, broader McRD networks admit scalar effective descriptions whose gradient sector is richer than the single square-gradient term considered here, and the duality places intermediate models between the pure reaction--diffusion form and the single-scalar form on the same footing as their endpoints. Recently, it has been shown that such networks support a non-equilibrium analog of surface tension: at curved interfaces, cyclic attachment and detachment produces a curvature-dependent mass redistribution potential obeying an effective Gibbs--Thomson relation, a non-equilibrium Neumann law, and Plateau vertex conditions, all derived without recourse to a free-energy functional~\citep{Weyer_Frey:2026}. Considerable effort has been devoted to extending equilibrium notions of surface tension to active systems where detailed balance is broken~\citep{Solon_Tailleur:2018a, Langford_Omar:2025, Li_Yang:2025}; the reaction--diffusion route opens a complementary, bottom-up approach to the same question, and any result established in the reaction--diffusion language transfers across the duality to every theory in the equivalence class. On the chemical-potential side, any insight gained from studying scalar field theories with active gradients will be likewise transferable to systems of the same equivalence class, including reaction-diffusion systems. Together, the two sides point toward a thermodynamic-style classification of non-equilibrium phase-separating systems on the basis of their reaction--diffusion realization alone.

\subsection*{Concluding remark}

Chemical-potential field theories and mass-conserving reaction--diffusion systems are two faces of a single dynamical structure.
At the level of the slow manifold, every mass-conserving reaction--diffusion system with attractive nullcline admits an exact chemical-potential representation, and every chemical-potential field theory with conserved order parameters embeds as the slow dynamics on an attracting manifold of an McRD system; the dictionary of Table~\ref{tab:dictionary} is the explicit form of that correspondence.
The structural origin of the long-noted overlap between the two frameworks is therefore neither analogy nor coincidence, but the projection of an underlying dynamics onto its passive shadow.
What the construction makes available is not only a translation but a procedure: features that resist analysis on one side can be computed on the other and pulled back, as the closed-form traveling-wave velocity for nonreciprocal Cahn--Hilliard dynamics demonstrates concretely.

\begin{acknowledgments}
We thank Natan Dominko Kobilica, Beatrice Nettuno, Davide Toffenetti, and Henrik Weyer for stimulating and inspiring discussions. This work was supported by the Deutsche Forschungsgemeinschaft (DFG, German Research Foundation) under Germany's Excellence Strategy through the Excellence Clusters ORIGINS (EXC-2094 -- 390783311) and BioSysteM (EXC3092/1-533751719), and by the European Research Council (ERC) under the European Union's Horizon Europe programme (CellGeom, Grant Agreement No.\ 101097810).
\end{acknowledgments}

\appendix


\section{Duality between reaction--diffusion and Cahn--Hilliard models}
\label{app:duality_McRD_CH}

In this Appendix we show how the $\kappa$-free dual reaction--diffusion system reduces, in the fast-reaction limit $\tau\to0$, to an effective dynamics for the conserved density $\phi=c+m$.
Adiabatic elimination of the fast field $m$ generates an emergent fourth-order regularisation term in the reduced $\phi$-equation, whose coefficient is fixed by steady-state matching to the Cahn--Hilliard model.
We then quantify the leading dynamical deviation from Cahn--Hilliard evolution, which vanishes as $O\!\bigl(\sqrt{\tau}\bigr)$, and briefly note how the construction extends to birth--death kinetics $s(\phi)$.

We start from the $\kappa$-free dual reaction--diffusion system
\begin{subequations}\label{eq:dual_system_app}
\begin{align}
  \partial_t c
    &= D_c\,\nabla^2 c
       - \frac{1}{\tau}\bigl[c-\hat\mu(c+m)\bigr],
  \label{eq:dual_c_tau_app}\\
  \partial_t m
    &= D_m\,\nabla^2 m
       + \frac{1}{\tau}\bigl[c-\hat\mu(c+m)\bigr].
  \label{eq:dual_m_tau_app}
\end{align}
\end{subequations}
The total density $\phi=c+m$ obeys
\begin{equation}
  \partial_t\phi
    = D_c\,\nabla^2 c + D_m\,\nabla^2 m
    = D_c\,\nabla^2 \eta\,,
  \label{eq:phi_eta_identity_app}
\end{equation}
with $\eta \equiv c+d\,m$ the mass-redistribution potential and $d \equiv D_m/D_c$.

\smallskip

\paragraph*{Adiabatic elimination of the fast field and gradient corrections.}
To reduce the dual dynamics to an effective equation for the conserved density
$\phi$, we eliminate the fast variable $m$ in the limit $\tau\to0$.
Multiplying Eq.~\eqref{eq:dual_m_tau_app} by $\tau$ and using $c=\phi-m$ yields the
exact constraint form
\begin{equation}
  \phi - m - \hat\mu(\phi)
    \;=\;
    \tau\,\partial_t m
    - \tau\,D_m\,\nabla^2 m\,.
  \label{eq:constraint_exact_app}
\end{equation}
To make the $\nabla^2 m$ term on the right-hand side explicit in terms of the dynamical variables $\phi$ and $\eta$, we use the kinematic identity $\phi - \eta = (1-d)\,m$, which gives
\begin{equation}
  m = \frac{\phi-\eta}{1-d}\,,
  \qquad
  \nabla^2 m = \frac{1}{1-d}\,\nabla^2(\phi-\eta)\,.
  \label{eq:m_from_phi_eta_app}
\end{equation}
Substituting into Eq.~\eqref{eq:constraint_exact_app} and solving for $m$ yields the exact decomposition
\begin{equation}
  m
    = \phi - \hat\mu(\phi)
      + \frac{\tau\,D_m}{1-d}\,\nabla^2(\phi-\eta)
      - \tau\,\partial_t m\,.
  \label{eq:m_exact_decomp_app}
\end{equation}
Because $m = (\phi-\eta)/(1-d)$ and $1-d = O(\sqrt{\tau})$ in the matching regime (see Eq.~\eqref{eq:kappa-match} and the discussion following it), while $\partial_t\phi = O(1)$ and $\partial_t\eta$ remains bounded, the time derivative $\partial_t m = O(\tau^{-1/2})$.
The last term in Eq.~\eqref{eq:m_exact_decomp_app} therefore contributes $\tau\,\partial_t m = O\!\bigl(\sqrt{\tau}\bigr)$.
Retaining the leading spatial correction and absorbing the remainder gives the slow-manifold relation
\begin{equation}
  m
    = \phi - \hat\mu(\phi)
      + \frac{\tau\,D_m}{1-d}\,\nabla^2(\phi-\eta)
      + O\!\bigl(\sqrt{\tau}\bigr).
  \label{eq:m_slaving_app}
\end{equation}
Although the gradient term is proportional to $\tau$, it must be retained: as shown below, steady-state matching fixes $\tau\,D_m/(1-d)=M \kappa/D_m$, so this term generates the effective interfacial regularisation.

\smallskip

\paragraph*{Reduced equation for the density field $\phi$ and emergent fourth-order term.}
Substituting $\eta=\phi-(1-d)\,m$ in the continuity equation~\eqref{eq:phi_eta_identity_app} yields
\begin{equation}
  \partial_t\phi
    = D_c\,\nabla^2\!\bigl[\phi-(1-d)\,m\bigr].
  \label{eq:phi_via_m_app}
\end{equation}
Inserting the adiabatic relation~\eqref{eq:m_slaving_app} into this expression, the bracket becomes
\begin{equation}
  \phi-(1-d)\,m
    = d\,\phi + (1-d)\,\hat\mu(\phi)
      - \tau\,D_m\,\nabla^2(\phi-\eta)
      + O\!\bigl(\sqrt{\tau}\bigr),
  \nonumber
\end{equation}
and hence
\begin{align}
  \partial_t\phi
    &= D_c\,\nabla^2\!\bigl[d\,\phi + (1-d)\,\hat\mu(\phi)\bigr]
    \nonumber\\
    &\quad
       - \tau\,D_c\,D_m\,\nabla^4(\phi-\eta)
       + O\!\bigl(\sqrt{\tau}\bigr).
  \label{eq:phi_pre_final_app}
\end{align}
To simplify the first term, recall the definition of the modified chemical potential, Eq.~\eqref{eq:def-of-mu-hat}: $\hat\mu(\phi)=\mu(\phi)-(D_m/M)\,(\phi-\phi_{\mathrm{eq}})$.
Using $1-d = M/D_c$ and expanding yields
\begin{equation}
  D_m\,\nabla^2\phi + (D_c-D_m)\,\nabla^2\hat{\mu}(\phi)
    = M\,\nabla^2\mu(\phi)\,,
  \label{eq:Dc_identity_app}
\end{equation}
since $D_c\bigl[d\,\phi+(1-d)\,\hat\mu(\phi)\bigr] = D_m\,\phi + M\,\hat\mu(\phi)$ and the $D_m\,\nabla^2\phi$ terms cancel.
Equation~\eqref{eq:phi_pre_final_app} therefore becomes
\begin{equation}
  \partial_t\phi
    = M\,\nabla^2\mu(\phi)
      - \tau\,D_m\,D_c\,\nabla^4(\phi-\eta)
      + O\!\bigl(\sqrt{\tau}\bigr).
  \label{eq:phi_reduced_app}
\end{equation}
Thus the $\kappa$-free reaction--diffusion system generates an effective fourth-order term through the spatial correction in the slow-manifold relation~\eqref{eq:m_slaving_app}.

\smallskip

\paragraph*{Steady-state matching and parameter relation.}
For time-independent states, $\partial_t\phi=0$ in Eq.~\eqref{eq:phi_eta_identity_app} implies $D_c\,\nabla^2\eta=0$, so that $\eta=\mathrm{const}$ on a connected domain with periodic or no-flux boundary conditions.
In particular, $\nabla^4\eta = 0$, and Eq.~\eqref{eq:phi_reduced_app} reduces to
\begin{equation}
  0 = M\,\nabla^2\mu(\phi)
      - \tau\,D_m\,D_c\,\nabla^4\phi\,.
\end{equation}
Comparing with the stationary Cahn--Hilliard equation, $0=M\,\nabla^2\mu(\phi)-M\kappa\,\nabla^4\phi$, yields the matching condition
\begin{equation}
  \tau
    = \frac{M\,\kappa}{D_m\,D_c}
    = \frac{M\,\kappa}{D_m\,(M+D_m)}\,.
  \label{eq:tau_match_app}
\end{equation}
With this identification, the $\kappa$-free dual reproduces the exact Cahn--Hilliard steady-state equation, including interfacial profiles.

\smallskip
\paragraph*{Reduction to the Cahn--Hilliard model.}
Imposing the matching condition~\eqref{eq:tau_match_app} in the reduced $\phi$-dynamics~\eqref{eq:phi_reduced_app} gives
\begin{equation}
  \partial_t\phi
    = \mathcal{L}[\phi]
      + M\kappa\,\nabla^4\eta
      + O\!\bigl(\sqrt{\tau}\bigr),
  \label{eq:phi_modelB_plus_eta_app}
\end{equation}
where
\begin{equation}
  \mathcal{L}[\phi]
    \equiv M\,\nabla^2\mu(\phi) - M\kappa\,\nabla^4\phi
\end{equation}
is the Cahn--Hilliard operator.
To bring this into closed form, we eliminate $\nabla^4\eta$ using the exact identity~\eqref{eq:phi_eta_identity_app}, $\partial_t\phi=D_c\,\nabla^2\eta$, which implies
\begin{equation}
  \nabla^4\eta
    = \frac{1}{D_c}\,\nabla^2\partial_t\phi\,.
  \label{eq:eta_elim_app}
\end{equation}
The $O\!\bigl(\sqrt{\tau}\bigr)$ remainder in Eq.~\eqref{eq:phi_modelB_plus_eta_app} arises from the omitted $-\tau\,\partial_t m$ term in the slaving relation~\eqref{eq:m_slaving_app}.
This omitted term contributes to $\partial_t\phi$ through Eq.~\eqref{eq:phi_via_m_app} as
\begin{equation}
  D_c\,(1-d)\,\tau\;\nabla^2\partial_t m\,,
  \label{eq:omitted_error_app}
\end{equation}
where the factor $(1-d)$ comes from the prefactor in Eq.~\eqref{eq:phi_via_m_app} and $\nabla^2$ from the Laplacian acting on it.
Using Eq.~\eqref{eq:m_from_phi_eta_app} to write $\partial_t m = (\partial_t\phi - \partial_t\eta)/(1-d)$, we rewrite this contribution as
\begin{equation}
  D_c\,\tau\,\bigl(\nabla^2\partial_t\phi - \nabla^2\partial_t\eta\bigr)
    = \frac{M\kappa}{D_m}\,\nabla^2\partial_t\phi
      \;-\;
      \frac{M\kappa}{D_m}\,\nabla^2\partial_t\eta\,,
  \label{eq:omitted_split_app}
\end{equation}
where we used $D_c\,\tau = M\kappa/D_m$ from Eq.~\eqref{eq:tau_match_app}.
The first term on the right-hand side of Eq.~\eqref{eq:omitted_split_app} is $O\!\bigl(\sqrt{\tau}\bigr)$ and must be retained; together with the contribution from Eq.~\eqref{eq:eta_elim_app}, it produces the implicit operator below.
The second term is of order $O(\tau)$: using Eq.~\eqref{eq:phi_eta_identity_app} one finds $\nabla^2\partial_t\eta = D_c^{-1}\,\partial_t^2\phi$, which is $O\!\bigl(\sqrt{\tau}\bigr)$ since $D_c^{-1}=O\!\bigl(\sqrt{\tau}\bigr)$ and $\partial_t^2\phi=O(1)$; multiplied by the prefactor $M\kappa/D_m = O\!\bigl(\sqrt{\tau}\bigr)$, the product is $O(\tau)$.

Collecting the $\nabla^2\partial_t\phi$ contributions from Eqs.~\eqref{eq:eta_elim_app} and~\eqref{eq:omitted_split_app} and substituting into Eq.~\eqref{eq:phi_modelB_plus_eta_app} gives the implicit form
\begin{equation}
  \bigl(1-\ell_m^2\,\nabla^2\bigr)\,\partial_t\phi
    = \mathcal{L}[\phi]
      + O(\tau)\,,
  \label{eq:implicit_eps_app}
\end{equation}
with the diffusion length
\begin{equation}
  \ell_m^2
    \;\equiv\;
    M\kappa\!\left(\frac{1}{D_c}+\frac{1}{D_m}\right)
    = (D_m+D_c)\,\tau\,.
  \label{eq:ell_def_app}
\end{equation}
For regular solutions on $\tau$-independent length scales and $\ell_m\ll 1$, we may invert
$1-\ell_m^2\,\nabla^2$ perturbatively to obtain
\begin{equation}
  \partial_t\phi
    = \mathcal{L}[\phi]
      + \ell_m^2\,\nabla^2\mathcal{L}[\phi]
      + O(\ell_m^4) + O(\tau)\,.
  \label{eq:explicit_dev_app}
\end{equation}
Thus the $\kappa$-free dual approaches the Cahn--Hilliard model as $\tau\to0$, with an explicit $O\!\bigl(\sqrt{\tau}\bigr)$ correction.

\smallskip

\paragraph*{Including birth--death kinetics.}
The reduction above also applies when the dual system is supplemented by a
nonconservative source term that depends only on the total density
$\phi=c+m$. A representative example is
\begin{align}
  \partial_t c
    &= D_c\,\nabla^2 c
       - \frac{1}{\tau}\bigl[c-\hat\mu(\phi)\bigr]
       + s(\phi),
  \label{eq:dual_c_tau_reac_app}\\
  \partial_t m
    &= D_m\,\nabla^2 m
       + \frac{1}{\tau}\bigl[c-\hat\mu(\phi)\bigr].
  \label{eq:dual_m_tau_reac_app}
\end{align}
Summing the two equations eliminates the fast reaction term and yields the
evolution equation for $\phi$,
\begin{align}
  \partial_t \phi
    &= D_c\,\nabla^2 c + D_m\,\nabla^2 m + s(\phi)
    \nonumber\\
    &= D_c\,\nabla^2\eta + s(\phi),
  \label{eq:phi_eta_identity_reac_app}
\end{align}
where $\eta\equiv c + d\,m$ as before. Thus $s(\phi)$ acts as a birth--death
term for the total density, while the diffusive contribution remains purely
redistributive. In particular, for periodic or no-flux boundary conditions on a connected domain
$\Omega$, integrating Eq.~\eqref{eq:phi_eta_identity_reac_app} gives
\begin{equation}
  \frac{\mathrm{d}}{\mathrm{d}t}\int_\Omega \phi\;\mathrm{d}\boldsymbol{x}
    \;=\;
    \int_\Omega s(\phi)\;\mathrm{d}\boldsymbol{x}\,,
  \label{eq:global_mass_balance_reac_app}
\end{equation}
since $\int_\Omega \nabla^2\eta\;\mathrm{d}\boldsymbol{x}=0$. Hence the source
term is the sole mechanism that changes the total mass, whereas diffusion
merely redistributes $\phi$ within $\Omega$.


\section{Multi-component Cahn--Hilliard dynamics}
\label{app:CH_multicomponent}

We consider a multi-component system within the potential--current class
[Eqs.~\eqref{eq:mu_n_continuity}--\eqref{eq:mu_n_potential_form}] and, for simplicity, take constant diagonal mobilities,
${M_{\alpha\beta}=M_\alpha\,\delta_{\alpha\beta}}$ with ${M_\alpha>0}$.
We assume that the chemical potentials are of the form
\begin{equation}
  \mu_{\alpha,\mathrm{tot}}[\boldsymbol{\phi}]
    = \mu_\alpha(\boldsymbol{\phi})
      - \kappa_\alpha\,\nabla^2 \phi_\alpha\,,
  \label{eq:mu_tot_multicomponent}
\end{equation}
where the local part ${\mu_\alpha(\boldsymbol{\phi})}$ may or may not exhibit Maxwell symmetry, and the term ${-\kappa_\alpha\,\nabla^2\phi_\alpha}$ provides the usual square-gradient (Cahn--Hilliard-type) regularisation that penalises spatial variations of component ${\alpha}$ (with ${\kappa_\alpha>0}$), thereby setting a finite interfacial width and ensuring well-posed continuum dynamics.
The resulting multi-component Cahn--Hilliard dynamics reads
\begin{equation}
  \partial_t \phi_\alpha
    = M_\alpha\,\nabla^2\!\bigl(
        \mu_\alpha(\boldsymbol{\phi})
        - \kappa_\alpha\,\nabla^2 \phi_\alpha\bigr).
  \label{eq:MB_multicomponent_kappa}
\end{equation}

\paragraph*{Dual variables and embedding.}
Building on the single-component construction presented earlier, we now seek a dual multi-component McRD model whose \emph{slow} dynamics, in the fast-reaction limit, reproduces the multi-component Cahn--Hilliard equation.
We introduce the dual variables
\begin{align}
  c_\alpha
    &\equiv \mu_\alpha(\boldsymbol{\phi}),
  \\
  m_\alpha
    &\equiv \phi_\alpha - \mu_\alpha(\boldsymbol{\phi})
    = \phi_\alpha - c_\alpha,
  \label{eq:def_cm_multicomponent}
\end{align}
so that the total conserved density is recovered as
\begin{equation}
  \phi_\alpha = c_\alpha + m_\alpha\,.
  \label{eq:phi_split}
\end{equation}
The map
$\Phi:\{\phi_\alpha\}\mapsto \{(c_\alpha,m_\alpha)\}$
defined by Eq.~\eqref{eq:def_cm_multicomponent} is injective:
if $\Phi(\boldsymbol{\phi})=\Phi(\boldsymbol{\phi}')$, then $c_\alpha=c'_\alpha$ and $m_\alpha=m'_\alpha$ for all $\alpha$,
hence $\phi_\alpha=m_\alpha+c_\alpha=m'_\alpha+c'_\alpha=\phi'_\alpha$.
Therefore, the image of $\Phi$ is a well-defined (non-self-intersecting) manifold in $(\mathbf{c},\mathbf{m})$-space.

\smallskip

\paragraph*{A two-state reaction--diffusion dual and the fast-reaction limit.}
A dual representation of Eq.~\eqref{eq:MB_multicomponent_kappa} is obtained by promoting
$c_\alpha$ and $m_\alpha$ to dynamical fields coupled by local interconversion reactions
\begin{subequations}\label{eq:dual_kappa_free_multicomponent}
\begin{align}
  \partial_t c_\alpha
    &= D_\alpha^c\,\nabla^2 c_\alpha
       - \hat A_\alpha(\mathbf{c},\mathbf{m}),
  \label{eq:dual_kappa_free_c_alpha}\\
  \partial_t m_\alpha
    &= D_\alpha^m\,\nabla^2 m_\alpha
       + \hat A_\alpha(\mathbf{c},\mathbf{m}),
  \label{eq:dual_kappa_free_m_alpha}
\end{align}
\end{subequations}
where $\hat A_\alpha$ is an internal conversion flux that exchanges mass between $c_\alpha$ and $m_\alpha$ while preserving $\phi_\alpha$.
In direct analogy to the scalar case, we choose the conversion to relax $c_\alpha$ toward a local target value $\hat\mu_\alpha(\boldsymbol{\phi})$,
\begin{equation}
  \hat A_\alpha(\mathbf{c},\mathbf{m})
    = \frac{1}{\tau_\alpha}\,
      \bigl[c_\alpha - \hat\mu_\alpha(\boldsymbol{\phi})\bigr],
  \label{eq:Ahat_def_multicomponent}
\end{equation}
with $\tau_\alpha>0$ the (small) reaction time scales.

Before establishing the duality, it is useful to recall the geometric structure of typical phase-separating patterns.
They are composed of extended plateau-like bulk regions, where the fields vary only weakly, separated by narrow interfacial zones in which the fields change rapidly and gradients are large.
This separation of regimes suggests a natural route to the duality: we construct the reaction--diffusion model such that, in the fast-reaction limit, its slow dynamics reproduces the Cahn--Hilliard equation both in the outer (bulk) regime and in the inner (interfacial) regime.
We then show that this reduction is in fact valid more generally, without relying on an explicit plateau--interface decomposition.

\smallskip
\paragraph*{Matching the bulk dynamics.}
We start with the plateau-like bulk regions, where the fields vary only weakly on a length scale $L$.
In the fast-reaction limit, the conversion dynamics introduces the reaction--diffusion length
$\ell_\alpha^c \equiv\sqrt{D_\alpha^{c}\,\tau_\alpha}$, and the outer regime corresponds to $\ell_\alpha^c \ll L$.
On these scales, Eq.~\eqref{eq:dual_kappa_free_c_alpha} enforces local reactive equilibrium between $c_\alpha$ and its setpoint:
\begin{equation}
  c_\alpha
    = \hat\mu_\alpha(\boldsymbol{\phi})
  \label{eq:c_slaving_bulk_lengthscale}
\end{equation}
up to terms of order $(\ell_{\alpha}^c/L)^2$.
To obtain the slow bulk dynamics of the conserved fields $\phi_\alpha=c_\alpha+m_\alpha$, we sum
Eqs.~\eqref{eq:dual_kappa_free_multicomponent} to eliminate the conversion flux:
\begin{equation}
  \partial_t \phi_\alpha
    = D_\alpha^c\,\nabla^2 c_\alpha
      + D_\alpha^m\,\nabla^2 m_\alpha\,.
  \label{eq:phi_dyn_dual_sum_multicomponent}
\end{equation}
Using $\phi_\alpha=c_\alpha+m_\alpha$, this becomes
\begin{equation}
  \partial_t \phi_\alpha
    = D_\alpha^m\,\nabla^2 \phi_\alpha
      + \bigl(D_\alpha^c - D_\alpha^m\bigr)\,\nabla^2 c_\alpha\,.
  \label{eq:phi_dyn_dual_rewrite_multicomponent}
\end{equation}
Substituting the local equilibrium condition~\eqref{eq:c_slaving_bulk_lengthscale} yields the leading-order bulk reduction
\begin{equation}
  \partial_t \phi_\alpha
    = D_\alpha^m\,\nabla^2 \phi_\alpha
      + \bigl(D_\alpha^c - D_\alpha^m\bigr)\,
        \nabla^2 \hat\mu_\alpha(\boldsymbol{\phi}).
  \label{eq:phi_reduced_bulk_multicomponent}
\end{equation}

To ensure that the slow bulk dynamics reproduces
$\partial_t\phi_\alpha=M_\alpha\,\nabla^2\mu_\alpha(\boldsymbol{\phi})$, we identify the effective mobility $M_\alpha$ with the diffusivity contrast,
\begin{equation}
  M_\alpha
    = D_\alpha^c - D_\alpha^m\,,
  \label{eq:Dc_choice_multicomponent}
\end{equation}
and absorb the remaining Fickian term $D_\alpha^m\,\nabla^2\phi_\alpha$ into the setpoint by defining
\begin{equation}
  \hat\mu_\alpha(\boldsymbol{\phi})
    = \mu_\alpha(\boldsymbol{\phi})
      - \frac{D_\alpha^m}{M_\alpha}\,
        (\phi_\alpha - \phi_{\alpha,\mathrm{eq}}),
  \label{eq:mu_hat_multicomponent}
\end{equation}
with arbitrary constants $\phi_{\alpha,\mathrm{eq}}$ (irrelevant since only gradients enter).
With Eqs.~\eqref{eq:Dc_choice_multicomponent} and~\eqref{eq:mu_hat_multicomponent}, Eq.~\eqref{eq:phi_reduced_bulk_multicomponent} reduces to
\begin{equation}
  \partial_t\phi_\alpha
    = M_\alpha\,\nabla^2 \mu_\alpha(\boldsymbol{\phi})
  \label{eq:MB_multicomponent_kappa0_app_matched}
\end{equation}
up to terms of order $(\ell_{\alpha}^c/L)^2$.

\smallskip

\paragraph*{Matching the interface profile.}
Having matched the bulk (long-wavelength) reduced dynamics to the $\kappa_\alpha=0$ Cahn--Hilliard equation, the next step
is to match the \emph{interfacial} structure, i.e., to identify the condition under which the $\kappa$-free dual
generates the effective regularising term $-M_\alpha\kappa_\alpha\nabla^4\phi_\alpha$ in
Eq.~\eqref{eq:MB_multicomponent_kappa}. To this end, we consider stationary profiles of the conserved totals
($\partial_t\phi_\alpha=0$).
Summing Eqs.~\eqref{eq:dual_kappa_free_multicomponent} and using $c_\alpha = \phi_\alpha -m_\alpha$ then gives the flux-balance relation
\begin{equation}
  0 = D_\alpha^c\,\nabla^2(\phi_\alpha - m_\alpha)
      + D_\alpha^m\,\nabla^2 m_\alpha\,.
\end{equation}
Using $D_\alpha^c - D_\alpha^m = M_\alpha$ implies
\begin{equation}
  \nabla^2 m_\alpha
    = \frac{D_\alpha^c}{M_\alpha}\,\nabla^2\phi_\alpha\,.
  \label{eq:lapm_fluxbalance_multicomponent}
\end{equation}
To extract the emergent interfacial stiffness, we now analyse stationary profiles.
In steady state, the profile equation for the membrane densities $m_\alpha$, Eq.~\eqref{eq:dual_kappa_free_m_alpha}, becomes
\begin{equation}
  0 = D_\alpha^m\,\nabla^2 m_\alpha
      + \frac{1}{\tau_\alpha}\,
        \bigl[\phi_\alpha - m_\alpha
              - \hat\mu_\alpha(\boldsymbol{\phi})\bigr].
  \label{eq:m_steady_eq_multicomponent}
\end{equation}
Applying $\tau_\alpha\,\nabla^2$ yields
\begin{equation}
  0 = D_\alpha^m\,\tau_\alpha\,\nabla^4 m_\alpha
      + \nabla^2\phi_\alpha
      - \nabla^2 m_\alpha
      - \nabla^2\hat\mu_\alpha(\boldsymbol{\phi}).
  \label{eq:lap_profile_multicomponent}
\end{equation}
Inserting the bulk-matching condition~\eqref{eq:mu_hat_multicomponent} and the stationary flux-balance identity~\eqref{eq:lapm_fluxbalance_multicomponent} yields
\begin{equation}
  0 = M_\alpha\,\nabla^2\mu_\alpha(\boldsymbol{\phi})
      - \tau_\alpha\,D_\alpha^m\,D_\alpha^c\,\nabla^4\phi_\alpha\,.
  \label{eq:phi_eff_multicomponent}
\end{equation}
Comparing with the steady-state form of the Cahn--Hilliard equation,
\begin{equation}
  0 = M_\alpha\,\nabla^2\mu_\alpha(\boldsymbol{\phi})
      - M_\alpha\,\kappa_\alpha\,\nabla^4\phi_\alpha\,,
  \label{eq:MB_steady_multicomponent}
\end{equation}
we identify the componentwise stiffness generated by the $\kappa$-free dual,
\begin{equation}
  \kappa_\alpha
    = \frac{\tau_\alpha\,D_\alpha^m\,D_\alpha^c}{M_\alpha}\,.
  \label{eq:kappa_match_multicomponent}
\end{equation}
Equivalently, in terms of $d_\alpha \equiv D_\alpha^m/D_\alpha^c\in[0,1)$,
\begin{equation}
  \kappa_\alpha
    = \frac{D_\alpha^m\,\tau_\alpha}{1-d_\alpha}
    = \frac{\tau_\alpha\,d_\alpha\,M_\alpha}{(1-d_\alpha)^2}\,.
  \label{eq:kappa_match_multicomponent_d}
\end{equation}

\smallskip

\paragraph*{Scaling of diffusivities in the fast-reaction limit.}
Requiring that the reduced coefficients $M_\alpha$ and $\kappa_\alpha$ remain finite as $\tau_\alpha\to0$
constrains how the microscopic diffusion constants scale in the McRD representation.
In particular, Eq.~\eqref{eq:kappa_match_multicomponent_d} implies that the diffusivity ratio
$d_\alpha\equiv D_\alpha^m/D_\alpha^c$ approaches unity as
\begin{equation}
  1 - d_\alpha
    = \mathcal{O}\!\bigl(\sqrt{\tau_\alpha}\bigr),
  \qquad (\tau_\alpha\to0).
  \label{eq:d_scaling_fast_reaction}
\end{equation}
Using $D_\alpha^c - D_\alpha^m = M_\alpha$ and $d_\alpha = D_\alpha^m/D_\alpha^c$, we obtain
\begin{equation}
  D_\alpha^c
    = \frac{M_\alpha}{1-d_\alpha}\,,
  \qquad
  D_\alpha^m
    = \frac{d_\alpha\,M_\alpha}{1-d_\alpha}\,,
  \label{eq:D_scaling_fast_reaction}
\end{equation}
so that both diffusivities diverge as
\begin{equation}
  D_\alpha^c \sim D_\alpha^m
    \sim \tau_\alpha^{-1/2}\,.
  \label{eq:D_asymp_fast_reaction}
\end{equation}
Physically, a finite emergent interfacial stiffness in the fast-reaction limit requires the auxiliary fields to become
rapidly diffusive while remaining nearly equally mobile, such that their difference
$D_\alpha^c - D_\alpha^m = M_\alpha$ stays finite.

\smallskip

\paragraph*{Reduction to Cahn--Hilliard dynamics.}
We now establish the equivalence between the McRD model,
Eqs.~\eqref{eq:dual_kappa_free_multicomponent}--\eqref{eq:Ahat_def_multicomponent}, and the multi-component Cahn--Hilliard equation in a controlled fast-reaction limit.
For convenience we introduce the reaction--diffusion length
\begin{equation}
  \ell_{m,\alpha}^2
    \equiv D_\alpha^m\,\tau_\alpha\,,
  \label{eq:ellm_def_multicomponent}
\end{equation}
and the associated Helmholtz operator $\mathcal{H}_\alpha\equiv 1-\ell_{m,\alpha}^2\,\nabla^2$.
Using $c_\alpha=\phi_\alpha-m_\alpha$ in Eq.~\eqref{eq:dual_kappa_free_m_alpha} and rearranging yields the exact identity
\begin{equation}
  \mathcal{H}_\alpha\,m_\alpha
    = \phi_\alpha - \hat\mu_\alpha(\boldsymbol{\phi})
      - \tau_\alpha\,\partial_t m_\alpha\,.
  \label{eq:Helmholtz_m_identity_multicomponent}
\end{equation}

To derive a closed equation for the conserved field $\phi_\alpha$, we apply the Helmholtz operator $\mathcal{H}_\alpha$ to the summed dynamics~\eqref{eq:phi_dyn_dual_sum_multicomponent}, rewritten using $D_\alpha^c - D_\alpha^m = M_\alpha$ as
\begin{equation}
  \partial_t\phi_\alpha
    = D_\alpha^c\,\nabla^2\phi_\alpha
      - M_\alpha\,\nabla^2 m_\alpha\,.
  \label{eq:phi_dyn_rewritten_multicomponent}
\end{equation}
Since $\mathcal{H}_\alpha$ is a time-independent operator, it commutes with $\partial_t$ and we obtain
\begin{align}
  \mathcal{H}_\alpha\,\partial_t\phi_\alpha
    &= D_\alpha^c\,\nabla^2\phi_\alpha
       - D_\alpha^c\,\ell_{m,\alpha}^2\,\nabla^4\phi_\alpha
    \nonumber\\
    &\quad
       - M_\alpha\,\nabla^2\!\bigl(\mathcal{H}_\alpha\,m_\alpha\bigr).
  \label{eq:H_applied_to_sum_multicomponent}
\end{align}
Substituting the Helmholtz identity~\eqref{eq:Helmholtz_m_identity_multicomponent} into the last term and using the bulk-matching relation~\eqref{eq:mu_hat_multicomponent}, which implies the differential identity
\begin{equation}
  -M_\alpha\,\nabla^2\!\bigl[\phi_\alpha - \hat\mu_\alpha(\boldsymbol{\phi})\bigr]
    = -D_\alpha^c\,\nabla^2\phi_\alpha
      + M_\alpha\,\nabla^2\mu_\alpha(\boldsymbol{\phi})\,,
  \label{eq:bulk_matching_diff_form_reuse}
\end{equation}
the $D_\alpha^c\,\nabla^2\phi_\alpha$ terms cancel, yielding the exact closed identity
\begin{align}
  \bigl(1 - \ell_{m,\alpha}^2\,\nabla^2\bigr)\,\partial_t\phi_\alpha
    &= M_\alpha\,\nabla^2\mu_\alpha(\boldsymbol{\phi})
       - D_\alpha^c\,\ell_{m,\alpha}^2\,\nabla^4\phi_\alpha
    \nonumber \\
    &\quad
       + M_\alpha\,\tau_\alpha\,\nabla^2\partial_t m_\alpha\,,
  \label{eq:phi_reduction_exact_multicomponent}
\end{align}
which isolates all residual dependence on the auxiliary field $m_\alpha$ in the last term.

We now take the fast-reaction limit $\tau_\alpha\to0$ at fixed reduced coefficients $M_\alpha$ and $\kappa_\alpha$.
With the scaling derived above, $\ell_{m,\alpha}^2 = D_\alpha^m\,\tau_\alpha = \mathcal{O}\!\bigl(\tau_\alpha^{1/2}\bigr)$, so the operator correction
$\ell_{m,\alpha}^2\,\nabla^2\partial_t\phi_\alpha$ is parametrically small on pattern scales.
At the same time, the stiffness-matching relation implies
\begin{equation}
  D_\alpha^c\,\ell_{m,\alpha}^2
    = D_\alpha^c\,D_\alpha^m\,\tau_\alpha
    = M_\alpha\,\kappa_\alpha\,,
  \label{eq:Dc_ellm2_equals_Mkappa}
\end{equation}
so the regularising $\nabla^4$ term remains finite.
To control the remainder $M_\alpha\,\tau_\alpha\,\nabla^2\partial_t m_\alpha$, we use $m_\alpha = (\phi_\alpha - \eta_\alpha)/(1-d_\alpha)$ (with $\eta_\alpha \equiv c_\alpha + d_\alpha\,m_\alpha$ the mass-redistribution potential) to write
\begin{equation}
  M_\alpha\,\tau_\alpha\,\nabla^2\partial_t m_\alpha
    = D_\alpha^c\,\tau_\alpha\,
      \nabla^2\partial_t\phi_\alpha
      \;-\;
      D_\alpha^c\,\tau_\alpha\,
      \nabla^2\partial_t\eta_\alpha\,,
  \label{eq:dtm_split_multicomponent}
\end{equation}
where we used $M_\alpha\,\tau_\alpha/(1-d_\alpha) = D_\alpha^c\,\tau_\alpha$.
Absorbing the first term on the right-hand side into the left-hand side of Eq.~\eqref{eq:phi_reduction_exact_multicomponent} modifies the effective diffusion length from $\ell_{m,\alpha}^2 = D_\alpha^m\,\tau_\alpha$ to
\begin{equation}
  \tilde\ell_\alpha^2
    \;\equiv\;
    \ell_{m,\alpha}^2 + D_\alpha^c\,\tau_\alpha
    = (D_\alpha^m + D_\alpha^c)\,\tau_\alpha\,.
  \label{eq:ell_full_multicomponent}
\end{equation}
The second term in Eq.~\eqref{eq:dtm_split_multicomponent} is of order $\mathcal{O}(\tau_\alpha)$: using $\partial_t\phi_\alpha = D_\alpha^c\,\nabla^2\eta_\alpha$ one finds $\nabla^2\partial_t\eta_\alpha = (D_\alpha^c)^{-1}\,\partial_t^2\phi_\alpha$, which is $\mathcal{O}\!\bigl(\tau_\alpha^{1/2}\bigr)$ since $D_\alpha^c = \mathcal{O}(\tau_\alpha^{-1/2})$; multiplied by the prefactor $D_\alpha^c\,\tau_\alpha = \mathcal{O}\!\bigl(\tau_\alpha^{1/2}\bigr)$, the product is $\mathcal{O}(\tau_\alpha)$.
The exact identity~\eqref{eq:phi_reduction_exact_multicomponent} therefore takes the implicit form
\begin{equation}
  \bigl(1 - \tilde\ell_\alpha^2\,\nabla^2\bigr)\,
  \partial_t\phi_\alpha
    = M_\alpha\,\nabla^2\!\bigl[
        \mu_\alpha(\boldsymbol{\phi})
        - \kappa_\alpha\,\nabla^2\phi_\alpha\bigr]
      + \mathcal{O}(\tau_\alpha)\,,
  \label{eq:implicit_form_multicomponent}
\end{equation}
with $\tilde\ell_\alpha^2 = (D_\alpha^m + D_\alpha^c)\,\tau_\alpha$ as defined in Eq.~\eqref{eq:ell_full_multicomponent}.
For regular solutions on $\tau_\alpha$-independent length scales and $\tilde\ell_\alpha\ll 1$, inverting $(1-\tilde\ell_\alpha^2\,\nabla^2)$ perturbatively yields, to leading order,
\begin{equation}
  \partial_t \phi_\alpha
    = M_\alpha\,\nabla^2\!\bigl[
        \mu_\alpha(\boldsymbol{\phi})
        - \kappa_\alpha\,\nabla^2\phi_\alpha\bigr]
      + \mathcal{O}\!\bigl(\tau_\alpha^{1/2}\bigr),
  \label{eq:MB_multicomponent_leading_order}
\end{equation}
which is the multi-component Cahn--Hilliard dynamics.

\section{Interface theory systems with broken mass-conservation}
\label{app:interface_theory_non-conserved}

This Appendix develops the sharp-interface description of Cahn--Hilliard dynamics with a weak source--sink term $s(\phi)$ [Eq.~\eqref{eq:CHR}] and of the corresponding dual McRD formulation with weakly broken mass conservation [Eq.~~\eqref{eq:dual_RD}]. We derive the plateau-scale equations and interfacial matching conditions in a unified framework, establish their equivalence in the fast-relaxation limit of the dual system, and then apply the resulting theory to the one-dimensional mesa geometry to obtain the selected droplet width.

\bigskip

\paragraph*{Sharp-interface limit and reduced dual system.---}
For the droplet calculation it is convenient to work \emph{directly} with the
sharp-interface form of the dual dynamics.  In this limit we set $D_m\to 0$,
so that the auxiliary field $m$ is immobile on the plateau scale, while $c$
remains the sole diffusive component.  The dual system then reduces to
\begin{subequations}
\label{eq:dual_RD_sharp}
\begin{align}
\partial_t c &= D_c \nabla^2 c - A(c,m) + s(\phi),
\label{eq:dual_RD_sharp_c}\\
\partial_t m &= A(c,m),
\label{eq:dual_RD_sharp_m}
\end{align}
\end{subequations}
with $\phi=c+m$.

\smallskip

\subsection*{Equivalence of the dual models \\ in the sharp-interface limit}

We consider the strong-segregation limit, where a droplet (a ``mesa'' in one
dimension) is well approximated by two nearly uniform phases---a dense interior
and a dilute exterior---separated by a narrow interface.  For
Cahn--Hilliard--reaction dynamics the interfacial width is set by the
square-gradient term and scales as $\ell_\kappa\sim\sqrt{\kappa}$; in the dual
formulation the corresponding microscopic width is controlled by the internal
conversion length.

Throughout, we assume \emph{weak} conservation breaking: reactions act only as a
mesoscopic perturbation of phase separation, leaving the plateau values close to
the (curvature-shifted) binodals while generating only slow spatial variations on
length scales large compared to $\ell_\kappa$.

\smallskip

\paragraph*{Stationary profiles in the mesa regimes (CHR).---}
In a steady state, the CHR equation reduces in the sharp interface limit ($\kappa = 0$) to
\begin{align}
0 = M \nabla^2 \mu (\phi) + s(\phi).
\label{eq:app_steady_outer_mu}
\end{align}
Within a given plateau region (inside or outside the droplet), let $\phi_\alpha$
denote the corresponding bulk value and write $\phi=\phi_\alpha+\delta\phi$.
Since gradients are weak on the plateau scale, we expand the \emph{local} chemical potential $\mu(\phi)$ 
as
\begin{align}
\mu(\phi)
=
\mu(\phi_\alpha) + \chi_\alpha\,\delta\phi + O(\delta\phi^2),
\label{eq:app_mu_lin}
\end{align}
with bulk susceptibility
\begin{align}
\chi_\alpha
\equiv
\left.\frac{\partial\mu}{\partial\phi}\right|_{\phi_\alpha}
=
\left.f''(\phi)\right|_{\phi_\alpha}.
\label{eq:app_chi_def}
\end{align}
Likewise, we linearize the reaction term about $\phi_\alpha$,
\begin{align}
s(\phi)
=
s_\alpha + s'_\alpha\,\delta\phi + O(\delta\phi^2),
\label{eq:app_s_lin}
\end{align}
where
\begin{align}
s_\alpha \equiv s(\phi_\alpha),
\qquad
s'_\alpha \equiv s'(\phi_\alpha).
\end{align}

It is convenient to work with the chemical-potential deviation from its plateau
value,
\begin{align}
\delta\mu \equiv \mu-\mu(\phi_\alpha),
\label{eq:app_delta_mu}
\end{align}
so that Eq.~\eqref{eq:app_mu_lin} gives the linear relation
\begin{align}
\delta\mu=\chi_\alpha\,\delta\phi.
\label{eq:app_delta_phi_from_mu}
\end{align}
Substituting Eqs.~\eqref{eq:app_s_lin} and \eqref{eq:app_delta_phi_from_mu} into
Eq.~\eqref{eq:app_steady_outer_mu} yields a closed equation for $\delta\mu$,
\begin{align}
0
=
M \nabla^2 \delta\mu
+
s_\alpha
+
\frac{s'_\alpha}{\chi_\alpha}\,\delta\mu .
\label{eq:app_screened_poisson_mu_1}
\end{align}
Rearranging, we obtain a screened Poisson (Helmholtz) form,
\begin{align}
\nabla^2 \delta\mu
-
\frac{1}{\ell_\alpha^2}\,\delta\mu
=
-\Gamma_\alpha,
\label{eq:app_screened_poisson_mu_2}
\end{align}
with
\begin{align}
\ell_\alpha^2
\equiv
\frac{M\chi_\alpha}{|s'_\alpha|},
\qquad
\Gamma_\alpha
\equiv
\frac{s_\alpha}{M}.
\label{eq:app_def_ell_Gamma}
\end{align}
Here $\ell_\alpha$ is the mesoscopic screening length over which the chemical potential
relaxes back toward its plateau value under restoring kinetics ($s'_\alpha<0$),
while $\Gamma_\alpha$ quantifies the local bias from a nonzero basal source
$s_\alpha$.

\smallskip

\paragraph*{Dual plateau-scale equation in the fast-relaxation limit.---}
We next derive the analogue of Eq.~\eqref{eq:app_screened_poisson_mu_2} from the dual mass-conserving reaction--diffusion formulation with a weak source--sink term, specializing to the limit $D_m=0$ and assuming that the auxiliary field $m$ relaxes
rapidly compared to the droplet-scale dynamics.

Consider a given plateau state $(c_\alpha,m_\alpha)$ (inside or outside the droplet)
and expand
\begin{align*}
c &= c_\alpha + \delta c,
\qquad
m = m_\alpha + \delta m,
\\
\phi &= \phi_\alpha + \delta\phi,
\qquad
\delta\phi = \delta c + \delta m .
\end{align*}
Linearizing the interconversion term $A(c,m)$ around $(c_\alpha,m_\alpha)$ yields
\begin{align}
A(c,m)
=
\hat{A}(c_\alpha,m_\alpha)
+
A_c\,\delta c + A_m\,\delta m,
\label{eq:app_A_lin}
\end{align}
with
\begin{align}
A_c \equiv \partial_c A\big|_\alpha,
\qquad
A_m \equiv \partial_m A\big|_\alpha .
\label{eq:app_Acm_def}
\end{align}
On a plateau, $(c_\alpha,m_\alpha)$ lies on the nullcline $A(c_\alpha,m_\alpha)=0$.
The source--sink term is expanded in the same way as before,
\begin{align}
s(\phi)= s_\alpha + s'_\alpha\,\delta\phi + O(\delta\phi^2),
\label{eq:app_s_lin_dual}
\end{align}
For $D_m=0$ (so that $D_c=M$), the linearized dynamics read
\begin{subequations}
\begin{align}
\partial_t \delta c
&=
M\nabla^2 \delta c
-
A_c\,\delta c
-
A_m\,\delta m
+
s_\alpha + s'_\alpha \delta \phi,
\label{eq:app_dual_lin_c}
\\
\partial_t \delta m
&=
A_c\,\delta c + A_m\,\delta m .
\label{eq:app_dual_lin_m}
\end{align}
\end{subequations}
Assuming fast relaxation of the auxiliary mode, we impose quasi-stationarity of $m$ on
the droplet scale,
\begin{align}
0 = A_c\,\delta c + A_m\,\delta m
\quad\Longrightarrow\quad
\delta m = -\frac{A_c}{A_m}\,\delta c,
\label{eq:app_delta_m_slaving}
\end{align}
so that
\begin{align}
\delta\phi
=
\delta c + \delta m
=
\Bigl(1-\frac{A_c}{A_m}\Bigr)\delta c .
\label{eq:app_delta_phi_delta_c}
\end{align}
Adding Eqs.~\eqref{eq:app_dual_lin_c} and \eqref{eq:app_dual_lin_m} eliminates the $A$-terms identically and gives the closed evolution equation for the total perturbation,
\begin{align}
\partial_t \delta\phi
=
M\nabla^2 \delta c
+
s_\alpha + s'_\alpha\,\delta\phi .
\label{eq:app_dual_sum}
\end{align}
In a steady mesa profile, $\partial_t\delta\phi=0$, and using
Eq.~\eqref{eq:app_delta_phi_delta_c} we obtain a screened-Poisson equation for $\delta c$,
\begin{align}
0
=
M\nabla^2 \delta c
+
s_\alpha
+
s'_\alpha\Bigl(1-\frac{A_c}{A_m}\Bigr)\delta c .
\label{eq:app_dual_steady_c_2}
\end{align}
This is the precise dual counterpart of Eq.~\eqref{eq:app_screened_poisson_mu_1}, with $c$ being the mass-redistribution potential.

\smallskip

\paragraph*{Identification with CHR via the nullcline geometry.---}
In the fast-interconversion limit, the nullcline defines a slow manifold
parameterized by $\phi=c+m$,
\begin{align}
\hat{A}(c,m)=0
\quad\Longrightarrow\quad
c=c_\star(\phi).
\label{eq:app_slow_manifold}
\end{align}
In the duality construction, $c_\star(\phi)$ coincides with the local chemical-potential map,
\begin{align}
c_\star(\phi)=\mu(\phi),
\label{eq:app_c_equals_mu}
\end{align}
so that $c$ is the chemical-potential variable.
Differentiating the constraint $\hat{A}(c,m)=0$ along the nullcline yields
\begin{align}
0
=
\mathrm{d} A
&=
A_c\,\mathrm{d}c + A_m\,\mathrm{d}m
\nonumber \\
&=
A_c\,\mathrm{d}c + A_m(\mathrm{d}\phi-\mathrm{d}c),
\end{align}
and therefore
\begin{align}
\frac{\mathrm{d}c}{\mathrm{d}\phi}
=
\frac{A_m}{A_m-A_c}
=
\frac{1}{1-A_c/A_m}.
\label{eq:app_dc_dphi_from_A}
\end{align}
Using Eq.~\eqref{eq:app_c_equals_mu} gives $\mathrm{d}c/\mathrm{d}\phi=\mu'(\phi)$ and hence
\begin{align}
1-\frac{A_c}{A_m}=\frac{1}{\mu'(\phi)}.
\label{eq:app_E6_derived}
\end{align}
In particular, on plateau $\alpha$ one has
\begin{align}
\delta c = \mu'(\phi_\alpha)\,\delta\phi.
\label{eq:app_delta_c_muprime}
\end{align}
Comparing Eq.~\eqref{eq:app_dual_steady_c_2} to the CHR result
Eq.~\eqref{eq:app_screened_poisson_mu_1} thus amounts to the identifications
\begin{align}
\delta c \equiv \delta\mu,
\qquad
\Bigl(1-\frac{A_c}{A_m}\Bigr)\equiv \frac{1}{\chi_\alpha},
\label{eq:app_identifications}
\end{align}
with $\chi_\alpha=\mu'(\phi_\alpha)$ [cf.\ Eq.~\eqref{eq:app_chi_def}].  Therefore, in the
sharp-interface and fast-relaxation limit, the dual plateau-scale equation is
\emph{identical} to that of CHR when expressed in the chemical-potential variable.

\subsection*{Isolated mesa}
We determine the selected half-width $R$ of a symmetric one-dimensional mesa.  
The dense phase occupies $|x|<R$ and the dilute phase occupies $|x|>R$.  
By symmetry it suffices to work on the half-line $x\ge 0$.  
At the mesa center the diffusive flux must vanish, which gives a no-flux condition for the chemical potential.  
On each plateau (inside/outside), the deviation of the chemical potential from its local plateau value,
$\delta\mu_\alpha\equiv \mu-\mu(\phi_\alpha)$ with $\alpha\in\{\mathrm{in},\mathrm{out}\}$, obeys the screened-Poisson equation derived above,
\begin{equation}
\partial_x^2 \delta\mu_\alpha
-\ell_\alpha^{-2}\,\delta\mu_\alpha
=
-\Gamma_\alpha,
\label{eq:app_1d_helmholtz}
\end{equation}
where $\ell_\alpha$ and $\Gamma_\alpha$ are defined in Eq.~\eqref{eq:app_def_ell_Gamma}.

In the sharp-interface limit, the interface remains locally equilibrated.  
Reactions do not contribute at leading order within the narrow interfacial region.  
In one dimension the curvature vanishes, so there is no Gibbs--Thomson shift of the coexistence chemical potential.  
With our convention $\delta\mu_\alpha\equiv \mu-\mu(\phi_\alpha)$ this implies the interfacial conditions
\begin{equation}
\delta\mu_{\mathrm{in}}(R)=0,
\qquad
\delta\mu_{\mathrm{out}}(R)=0.
\label{eq:app_bc_mu_at_R}
\end{equation}
Symmetry at the mesa center gives
\begin{equation}
\partial_x \delta\mu_{\mathrm{in}}(0)=0,
\label{eq:app_bc_sym_far_in}
\end{equation}
and boundedness in the dilute phase requires $\delta\mu_{\mathrm{out}}(x)$ to remain finite as $x\to\infty$.  

\smallskip

\paragraph*{Solution and radius selection.---}
We solve Eq.~\eqref{eq:app_1d_helmholtz} separately in the dense region $0\le x\le R$ and the dilute region $x\ge R$.  

Inside the mesa, a constant particular solution is $\delta\mu_{\mathrm{in}}^{(p)}=\Gamma_{\mathrm{in}}\ell_{\mathrm{in}}^2$.  
The general solution can be written as
\begin{align*}
&\delta\mu_{\mathrm{in}}(x)
=
\Gamma_{\mathrm{in}}\ell_{\mathrm{in}}^2
+
A_{\mathrm{in}}\cosh\!\left(\frac{x}{\ell_{\mathrm{in}}}\right)
+
B_{\mathrm{in}}\sinh\!\left(\frac{x}{\ell_{\mathrm{in}}}\right),
\\
&\partial_x \delta\mu_{\mathrm{in}}(0)=0
\Longrightarrow
B_{\mathrm{in}}=0,
\\
&\delta\mu_{\mathrm{in}}(R)=0
\Longrightarrow
A_{\mathrm{in}}
=
-\frac{\Gamma_{\mathrm{in}}\ell_{\mathrm{in}}^2}
{\cosh\!\left(R/\ell_{\mathrm{in}}\right)}.
\end{align*}
Substituting back yields the compact form
\begin{equation}
\delta\mu_{\mathrm{in}}(x)
=
\Gamma_{\mathrm{in}}\ell_{\mathrm{in}}^2
\left[
1-\frac{\cosh\!\left(x/\ell_{\mathrm{in}}\right)}
{\cosh\!\left(R/\ell_{\mathrm{in}}\right)}
\right].
\label{eq:app_sol_in_final}
\end{equation}

Outside the mesa, a constant particular solution is $\delta\mu_{\mathrm{out}}^{(p)}=\Gamma_{\mathrm{out}}\ell_{\mathrm{out}}^2$.  
Imposing boundedness as $x\to\infty$ selects the decaying exponential, so that
\begin{align*}
&\delta\mu_{\mathrm{out}}(x)
=
\Gamma_{\mathrm{out}}\ell_{\mathrm{out}}^2
+
A_{\mathrm{out}}
\exp\!\left(-\frac{x-R}{\ell_{\mathrm{out}}}\right),
\\
&\delta\mu_{\mathrm{out}}(R)=0
\Longrightarrow
A_{\mathrm{out}}
=
-\Gamma_{\mathrm{out}}\ell_{\mathrm{out}}^2,
\end{align*}
and therefore
\begin{equation}
\delta\mu_{\mathrm{out}}(x)
=
\Gamma_{\mathrm{out}}\ell_{\mathrm{out}}^2
\left[
1-\exp\!\left(-\frac{x-R}{\ell_{\mathrm{out}}}\right)
\right].
\label{eq:app_sol_out_final}
\end{equation}

\smallskip

\paragraph*{Flux matching and selected width.---}
The diffusive transport current is $J(x)=-M\,\partial_x\delta\mu(x)$.  
Since the interface carries no singular source at leading order, the normal flux is continuous at $x=R$.  
Equivalently,
\begin{equation}
\partial_x\delta\mu_{\mathrm{in}}(R^-)
=
\partial_x\delta\mu_{\mathrm{out}}(R^+).
\label{eq:app_flux_match_deriv}
\end{equation}
Differentiating Eqs.~\eqref{eq:app_sol_in_final} and \eqref{eq:app_sol_out_final} gives
\begin{subequations}
\label{eq:app_derivs_at_R}
\begin{align}
\partial_x\delta\mu_{\mathrm{in}}(R^-)
&=
-\Gamma_{\mathrm{in}}\ell_{\mathrm{in}}
\tanh\!\left(\frac{R}{\ell_{\mathrm{in}}}\right),
\label{eq:app_deriv_in_R}
\\
\partial_x\delta\mu_{\mathrm{out}}(R^+)
&=
\Gamma_{\mathrm{out}}\ell_{\mathrm{out}}.
\label{eq:app_deriv_out_R}
\end{align}
\end{subequations}
Inserting into Eq.~\eqref{eq:app_flux_match_deriv} yields the selection condition
\begin{equation}
-\Gamma_{\mathrm{in}}\ell_{\mathrm{in}}
\tanh\!\left(\frac{R}{\ell_{\mathrm{in}}}\right)
=
\Gamma_{\mathrm{out}}\ell_{\mathrm{out}}.
\label{eq:app_tanh_condition}
\end{equation}
It is useful to introduce the dimensionless ratio
\begin{equation}
\rho
\equiv
\frac{\ell_{\mathrm{out}}}{\ell_{\mathrm{in}}}\,
\frac{\Gamma_{\mathrm{out}}}{-\Gamma_{\mathrm{in}}},
\label{eq:app_rho_def}
\end{equation}
so that Eq.~\eqref{eq:app_tanh_condition} becomes $\tanh(R/\ell_{\mathrm{in}})=\rho$.  
When $0<\rho<1$, the selected half-width is
\begin{equation}
R
=
\ell_{\mathrm{in}}\,
\operatorname{artanh}(\rho).
\label{eq:app_R_artanh}
\end{equation}

\smallskip

\paragraph*{Existence and sign constraints.---}
A finite, stationary mesa requires opposing net source terms in the two phases.  
In the present notation this means
\begin{equation}
\Gamma_{\mathrm{in}}<0,
\qquad
\Gamma_{\mathrm{out}}>0,
\label{eq:app_sign_conditions}
\end{equation}
so that the dense phase is a net sink and the dilute phase a net source.  
This sign structure produces an outward diffusive flux in the dense phase and an inward diffusive flux in the dilute phase, which can balance at the interface.  

Equation~\eqref{eq:app_tanh_condition} admits a solution only if the required flux ratio lies within the range of $\tanh$.  
Equivalently, a finite $R$ exists if and only if
\begin{equation}
0
<
\frac{\ell_{\mathrm{out}}}{\ell_{\mathrm{in}}}\,
\frac{\Gamma_{\mathrm{out}}}{-\Gamma_{\mathrm{in}}}
<1.
\label{eq:app_existence_ineq}
\end{equation}
The upper bound has a direct physical meaning.  
As $\rho\to 1^-$, the interior profile must sustain an interface flux that is as large as the maximal diffusive flux compatible with the screened plateau solution, and the selected width diverges as $R\simeq \tfrac{\ell_{\mathrm{in}}}{2}\ln\!\bigl[2/(1-\rho)\bigr]$.  
For $\rho\ge 1$, the required interfacial flux cannot be supported and the steady isolated mesa ceases to exist (formally $R\to\infty$). 

\smallskip

\paragraph*{Selected width of an isolated mesa under linear reactions.---}
For linear (Langmuir-type) reactions $s(\phi)=-k(\phi-\phi_\ast)$, an isolated one-dimensional mesa selects a finite half-width $R$ through a balance between reaction-driven source--sink fluxes and diffusive transport of chemical potential. Solving the screened-Poisson problem inside and outside the plateau and matching the diffusive flux across the sharp interface gives
\begin{equation}
R
=
\sqrt{\frac{M\chi_+}{k}}\,
\operatorname{artanh}\!\left[
\sqrt{\frac{\chi_-}{\chi_+}}\,
\frac{\phi_\ast-\phi_-}{\phi_+-\phi_\ast}
\right],
\label{eq:app_R_final_summary}
\end{equation}
with $\chi_\pm\equiv\mu'(\phi_\pm)$. In contrast to conservative Cahn--Hilliard dynamics, linear turnover therefore introduces a reaction-controlled length scale, with the prefactor $\sqrt{M\chi_+/k}$ corresponding to the screening length in the dense phase and setting the scaling $R\sim k^{-1/2}$: faster reactions require steeper chemical-potential gradients and thus support only narrower mesas. A finite stationary solution exists only when the argument of $\operatorname{artanh}$ lies strictly between $0$ and $1$. The condition $\phi_-<\phi_\ast<\phi_+$ ensures the required source--sink structure, so that the dilute phase acts as a net producer and the dense phase as a net consumer, while the upper bound represents the maximal diffusive flux that the screened plateau solution can sustain. As this bound is approached, $R$ diverges logarithmically, signalling the loss of a localized mesa and crossover to macroscopically separated domains. For the symmetric Ginzburg--Landau double well with $\phi_\pm=\pm\phi_0$ and $\chi_+=\chi_-=2r$, Eq.~\eqref{eq:app_R_final_summary} reduces to
\begin{equation}
R
=
\sqrt{\frac{2Mr}{k}}\,
\operatorname{artanh}\!\left(
\frac{\phi_\ast+\phi_0}{\phi_0-\phi_\ast}
\right),
\label{eq:app_R_GL_summary}
\end{equation}
which makes the bias requirement explicit: only for $\phi_\ast<0$ does the dilute background supply mass to the degrading droplet and select a finite width, whereas $\phi_\ast=0$ is marginal and yields $R\to\infty$. This provides the one-dimensional sharp-interface realization of externally maintained droplets, where reactions arrest coarsening by coupling source and sink regions through diffusion.

\subsection*{Mesa splitting}

We estimate the mesa half-width $R$ at which a symmetric one-dimensional mesa becomes unstable to splitting.  
The mechanism is that reactions generate a weak but finite spatial variation of the dense-phase plateau, producing a minimum at the mesa center $x=0$.  
If this minimum enters the (passive) spinodal region of the dense phase, the local state becomes linearly unstable and a dilute gap nucleates near $x=0$, leading to splitting into two mesas.  

We employ a quasi-steady approximation.  
On the time scale relevant for the onset of the local instability, we treat the interface positions $x=\pm R$ as fixed and compute the quasi-stationary interior profile.  
If $R$ changes significantly (for instance by shrinkage) before the instability grows, the system can leave the splitting regime without splitting, cf.\ Fig.~\ref{fig3:cont-multi-stab-PD}.  

\paragraph*{Spinodal threshold.}
Let $\phi_+$ denote the dense-phase binodal (plateau) value.  
Let $\phi_{\mathrm{sp}}^+$ denote the upper spinodal density of the passive free-energy density $f(\phi)$,
\begin{subequations}
\label{eq:app_spinodal_def}
\begin{align}
f''(\phi_{\mathrm{sp}}^+) &= 0,
\\
\phi_{\mathrm{sp}}^+ &< \phi_+.
\end{align}
\end{subequations}
We estimate the onset of splitting by the condition that the minimum of the quasi-stationary interior profile reaches $\phi_{\mathrm{sp}}^+$.  

\paragraph*{Quasi-stationary inner profile.}
Inside the mesa ($0\le x\le R$), the chemical-potential deviation satisfies the plateau-scale Helmholtz equation
\begin{align}
\partial_x^2 \delta\mu_{\mathrm{in}}
-\ell_{\mathrm{in}}^{-2}\,\delta\mu_{\mathrm{in}}
=
-\Gamma_{\mathrm{in}}.
\label{eq:app_inner_helmholtz_mu_split}
\end{align}
The boundary conditions are symmetry at the mesa center and local equilibrium at the sharp interface,
\begin{subequations}
\label{eq:app_inner_bc_split}
\begin{align}
\partial_x\delta\mu_{\mathrm{in}}(0) &= 0,
\\
\delta\mu_{\mathrm{in}}(R) &= 0.
\end{align}
\end{subequations}
Solving yields
\begin{align}
\delta\mu_{\mathrm{in}}(x)
=
\Gamma_{\mathrm{in}}\ell_{\mathrm{in}}^2
\left[
1-\frac{\cosh(x/\ell_{\mathrm{in}})}
{\cosh(R/\ell_{\mathrm{in}})}
\right].
\label{eq:app_inner_solution_mu_split}
\end{align}
Using the linear plateau relation $\delta\mu_{\mathrm{in}}=\chi_+\,\delta\phi_{\mathrm{in}}$ with
$\chi_+=\mu'(\phi_+)=f''(\phi_+)$, we obtain
\begin{align}
\delta\phi_{\mathrm{in}}(x)
=
\frac{\Gamma_{\mathrm{in}}\ell_{\mathrm{in}}^2}{\chi_+}
\left[
1-\frac{\cosh(x/\ell_{\mathrm{in}})}
{\cosh(R/\ell_{\mathrm{in}})}
\right].
\label{eq:app_inner_solution_phi_split}
\end{align}
The minimum occurs at $x=0$,
\begin{align}
\phi_{\min}
=
\phi_+ + \delta\phi_{\mathrm{in}}(0)
=
\phi_+
+
\frac{\Gamma_{\mathrm{in}}\ell_{\mathrm{in}}^2}{\chi_+}
\left[
1-\frac{1}{\cosh(R/\ell_{\mathrm{in}})}
\right].
\label{eq:app_phi_min_split}
\end{align}

\paragraph*{Splitting condition and threshold width.}
We impose the splitting threshold $\phi_{\min}=\phi_{\mathrm{sp}}^+$.  
Introduce the positive quantity (for $\Gamma_{\mathrm{in}}<0$)
\begin{align}
\mathcal{A}
\equiv
\frac{-\Gamma_{\mathrm{in}}\ell_{\mathrm{in}}^2}{\chi_+}
=
\frac{-s(\phi_+)}{|s'_+|},
\qquad
\mathcal{A}>0,
\label{eq:app_A_def_split}
\end{align}
where we used $\Gamma_{\mathrm{in}}=s(\phi_+)/M$ and
$\ell_{\mathrm{in}}^2=M\chi_+/|s'_+|$.  
Then Eq.~\eqref{eq:app_phi_min_split} becomes
\begin{align}
\phi_{\min}
=
\phi_+ - \mathcal{A}
\left[
1-\frac{1}{\cosh(R/\ell_{\mathrm{in}})}
\right].
\label{eq:app_phi_min_Aform}
\end{align}
The threshold condition $\phi_{\min}=\phi_{\mathrm{sp}}^+$ gives
\begin{align}
\phi_+ - \phi_{\mathrm{sp}}^+
=
\mathcal{A}
\left[
1-\frac{1}{\cosh(R_{\mathrm{split}}/\ell_{\mathrm{in}})}
\right].
\label{eq:app_split_condition_A}
\end{align}
A finite splitting threshold exists iff the right-hand side can reach the left-hand side, i.e.
\begin{align}
\mathcal{A}
>
\phi_+ - \phi_{\mathrm{sp}}^+.
\label{eq:app_split_existence_A}
\end{align}
Solving Eq.~\eqref{eq:app_split_condition_A} for $R_{\mathrm{split}}$ yields
\begin{subequations}
\label{eq:app_R_split_general_twocol}
\begin{align}
\cosh\!\left(\frac{R_{\mathrm{split}}}{\ell_{\mathrm{in}}}\right)
&=
\frac{\mathcal{A}}{\mathcal{A}-(\phi_+ - \phi_{\mathrm{sp}}^+)},
\label{eq:app_R_split_cosh_twocol}
\\
R_{\mathrm{split}}
&=
\ell_{\mathrm{in}}\,
\operatorname{arcosh}\!\left(
\frac{\mathcal{A}}{\mathcal{A}-(\phi_+ - \phi_{\mathrm{sp}}^+)}
\right).
\label{eq:app_R_split_final_twocol}
\end{align}
\end{subequations}
The corresponding full mesa width at splitting is $2R_{\mathrm{split}}$.  

\medskip

\paragraph*{Specialization to linear (Langmuir-type) kinetics.}
For Langmuir-type conversion,
\begin{subequations}
\label{eq:app_langmuir_kinetics_k}
\begin{align}
s(\phi)
&=
-k(\phi-\phi_\ast),
\label{eq:app_langmuir_s_k}
\\
k
&\equiv
k_+ + k_-,
\qquad
\phi_\ast
=
\frac{k_+}{k}.
\label{eq:app_langmuir_k_phi_star}
\end{align}
\end{subequations}
Evaluated at the dense-phase plateau $\phi_+$ this gives
\begin{subequations}
\label{eq:app_langmuir_plateau_quantities_k}
\begin{align}
s'_+
&=
-k,
\label{eq:app_langmuir_sprime_k}
\\
s(\phi_+)
&=
-k(\phi_+-\phi_\ast).
\label{eq:app_langmuir_s_at_phi_plus}
\end{align}
\end{subequations}
Hence the inner screening length and the driving combination become
\begin{subequations}
\label{eq:app_langmuir_split_simplifications_k}
\begin{align}
\ell_{\mathrm{in}}^2
&=
\frac{M\chi_+}{k},
\label{eq:app_langmuir_ell_in_k}
\\
-\Gamma_{\mathrm{in}}\ell_{\mathrm{in}}^2
&=
-\frac{s(\phi_+)}{M}\,\ell_{\mathrm{in}}^2
=
\chi_+(\phi_+ - \phi_\ast).
\label{eq:app_langmuir_Gamma_ell_k}
\end{align}
\end{subequations}
Inserting into Eq.~\eqref{eq:app_R_split_final_twocol} yields a compact expression for
the splitting threshold,
\begin{align}
R_{\mathrm{split}}
&=
\sqrt{\frac{M\chi_+}{k}}\,
\operatorname{arcosh}\!\Biggl(
\frac{\phi_+ - \phi_\ast}
{\phi_{\mathrm{sp}}^+ - \phi_\ast}
\Biggr).
\label{eq:app_R_split_k_two_column}
\end{align}
A finite threshold exists iff
\begin{align}
\phi_\ast
<
\phi_{\mathrm{sp}}^+.
\label{eq:app_R_split_existence_k_two_column}
\end{align}

This result is different from \cite{Brauns_Frey:2021}, because in their paper, they expanded in small source strength $\epsilon\equiv k_-$, while we only linearise in the perturbed profile as implied by small $k_-$.

\medskip

\paragraph*{Next order calculation}
As seen in figure \ref{fig3:cont-multi-stab-PD}, the curve given by Eq.~\eqref{eq:app_R_split_k_two_column} underestimates the region of splitting. This is due to the fact that our linearization does not approximate the true profile of the splitting droplet very well.  Instead, one may hope that there exists an expansion in perturbation of the constant inner profile that improves the approximation with each order, up to some potential maximal order. In Cahn-Hilliard, this expansion is not easily tractable:
\begin{align*}
    \delta\dot{\phi} &= M\nabla^2(\mu'(\phi_+)\delta \phi + \frac{1}{2}\mu''(\phi_+)\delta \phi^2) + s(\phi)\\
    &= M\mu'(\phi_+)\nabla^2\delta \phi + \mu''(\phi_+)\left((\nabla\delta \phi)^2 + \delta \phi \nabla^2 \delta\phi \right) + s(\phi).
\end{align*}
No clear information is gained when solving this - potentially numerically.

Instead, the dual system provides a simple alternative. For illustration purposes only, we work with $c$, $m$: Choosing $\phi$ instead of $m$ is possible as well when we eliminate everything except $c$. We expand the constraint equation in $m$, in the sharp interface limit, to quadratic order:
\begin{align}
    \delta \dot{m} = 0 = A_m \delta m + A_c \delta c + \frac{1}{2}A''(\delta m^2 + 2 \delta m \delta c + \delta c^2),
\end{align}
where we used in the last term that for our choice of $A$, $A_{mm}= A_{cc} = A_{cm}$. The index $A_i$ denotes derivatives w.r.t. $i$, while a prime denotes a derivative w.r.t. $\phi$, if no confusion can arise. Solving this quadratic equation for $m$ yields
\begin{align}
    \delta m = -(\beta + \delta c) \pm \sqrt{\beta^2 + \frac{2\delta c}{\mu''}},
\end{align}
where we defined $\beta = \frac{\mu'}{\mu''}$ and inserted all derivatives of $A$ in terms of the chemical potential. The solution that is physical is the one that fulfills $\delta m = 0 \implies \delta c = 0$ (more generally, $\delta c + \delta m = 0 \implies \delta c = 0$). This means the $+$ branch. Note that this condition uniquely selects the branch even for an $n$-th order expansion, so long the coefficients do not yield degenerate roots of the original polynomial in $c(\phi)$. If it does, a regulator can make it non-degenerate again.

Inserting this into the equation for $\delta \phi$ gives (note that our $s$ is linear in $\phi$. Otherwise, an expansion to quadratic order in $\delta \phi$ is necessary. No form of $s$ necessitates a change in approach.):
\begin{align}
    \delta \dot{\phi} &= 0 = M \nabla^2 \delta c + s_0 + s_1 (\delta c + \delta m)\nonumber \\
    &= M \nabla^2 \delta c + s_0 + s_1 \left(\delta c -(\beta + \delta c) + \sqrt{\beta^2 + \frac{2\delta c}{\mu''}}\right).
\end{align}
In 1D, this can be interpreted as a force equation, with a corresponding energy (constant of motion)
\begin{equation}
    E = \frac{M}{2}\delta c_x^2 + V(\delta c),
\end{equation}
\begin{equation}
     V(\delta c) = (s_0 - \beta s_1)\delta c + s_1 \frac{\mu''}{3}\left(\beta^2 + 2\frac{\delta c}{\mu''}\right)^{3/2}.
\end{equation}
(In more than 1D, we instead get a 'time-dependent damping term'. This is fairly general for soliton problems in more than one dimension.)
The boundary conditions (as before) imply
\[
E = (s_0 - \beta s_1)\delta c(0) + s_1 \frac{\mu''}{3}\left(\beta^2 + 2\frac{\delta c(0)}{\mu''}\right)^{3/2}.
\]
and $\delta c(R) = 0$. Our splitting condition $\phi_+ + \delta \phi(0) < \phi_\text{sp}^+$ implies for $\delta c(0)$:
\[
    \delta c = \frac{\mu''}{2}\left[(\phi_\text{sp}^+ - \phi_+ + \beta)^2 - \beta^2 \right].
\]
A separation of variables immediately yields $R_\text{split}(k)$:
\begin{equation}
    R_\text{split} = \int^{\delta c(R) = 0}_{\delta c(0)} \frac{\d \delta c}{\sqrt{\frac{2}{M}(E - V(\delta c)}} \propto \frac{1}{\sqrt{k}}\label{eq:R_split_NO}.
\end{equation}
As can be seen, after decoupling the shrinkage from splitting, this curve much better represents the phase boundary of splitting than the zeroth order formula, Eq.~\eqref{eq:app_R_split_k_two_column}.

One may have noticed that what we are doing is essentially (in the sharp interface limit, if we eliminate $\delta \phi$ instead) inverting the relation $\mu(\phi)$ again. This begs the question why it is possible here, but not in the construction before. The answer is that we are expanding around $\phi_\pm$, with the perturbation \emph{at most} being defined up to the unstable regime, i.e. the point where it becomes non-invertible, but no further. This means we use local invertibility to do perturbation theory, where we know that our domain of interest for the perturbation is exactly the maximal domain where $\mu$ is injective.

\section{Linear stability analysis of the nonreciprocal Cahn--Hilliard system}
\label{app:TW-LSA}

For completeness we summarize the linear stability analysis (LSA) of
the nonreciprocal two-field system~\eqref{eq:tw:orig}, adapting the
notation to the present paper.  The dispersion relation and the
oscillatory-onset conditions derived below are well established; see
Refs.~\citep{Saha_Golestanian:2020,Saha.etal2020, Frohoff_Thiele:2023,Brauns_Marchetti:2024} for earlier presentations.  
The new element is
the explicit connection to the dual McRD formulation introduced in
Sec.~\ref{sec:duality-transformation_CH}.

\subsection*{Dispersion relation}
\label{app:LSA-dispersion}

Because Eqs.~\eqref{eq:tw:orig} are conserved, every spatially uniform
state is a steady state.  We linearize about the symmetric homogeneous
state $(\phi_1,\phi_2)=(0,0)$ by writing $\phi_i=\delta\phi_i$ and
retaining only first-order terms.  From the chemical
potentials~\eqref{eq:tw:pot} at first order,
\begin{equation}
\delta\mu_1=-r\,\delta\phi_1+\alpha_{12}\,\delta\phi_2,
\qquad
\delta\mu_2=\delta\phi_2+\alpha_{21}\,\delta\phi_1.
\label{eq:lsa_dmu}
\end{equation}
Inserting into Eqs.~\eqref{eq:tw:orig} yields
\begin{subequations}
\label{eq:lsa_lin}
\begin{align}
\partial_t\delta\phi_1
&= D_1\,\nabla^2\!\bigl(-r\,\delta\phi_1
   +\alpha_{12}\,\delta\phi_2
   -\kappa\,\nabla^2\delta\phi_1\bigr),
\label{eq:lsa_lin1}\\[4pt]
\partial_t\delta\phi_2
&= D_2\,\nabla^2\!\bigl(\delta\phi_2
   +\alpha_{21}\,\delta\phi_1\bigr).
\label{eq:lsa_lin2}
\end{align}
\end{subequations}
Substituting the normal-mode ansatz
$\delta\phi_i=\hat\phi_i\,\mathrm{e}^{\sigma t+i\mathbf{q}\cdot\mathbf{x}}$
(so that $\nabla^2\to -q^2$) reduces the system to the
$2\times 2$ eigenvalue problem
\begin{equation}
\sigma\,\hat{\boldsymbol\phi}
= -q^2\,A(q)\,\hat{\boldsymbol\phi},
\qquad
\hat{\boldsymbol\phi}
\equiv
\begin{pmatrix}\hat\phi_1\\[2pt] \hat\phi_2\end{pmatrix},
\label{eq:lsa_eig}
\end{equation}
with the wave-number-dependent matrix
\begin{equation}
A(q)=
\begin{pmatrix}
D_1(\kappa q^2-r) & D_1\,\alpha_{12}\\[4pt]
D_2\,\alpha_{21}  & D_2
\end{pmatrix}.
\label{eq:lsa_A}
\end{equation}
The two growth rates are therefore
$\sigma_\pm(q)=-q^2\,\lambda_\pm(q)$, where $\lambda_\pm$ are the
eigenvalues of~$A(q)$.  From the characteristic polynomial
$\lambda^2 - (\mathrm{Tr}\,A)\,\lambda + \mathrm{Det}\,A = 0$ one
obtains
\begin{equation}
\sigma_\pm(q)
= -\frac{q^2}{2}
  \Bigl[\mathrm{Tr}\,A
  \pm\sqrt{(\mathrm{Tr}\,A)^2 - 4\,\mathrm{Det}\,A}\,\Bigr],
\label{eq:lsa_sigma}
\end{equation}
with
\begin{subequations}
\label{eq:lsa_trdet}
\begin{align}
\mathrm{Tr}\,A(q)
  &= D_1(\kappa q^2 - r) + D_2,
\label{eq:lsa_trA}\\[4pt]
\mathrm{Det}\,A(q)
  &= D_1 D_2\bigl(\kappa q^2 - r
     - \alpha_{12}\alpha_{21}\bigr).
\label{eq:lsa_detA}
\end{align}
\end{subequations}

\subsection*{Oscillatory (conserved-Hopf) onset}
\label{app:LSA-onset}

The character of the modes is governed by the discriminant
\begin{equation}
\Delta(q)
\equiv (\mathrm{Tr}\,A)^2 - 4\,\mathrm{Det}\,A.
\label{eq:lsa_disc_def}
\end{equation}
For $\Delta>0$ both growth rates are real (stationary growth or
decay).  For $\Delta<0$ the eigenvalues form a complex-conjugate pair,
and the real part is controlled solely by the trace,
\begin{equation}
\mathrm{Re}\,\sigma_\pm(q)
= -\frac{q^2}{2}\,\mathrm{Tr}\,A(q).
\label{eq:lsa_Re_sigma}
\end{equation}
An oscillatory instability (conserved-Hopf bifurcation) at a finite
wave number~$q_H$ requires two simultaneous conditions:

\emph{(i)~Marginal real part.}
Setting $\mathrm{Re}\,\sigma=0$ demands
$\mathrm{Tr}\,A(q_H)=0$, which by Eq.~\eqref{eq:lsa_trA} gives
\begin{equation}
\kappa\,q_H^2 = r - \frac{D_2}{D_1}\,.
\label{eq:lsa_qH}
\end{equation}
A real solution exists only for $r > D_2/D_1$ (with $\kappa>0$).

\emph{(ii)~Complex eigenvalues.}
The discriminant must be negative at
$q=q_H$.  Substituting
Eqs.~\eqref{eq:lsa_trdet} into
Eq.~\eqref{eq:lsa_disc_def} and using $\mathrm{Tr}\,A(q_H)=0$
yields
\begin{equation}
\Delta(q_H)
= -4\,\mathrm{Det}\,A(q_H)
= 4\,D_2\bigl(D_2 + D_1\,\alpha_{12}\alpha_{21}\bigr),
\label{eq:lsa_disc_qH}
\end{equation}
where in the second step we used
$\kappa q_H^2 - r = -D_2/D_1$ from Eq.~\eqref{eq:lsa_qH}.
Hence $\Delta(q_H)<0$ is equivalent to
\begin{equation}
\alpha_{12}\alpha_{21} < -\frac{D_2}{D_1}\,.
\label{eq:lsa_hopf_cond}
\end{equation}
Rewriting in terms of the dimensionless nonreciprocity
$\gamma\equiv -(D_1/D_2)\,\alpha_{12}\alpha_{21}$ introduced in
Eq.~\eqref{eq:tw:gamma_def}, the Hopf condition becomes $\gamma>1$,
reproducing the onset~\eqref{eq:tw:onset_condition} obtained from the
sharp-interface analysis.  For the antisymmetric parametrization
$\alpha_{12}=-\alpha_{21}\equiv\alpha$, this reduces to
$|\alpha|>\sqrt{D_2/D_1}$.

At threshold the growth rates are purely imaginary,
$\sigma_\pm(q_H)=\pm i\,\omega_H$, with the Hopf frequency
\begin{equation}
\omega_H
= q_H^2
  \sqrt{D_2\bigl(-D_2 - D_1\,\alpha_{12}\alpha_{21}\bigr)}\,.
\label{eq:lsa_omegaH}
\end{equation}
The degenerate boundary $\Delta(q_H)=0$, at which the
complex-conjugate pair collapses into a repeated real eigenvalue, lies
at $\gamma=1$
(equivalently $\alpha_{12}\alpha_{21}=-D_2/D_1$)
and coincides with the codimension-two point where the conserved-Hopf
and the conserved-Turing thresholds merge.

\subsection*{Connection to the dual McRD formulation}
\label{app:LSA-dual}

    In the dual McRD representation~\eqref{eq:tw:dual}, the total
densities $\phi_i=c_i+m_i$ obey
$\partial_t\phi_i = D_i\,\nabla^2 c_i$, and local reactive
equilibrium enforces $c_i=\mu_i$ at leading order.  Linearizing the
dual system about the same homogeneous state and taking
the fast-reaction limit ($A_i\to 0$ faster than any
diffusion time scale) recovers
$\delta c_i = \delta\mu_i$ and hence exactly
Eqs.~\eqref{eq:lsa_lin}; the matrix~$A(q)$ and the
dispersion relation~\eqref{eq:lsa_sigma} are therefore identical.
This equivalence is a linear manifestation of the
slow-manifold reduction discussed in
Sec.~\ref{sec:dual_system}: departures from the constitutive manifold
are damped on the fast time scale~$\tau$, and the remaining slow modes
coincide with those of the chemical-potential dynamics.

\subsection*{LSA estimate for the traveling-wave speed}
\label{app:LSA-velocity}
In the oscillatory regime ($\Delta(q)<0$) the imaginary part of the
growth rate is
\begin{equation}
\mathrm{Im}\,\sigma_\pm(q)
= \mp\,\frac{q^2}{2}\,\sqrt{-\Delta(q)}\,.
\label{eq:lsa_Im_sigma}
\end{equation}
The associated phase velocity of the unstable mode reads
\begin{equation}
v_{\mathrm{LSA}}(q)
= \frac{\bigl|\mathrm{Im}\,\sigma(q)\bigr|}{q}
= \frac{q}{2}\,\sqrt{-\Delta(q)}\,.
\label{eq:lsa_vLSA}
\end{equation}
Evaluating at the Hopf wave number $q_H$ gives
\begin{equation}
v_H
= \frac{\omega_H}{q_H}
= q_H\,\sqrt{D_2\bigl(-D_2 - D_1\,\alpha_{12}\alpha_{21}\bigr)}\,.
\label{eq:lsa_vH}
\end{equation}

For a periodic domain of length~$L$, one evaluates
Eq.~\eqref{eq:lsa_vLSA} at the fundamental Fourier mode
$q=2\pi/L$ to obtain a harmonic-limit estimate for the propagation
speed.  This estimate becomes exact as the mesa width shrinks and the
traveling-wave profile approaches a sinusoidal shape.
Following Ref.~\citep{Brauns_Marchetti:2024}, we extrapolate the LSA velocity using a Stefan-type argument to mesas with finite size.
\[
(\phi_+ - \phi_-)v_H = J_+ + J_- \sim \frac{1}{\Lambda_+} + \frac{1}{\Lambda_-},
\]
such that in the no bulk limit $\Lambda_+ = \Lambda_- = \pi/q_H$, we get Eq.~\eqref{eq:lsa_vH} again. This yields the formula in the main text, Eq.~\eqref{eq:tw:veloc_LSA}.
The sharp-interface
formula~\eqref{eq:tw:v_final} derived in the main text extends this
prediction into the complementary wide-mesa regime, where the profile
is far from sinusoidal.  As shown in
Fig.~\ref{fig:TW-velocity}, the numerically measured speed
interpolates between the two limits: it converges to
$v_{\mathrm{LSA}}$ for narrow mesas and to
Eq.~\eqref{eq:tw:v_final} for wide mesas.

\section{Finite element simulations}
\label{app:all_numerics}

Finite-element simulations were performed with COMSOL Multiphysics v6.3~\cite{COMSOL}, using the PARDISO solver with adaptive time stepping, an implicit fifth-order backward differentiation formula, and a relative tolerance of $10^{-4}$.
All remaining solver settings were kept at COMSOL's defaults.
Complete simulation files are available in a dedicated Zenodo repository, to be made available.

Throughout this appendix we use the nondimensionalization introduced in Sec.~\ref{sec:num:passive-coarsen}: lengths are measured in units of the system size~$L$, time in units of $\tau_\text{diff} \equiv L^2/(Mr)$, the order parameter in units of the binodal amplitude $\phi_0 = \sqrt{r/\lambda}$, and diffusion coefficients in units of $Mr$, so that $\tilde D_{c,m} = D_{c,m}/(Mr)$ and $\tilde\tau = \tau/\tau_\text{diff}$.
The dimensionless interfacial stiffness is $\tilde\kappa = \kappa/(rL^2)$.
The local free-energy density is $f(\phi) = -\tfrac{r}{2}(\phi-\phi_\mathrm{eq})^2 + \tfrac{\lambda}{4}(\phi-\phi_\mathrm{eq})^4$, with chemical potential $\mu(\phi) = -r(\phi-\phi_\mathrm{eq}) + \lambda(\phi-\phi_\mathrm{eq})^3$ and binodals $\phi_\pm = \phi_\mathrm{eq} \pm \sqrt{r/\lambda}$.
The traveling-wave subsection is the only exception: there we hold the gradient length $\ell_\kappa \equiv \sqrt{\kappa/r}$ fixed rather than the system size, and rescale lengths by $\ell_\kappa$ instead.

\subsection{Numerical comparison: Cahn--Hilliard vs.\ $\kappa$-free dual McRD dynamics (Sec.~\ref{sec:num:passive-coarsen})}
\label{app:numerics:passive}

\paragraph*{Equations solved.}
For Figs.~\ref{fig:modelB_vs_dual_spinodal_squarepanels}, \ref{fig:char-length}, and \ref{fig:phase_space_dynamics}, we solve two systems in parallel:
\begin{enumerate}
\item[(i)] the nondimensionalized Cahn--Hilliard equation [Eq.~\eqref{eq:CH_model}],
\begin{equation}
    \partial_t \phi
    = \nabla^{2}\!\left(\phi^{3} - \phi - \tilde\kappa\,\nabla^{2}\phi\right);
\end{equation}
\item[(ii)] the $\kappa$-free dual McRD system [Eq.~\eqref{eq2:dual-kappa-free}], with parameters chosen to satisfy the matching conditions of Eq.~\eqref{eq:kappa-match}.
\end{enumerate}

\paragraph*{Parameters.}
The default values are listed in Table~\ref{tab:params_passiveCH}and apply unless otherwise indicated in the main text or figure captions.
Simulations are performed in two dimensions on a square domain of side $L=200$ with periodic boundary conditions and a maximum mesh element size of~$1$.

\begin{table}[h]
\caption{Parameter values for the passive-coarsening simulations of Figs.~\ref{fig:modelB_vs_dual_spinodal_squarepanels}, \ref{fig:char-length}, and \ref{fig:phase_space_dynamics}.
The third column gives the corresponding dimensionless quantities, with $\tilde\kappa = \kappa/(rL^2)$, $\tilde D_{c,m} = D_{c,m}/(Mr)$, and $\tilde\tau = \tau\,Mr/L^2$.
The dual diffusivity and relaxation time follow from the matching conditions $D_c = D_m + M$ and $\tau = \kappa M/(D_m D_c)$ [Eq.~\eqref{eq:kappa-match}].}
\label{tab:params_passiveCH}
\begin{ruledtabular}
\begin{tabular}{lll}
Parameter & Value & Dimensionless \\
\colrule
$r$               & $1$                  & --- \\
$\lambda$         & $1$                  & --- \\
$\kappa$          & $2$                  & $\tilde\kappa = 5\times 10^{-5}$ \\
$\phi_\text{eq}$  & $2$                  & --- \\
$M$               & $1$                  & --- \\
$D_m$             & $10$                 & $\tilde D_m = 10$ \\
$D_c$             & $11$                 & $\tilde D_c = 11$ \\
$\tau$            & $1/55$               & $\tilde\tau \approx 4.5\times 10^{-7}$ \\
$\phi_0$          & $\phi_\text{eq}-0.5$ & --- \\
$L$               & $200$                & --- \\
\end{tabular}
\end{ruledtabular}
\end{table}

\paragraph*{Initial conditions.}
The homogeneous background $\phi_0 = \phi_\mathrm{eq} - 0.5 = 1.5$ lies in the spinodal regime, between the binodals $\phi_- = 1$ and $\phi_+ = 3$.
Both the Cahn--Hilliard and dual McRD simulations are initialized with the same realization of a spatially random perturbation $\delta\phi(\mathbf{x})$ about $\phi(\mathbf{x},0) = \phi_0$, drawn uniformly from $[-0.5,\,0.5]$ on the simulation grid (identical seed for both systems).
Strictly speaking the Cahn--Hilliard initial condition does not uniquely determine the dual one, since the duality fixes only the total mass $\phi = c+m$.
We checked that the late-time behavior is insensitive to this freedom: initializing the dual system on the nullcline (the canonical choice) and away from it both yield the same coarsening trajectory after a short transient that relaxes onto the nullcline (Fig.~\ref{fig:phase_space_dynamics}, first panel).
The transient is longer for initial conditions farther from the nullcline.

\paragraph*{Sweep for Fig.~\ref{fig:char-length}.}
Figure~\ref{fig:char-length} shows the characteristic length $\Lambda(t)$ for a sweep over the auxiliary diffusivity, $\tilde D_m \in \{0.1,\,1,\,5,\,10,\,20,\,50\}$.
For each $\tilde D_m$, the remaining dual parameters are fixed by the matching conditions $\tilde D_c - \tilde D_m = 1/r$ and $\tilde\kappa = \tilde\tau\,\tilde D_m\,\tilde D_c$  [Eq.~\eqref{eq:kappa-match}].
Increasing $\tilde D_m$ at fixed $\tilde\kappa$ reduces the relaxation time $\tilde\tau$ and brings the dual dynamics closer to the Cahn--Hilliard limit.

\subsection{Dual reaction--diffusion model of Cahn–Hilliard with reactive turnover (Sec.~\ref{sec:num:nmcRD})}
\label{app:numerics:nmcRD}

\paragraph*{Equations solved.}
In Figs.~\ref{fig:active_time_slides} and \ref{fig3:cont-multi-stab-PD} we compare two systems side by side:
\begin{enumerate}
\item[(i)] the Cahn–Hilliard with reactive turnover equation [Eq.~\eqref{eq:CHR}] with a linear source,
\begin{equation}
\begin{split}
\partial_t\phi
&= M\,\nabla^2\!\bigl[-r(\phi-\phi_\mathrm{eq}) + (\phi-\phi_\mathrm{eq})^3 - \kappa\,\nabla^2\phi\bigr] \\
&\quad + k_2 - k_1\phi;
\end{split}
\end{equation}
\item[(ii)] the dual two-field system [Eq.~\eqref{eq:dual_RD}], with parameters satisfying the matching conditions of Eq.~\eqref{eq:kappa-match},
\begin{subequations}
\begin{align}
\partial_t c &= D_c\,\nabla^2 c - \hat A(c,m) + s(\phi),  \\
\partial_t m &= D_m\,\nabla^2 m + \hat A(c,m), 
\end{align}
\end{subequations}
with $\phi = c+m$, $s(\phi) = k_2 - k_1\phi$, and exchange term $\hat A(c,m) = \tau^{-1}\bigl[c - \hat\mu(c+m)\bigr]$ with $\hat\mu$ as in Eq.~\eqref{eq:def-of-mu-hat}.
\end{enumerate}

\paragraph*{Parameters.}
The default values are listed in Table~\ref{tab:params_CHR} and apply unless otherwise indicated in the main text or figure captions.
Simulations are performed on a one-dimensional domain $[-L,\,L]$ (system size $2L=200$ for Fig.~\ref{fig:active_time_slides}) with periodic boundary conditions and a maximum mesh element size of $0.1$.
The dual parameters follow from the matching conditions $D_c = D_m + M = 11$ and $\tau = \kappa M/(D_m D_c) = 1/110$ [Eq.~\eqref{eq:kappa-match}].
The homogeneous fixed point $\phi^* = k_+/k_- = 1.2$ lies between the conservative binodals $\phi_- = 1$ and $\phi_+ = 3$, and the overall reaction strength is controlled by $k_-$.

\begin{table}[h]
\caption{Parameter values for the Cahn–Hilliard with reactive turnover simulations of Figs.~\ref{fig:active_time_slides} and~\ref{fig3:cont-multi-stab-PD}.
The third column gives dimensionless values; $\tilde\kappa$ and $\tilde\tau$ depend on $L$.
$D_c$ and $\tau$ follow from the matching conditions [Eq.~\eqref{eq:kappa-match}].
The source parameters $k_+$, $k_-$ are the baseline values; in the phase-diagram sweep, $k_-$ is varied at fixed $k_+/k_-$.}
\label{tab:params_CHR}
\begin{ruledtabular}
\begin{tabular}{lll}
Parameter & Value & Dimensionless \\
\colrule
$r$               & $1$                       & --- \\
$\kappa$          & $1$                       & $\tilde\kappa = 10^{-4}$ (at $L=100$) \\
$\phi_\mathrm{eq}$ & $2$                       & --- \\
$M$               & $1$                       & --- \\
$D_m$             & $10$                      & $\tilde D_m = 10$ \\
$D_c$             & $11$                      & $\tilde D_c = 11$ \\
$\tau$            & $1/110$                   & $\tilde\tau \approx 9.1\times 10^{-7}$ (at $L=100$) \\
$\phi_0$          & $\phi_\mathrm{eq}-0.1$    & --- \\
$L$               & $100$                     & --- \\
$k_+$             & $1.2\times 10^{-5}$       & --- \\
$k_-$             & $1.0\times 10^{-5}$       & --- \\
\end{tabular}
\end{ruledtabular}
\end{table}

\paragraph*{Sweep for Fig.~\ref{fig:active_time_slides}.}
Figure~\ref{fig:active_time_slides} compares Cahn–Hilliard with reactive turnover profiles (solid) with dual two-field profiles (dashed) at three values of the auxiliary diffusivity, $\tilde D_m \in \{0.1,\,1,\,10\}$.

\paragraph*{Phase-diagram protocol (Fig.~\ref{fig3:cont-multi-stab-PD}).}
For the phase diagram, we scan over two control parameters: the reaction strength $k_-$ (horizontal axis, with $k_+/k_- = 1.2$ held fixed, so that the fixed point $\phi^*$ does not shift) and the initial mesa width $\Lambda_\pm$ (vertical axis, in units of $L$).
We vary the source strength as $k_- \in \{10^{-8},\,10^{-6},\,10^{-4},\,10^{-2}\}$.
The system size is chosen to ensure adequate room for both coarsening and splitting at every parameter point: $L \geq 2\Lambda_+$ (i.e., total system size $2L$ accommodates at least two initial mesas), reaching up to $2L = 4000$ for the largest mesas.
For the hatched region of Fig.~\ref{fig3:cont-multi-stab-PD}, we use a different protocol that decouples droplet shrinkage from droplet splitting: we set the system size to $\Lambda_+$ and impose Dirichlet boundary conditions at $\phi_+$.

\paragraph*{Initial conditions for the phase diagram.}
We use periodically extended step functions with equal high- and low-value lengths set by the vertical axis of Fig.~\ref{fig3:cont-multi-stab-PD}.
The high-density plateau is initialized at the upper binodal $\phi_+$, the low-density plateau at $\phi_- + 0.1$; the latter offset is chosen so that the background coincides neither with the binodal nor with the nullcline value, since the true steady-state background lies between these.
The initial smoothing across each step is $0.1$ (one mesh element), corresponding to a piecewise-linear interpolation.
For the parameters of Table~\ref{tab:params_CHR}, $\phi_+ = 3$ and $\phi_- = 1$, so the low-density plateau is initialized at $1.1$.
For the hatched-region protocol, we instead initialize a homogeneous state at $\phi_+ = 3$ and observe the developing indent until the system-spanning droplet splits.

\paragraph*{Outcome classification.}
Long-time outcomes are classified by tracking the number of high-density mesas over time via visual inspection.
Coarsening corresponds to a net decrease in mesa number (mesas merge or disappear via mass competition).
Mesa splitting corresponds to a net increase (nucleation of a low-density trough within a high-density plateau); in the hatched-region protocol it manifests as a low-density plateau forming between two high-density plateaus at the boundary.
The intermediate regime is defined by conserved mesa number and relaxation toward a stationary mesa pattern.
Stationarity is typically reached at $t \approx 10^7$; we simulate to $t = 10^8$, by which time the adaptive step size has fallen to ${\sim}10^{-6}$, confirming convergence to a stationary state.

\paragraph*{Agreement between the two formulations.}
For every parameter point, simulations were performed for both the Cahn–Hilliard with reactive turnover equation [Eq.~\eqref{eq:CHR}] and the dual system [Eq.~\eqref{eq:dual_RD}].
The resulting phase boundaries and late-time mesa numbers agree, and Fig.~\ref{fig3:cont-multi-stab-PD} therefore shows a single set of symbols.
Quantitatively, the relative $L^2$ difference between the two profiles,
\begin{equation}
e \;\equiv\; \frac{\int_\Omega (c+m-\phi_\text{CH})^2}{\int_\Omega \phi_\text{CH}^2},
\end{equation}
is of order $10^{-12}$ at the chosen $\tilde D_m = 10$ in the late-time stationary state, indicating agreement not only in mesa number but also in mesa position to within numerical precision.

\subsection{Comparison to numerical simulations: nonreciprocal traveling waves (Sec.~\ref{sec:num:TW})}
\label{app:numerics:TW}

\paragraph*{Equations solved.}
We integrate the original nonreciprocal two-component conserved dynamics [Eqs.~\eqref{eq:tw:orig}] directly; the dual McRD representation [Eqs.~\eqref{eq:tw:dual}] is used only for the analytical sharp-interface construction and not for numerical integration. We did verify its matching dynamics numerically, however.

\paragraph*{Parameters.}
The default values are listed in Table~\ref{tab:params_TW} and apply unless otherwise indicated in the main text or figure captions.
Simulations are performed on a one-dimensional domain of length $L = 400$ with periodic boundary conditions and mesh element size $0.1$--$0.05$ (the smaller value used for narrow mesas; see the *Mesh* paragraph below).

\begin{table}[h]
\caption{Parameter values for the traveling-wave simulations of Figs.~\ref{fig:TW-phi-profile}, \ref{fig:TW-velocity}, and \ref{fig:TW-velocity_vs_alpha}.
Figures~\ref{fig:TW-velocity} and~\ref{fig:TW-velocity_vs_alpha} sweep over $\alpha_{21}$ at fixed $\alpha_{12}$.
The default antisymmetric parametrization is $\alpha_{12} = -\alpha_{21} \equiv \alpha$ with $\alpha=1$.}
\label{tab:params_TW}
\begin{ruledtabular}
\begin{tabular}{lll}
Parameter & Value & Dimensionless\\
\colrule
$r$              & $16$ & $1$ \\
$\kappa$         & $1$  & ---\\
$D_1$            & $1$  & ---\\
$D_2$            & $0.1$ & $6.25\times 10^{-3}$\\
$\alpha_{12}$    & $1$   & $1/64$\\
$\alpha_{21}$    & $-1$  & $-4$\\
$\bar\phi_1$     & $0$   & ---\\
$\bar\phi_2$     & $0.2$ & ---\\
\end{tabular}
\end{ruledtabular}
\end{table}

\paragraph*{Derived quantities at default parameters.}
With the antisymmetric parametrization $\alpha_{12} = -\alpha_{21} = 1$, the effective control parameter is $\tilde r = r + \alpha_{12}\alpha_{21} = 16 - 1 = 15$, giving coexistence plateau values $\phi_1^\pm = \pm\sqrt{\tilde r} = \pm\sqrt{15} \approx \pm 3.87$ [Eq.~\eqref{eq:tw:binodals}].
The dimensionless nonreciprocity is $\gamma = D_1\alpha^2/D_2 = 10$ [Eq.~\eqref{eq:tw:gamma_def}], well above the traveling-wave onset at $\gamma=1$ (equivalently $|\alpha| > \sqrt{D_2/D_1} \approx 0.316$ [Eq.~\eqref{eq:tw:onset_condition}]).
The interface width is $\ell_\kappa = \sqrt{\kappa/r} = 0.25$ in simulation units.

\paragraph*{Initial conditions.}
Simulations are initialized with a periodic mesa profile for the phase-separating field,
\begin{equation*}
\phi_1(x,0) \approx
\begin{cases}
+\sqrt{r} & (\text{high-density plateau of width } \Lambda_+),\\
-\sqrt{r} & (\text{low-density plateau of width } \Lambda_-),
\end{cases}
\end{equation*}
smoothed at each interface by a $C^2$ interpolation over ten mesh elements (default mesh size $0.1$).
The coupled field is initialized uniformly at $\phi_2(x,0) = \bar\phi_2 = 0.2$; the choice of nonzero $\bar\phi_2$ is for generality but does not affect the dynamics, as is evident from Eqs.~\eqref{eq:tw:orig} beyond the sharp-interface limit.
The system passes through a transient before locking into a long-lived traveling state, identified as the regime in which the adaptive step size has converged to a constant value; visual inspection of the real-space pattern then confirms uniform translation without change of profile in the comoving frame.

The mesa width is determined from the inflection points of $\phi_1$, as illustrated in Fig.~\ref{fig:TW-phi-profile}.
In the wide-mesa regime $\Lambda_+ \gg \ell_\text{int}$ we track the distinct peak of $\phi_2$ across time frames; in the narrow-mesa regime, where there is no spatial interval over which $\phi_1 = \phi_\pm$, we track the inflection points of $\phi_1$ directly.
Speeds are averaged over the last $21$ time frames; the final time, which ranges from $10^4$ to $10^6$ depending on parameters (faster asymptotic speeds reach the steady state earlier), is chosen so that the system is well within the locked traveling state.
Error bars (typically too small to see in the main-text plots) are estimated from the variance of frame-to-frame speeds.
Time-step choices are similarly parameter-dependent: $\Delta t = 1000$ for slow speeds (to keep discretization errors negligible) and $\Delta t = 100$ for fast speeds (to avoid aliasing across more than half a period and the resulting spurious velocity readout).

The mesh size varies from $0.1$ to $0.05$ depending on $\Lambda_+$; finer meshes are used at low $\Lambda_+/\ell_\text{int}$ to resolve the inflection point accurately.
At the default $\kappa=1$, $r=16$, the interface width is $\ell_\text{int} \approx 2$ (heuristic estimate), giving $\ell_\text{int}/\Delta x = 20$ mesh elements per interface in the wide-mesa regime and twice that in the narrow-mesa regime.

\paragraph*{Sweep for Fig.~\ref{fig:TW-velocity}.}
Figure~\ref{fig:TW-velocity} shows the measured speed as a function of mesa width $\Lambda_+$ for several values of the nonreciprocal coupling $\alpha_{21}$ (equivalently $-\alpha$).
Each curve corresponds to a fixed $\alpha_{21} \in \{-0.225,\, -0.450,\, -0.675,\, -0.900\}$, swept over $\Lambda_+ \in [1.5,\, 50]\,\ell_\text{int}$.

\paragraph*{Sweep for Fig.~\ref{fig:TW-velocity_vs_alpha}.}
Figure~\ref{fig:TW-velocity_vs_alpha} fixes the mesa geometry and sweeps the nonreciprocity magnitude $|\alpha|$.
Initial mesa widths depend on the total mass: $\Lambda_+ = 50$ for $\bar\phi_1 = 0$, $\Lambda_+ = 25$ for $\bar\phi_1 = -2$, $\Lambda_+ = 87.5$ for $\bar\phi_1 = 3$, chosen so that the period $\Lambda = 100$ is the same in all three cases.
The system size is $L = 400$ for moderate $\alpha > 0.2$; for $\alpha < 0.2$ we halve the system width to mitigate critical slowing down (the regime where the square-root scaling in Fig.~\ref{fig:TW-velocity_vs_alpha} becomes dominant).
We verified that the final mesa width matches the initialized one to within numerical precision.
The probed range is $\alpha \in [0.11,\,2]$, sampled at step size $0.02$ on $[0.11,\,0.19]$ and $0.1$ on $[0.2,\,2]$.

\paragraph*{Note on Fig.~\ref{fig:TW-analytical-profiles}.}
The curves in Fig.~\ref{fig:TW-analytical-profiles} are analytical and follow the piecewise-linear construction of the main text; no simulation data are involved.

\section{The converse direction}\label{app:converse_direction}
Here, we show the converse of the statement that every Cahn-Hilliard type dynamics admits a two component reaction diffusion dual; namely that every system of conserved species with two states (components $c$ and $m$) admits a chemical potential description in the fast reaction limit. This makes the dual bidirectional, as the word suggests. To this end we look at the general two component reaction diffusion system
\begin{align}
    \partial_t c &= D_c \nabla^2 c - f(c,m)\\
    \partial_t m &= D_m \nabla^2 m + f(c,m).
\end{align}
We want the nullcline to be a stable manifold everywhere, which is to say
\begin{equation}
    \text{Eigenvalues}\left[ \begin{pmatrix}
    -f_c & -f_m\\
    f_c & f_m
    \end{pmatrix}\right]
    \leq 0
\end{equation}
on all points of the manifold. Here, indices indicate a derivative. One eigenvalue, in the direction of the tangent of the manifold, is 0. The other is $f_m - f_c$ and we want this to be smaller than zero everywhere:
\begin{align}
    f_m - f_c < 0.
\end{align}
What we need to show is that the manifold $f=0$ incorporates the same information as $c-\mu(\phi)=0$ for some function $\mu$, where $\phi$ is again the total mass. Then we can argue along the lines of Sec.~\ref{sec:kappa_removal} to show the existence of a Cahn-Hilliard system. To this end, we note
\begin{equation}
    \frac{\d f(c,\phi-c)}{\d c} = f_c - f_m > 0, 
\end{equation}
by the stability assumption. By the implicit function theorem then, a function $\mu(\phi)$ exists everywhere, such that $f(\mu(\phi), \phi - \mu(\phi))=0$ for all $\phi$. This means we can equivalently rewrite our system of the form
\begin{align}
    \partial_t c &= D_c \nabla^2 c - A(c,\phi)\\
    \partial_t m &= D_m \nabla^2 m + A(c,\phi)\\
    A(c,\phi) &= c - \mu(\phi).
\end{align}
$\mu$ is, up to the usual shift, the chemical potential of the dual Cahn-Hilliard system.

We can extend this to $n$ components. Though here, we will only prove that there exists a bulk chemical potential - the gradient terms are out of scope. To this end, we assume again that the $n$ functions $\{f_i\}_n$ fulfill the properties:
\begin{equation}
    \sum_i^n f_i = 0 \;\land\;
    \text{Eigenvalues}[f_{ij}] \leq 0
\end{equation}
everywhere. Here, $f_{ij} = \partial_{\phi_j}f_i$, where $\phi_j$ are the $n$ components. We assume $f_i=0$ for all $i$ defines a 1d submanifold. (If it is not, then one chooses a subsystem such that it is. We may call such a minimal system where no subsystem is conserved irreducible.)
The first one enforces mass conservation, the second condition is stability. The left eigenvector $(1,\, ...,\,1)$ is the 0 eigenvector of the matrix $f_{ij}$, as seen using mass conservation. A convenient basis of the cokernel $U$ of $f_{ij}$ is
\begin{equation}
    u^{(a)} = (0,\,...,\,-1,\,0,\,...,\,1),
\end{equation}
where the $-1$ is on the $a$-th position, $a<n$. Then, by construction, we know that the matrix $\tilde B$, which is the matrix $f_{ij}$ constrained on $U$, is invertible, because we projected out the only 0 direction. The matrix $\tilde B$ is given by (implied sum)
\begin{equation}
    \tilde B_{ab} = u^{(a)}_i f_{ij} u^{(b)}_j = f_{ab} - f_{an} - (f_{nj} - f_{nn}).
\end{equation}
The second equality comes from explicit calculation. Now we use mass conservation again to recast the second term to obtain
\begin{equation}
    \tilde B_{ab} = B_{ab} + \sum_c^{n-1} B_{cb},
\end{equation}
with
\begin{equation}
    B_{ab} = f_{ab} - f_{an}.
\end{equation}
The matrix $B_{ab}$ has a special meaning: This is precisely the Jacobian of
\begin{equation}
    B_{ab} = \partial_{\phi_b} f_a\left(\{\phi_c\},\phi - \sum_c \phi_c\right),
\end{equation}
where all indices only go up to $n-1$, since the $n$-th $f$ is linearly dependent on the others. $\phi$ is the conserved total mass. Now our goal is to use the implicit function theorem again. That is, we want to proof that $B_{ab}$ is invertible. We already know that $\tilde B$ is invertible. So if there exists an invertible matrix $M$, such that
\begin{equation}
    B = M\tilde B,
\end{equation}
then we are done. We see that
\begin{equation}
    \tilde B = (\mathbbm{1}+\Lambda) B,
\end{equation}
with $\mathbbm{1}$ the $n-1$ unit matrix and $\Lambda$ the matrix with every entry being 1. An inverse is readily found as
\begin{equation}
    M \equiv (\mathbbm{1}+\Lambda)^{-1} = 1 - \frac{1}{n}\Lambda.
\end{equation}
This proves our claim.

Note that these statements, stability of nullcline everywhere and existence of chemical potential, is an equivalence. This allows one to classify algebraically different nullclines by the amount of zero crossings the determinant has. One can further resolve details by looking at which direction admits such zero crossings.

\end{document}